\documentclass[11pt,twoside]{article}
\usepackage{newpasp}
\usepackage{epsf}
\usepackage{graphics}

\newcommand{\pcad}{PCAD}
\newcommand{\chandra}{{\em Chandra}}
\newcommand{\ccd}{CCD}
\newcommand{\psf}{PSF}
\newcommand{\aca}{ACA}

\newcommand{\aciss}{ACIS-S}
\newcommand{\acis}{ACIS}
\newcommand{\hrc}{HRC}

\begin{document}

\title {An Overview of the Performance of the \chandra\ X-ray Observatory\\
}

\author{M.C.Weisskopf}
\affil{NASA/MSFC, SD50, MSFC AL 35812}
\author{T. L. Aldcroft}
\affil{SAO, 60 Garden Street, Cambridge MA 02138}
\author{M. Bautz}
\affil{MIT, Cambridge MA 02139}
\author{R. A. Cameron} 
\affil{SAO, 60 Garden Street, Cambridge MA 02138}
\author{D. Dewey}
\affil{MIT, Cambridge MA 02139}
\author{J. J. Drake}
\affil{SAO, 60 Garden Street, Cambridge MA 02138}
\author{C. E. Grant}
\affil{MIT, Cambridge MA 02139}
\author{H. L.  Marshall}
\affil{MIT, Cambridge MA 02139}
\author{S. S. Murray}
\affil{SAO, 60 Garden Street, Cambridge MA 02138}
 
\begin{abstract}
The \chandra\ X-ray Observatory is the X-ray component of NASA's Great
Observatory
Program which includes the recently launched Spitzer Infrared Telescope, the
Hubble Space Telescope (HST) for
observations in the visible, and the Compton Gamma-Ray Observatory (CGRO)
which, after providing years of useful data has reentered the atmosphere.
All these facilities provide, or provided, scientific data to the international
astronomical community in response to peer-reviewed proposals for their use. 
The \chandra\ X-ray Observatory was the result of the efforts of many academic,
commercial, and government organizations primarily in the United States but also
in Europe. 
NASA's Marshall Space Flight Center (MSFC) manages the Project and provides
Project Science; Northrop Grumman Space Technology (NGST -- formerly TRW) served
as prime contractor responsible for providing the spacecraft, the telescope,
and assembling and testing the Observatory; and the Smithsonian Astrophysical
Observatory (SAO) provides technical support and is responsible for ground
operations including the \chandra\ X-ray Center (CXC).
Telescope and instrument teams at SAO, the Massachusetts Institute of Technology
(MIT), the Pennsylvania State University (PSU), the Space Research Institute of
the Netherlands (SRON), the Max-Planck Instit\"ut f\"ur extraterrestrische
Physik (MPE), and the University of Kiel also provide technical support
to the \chandra\ Project. 
We present here a detailed description of the hardware, its on-orbit
performance, and a brief overview of some of the remarkable discoveries that
illustrate that performance.

\end{abstract}

\section{The Observatory}

\begin{figure} 
\begin{center} 
\epsfysize=8cm
\epsfbox{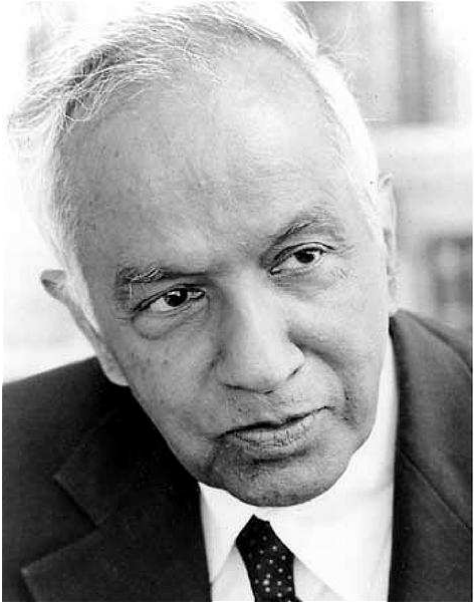} 
\caption{Professor Subramanyan Chandrasekhar.
\label{f:SC}}  
\end{center}
\end{figure}

\begin{figure}
\begin{center} 
\epsfysize=8cm
\epsfbox{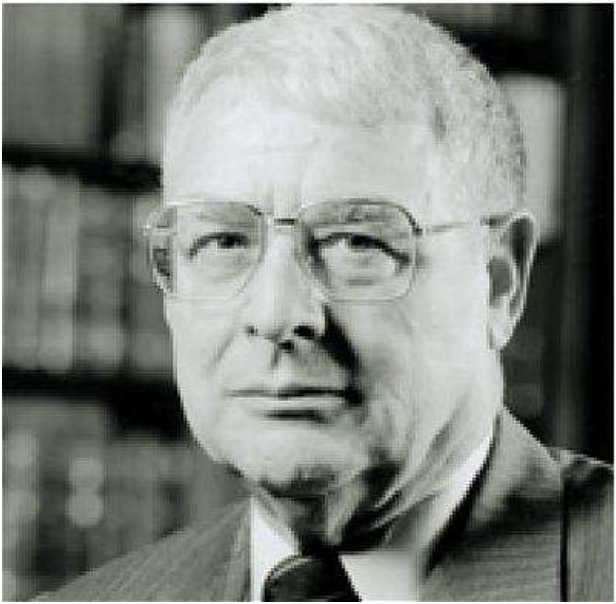} 
\caption{Professor Riccardo Giacconi.
\label{f:RG}}
\end{center}
\end{figure}

In 1977, NASA/MSFC and SAO began a study which led to the definition of the
then named Advanced X-ray Astrophysics Facility. 
This study had been initiated as a result of an unsolicited proposal submitted
to NASA in 1976 by Prof. R. Giacconi (Harvard University and SAO) and Dr. H.
Tananbaum (SAO). 
Subsequently: the project received the highest recommendation by the National
Academy of Sciences Astronomy Survey Committee in the report, "Astronomy and
Astrophysics for the 1980's"; instruments were selected in 1985; the
prime contractor (NGST) was selected in 1988; a demonstration of the ability to
build the flight optics was accomplished in 1991; in 1992 the scope of the
mission was restructured to reduce cost, including eliminating servicing;
the Observatory was named in honor of the Nobel Prize winner Subramanyan
Chandrasekhar (Figure~\ref{f:SC}) in 1998; and the launch occurred the following
year. 
In 2002, Prof. Giacconi (Figure ~\ref{f:RG}) was awarded the Nobel Prize in
Physics for his pioneering work in X-ray astronomy.

\begin{figure}
\begin{center} 
\epsfysize=8cm
\epsfbox{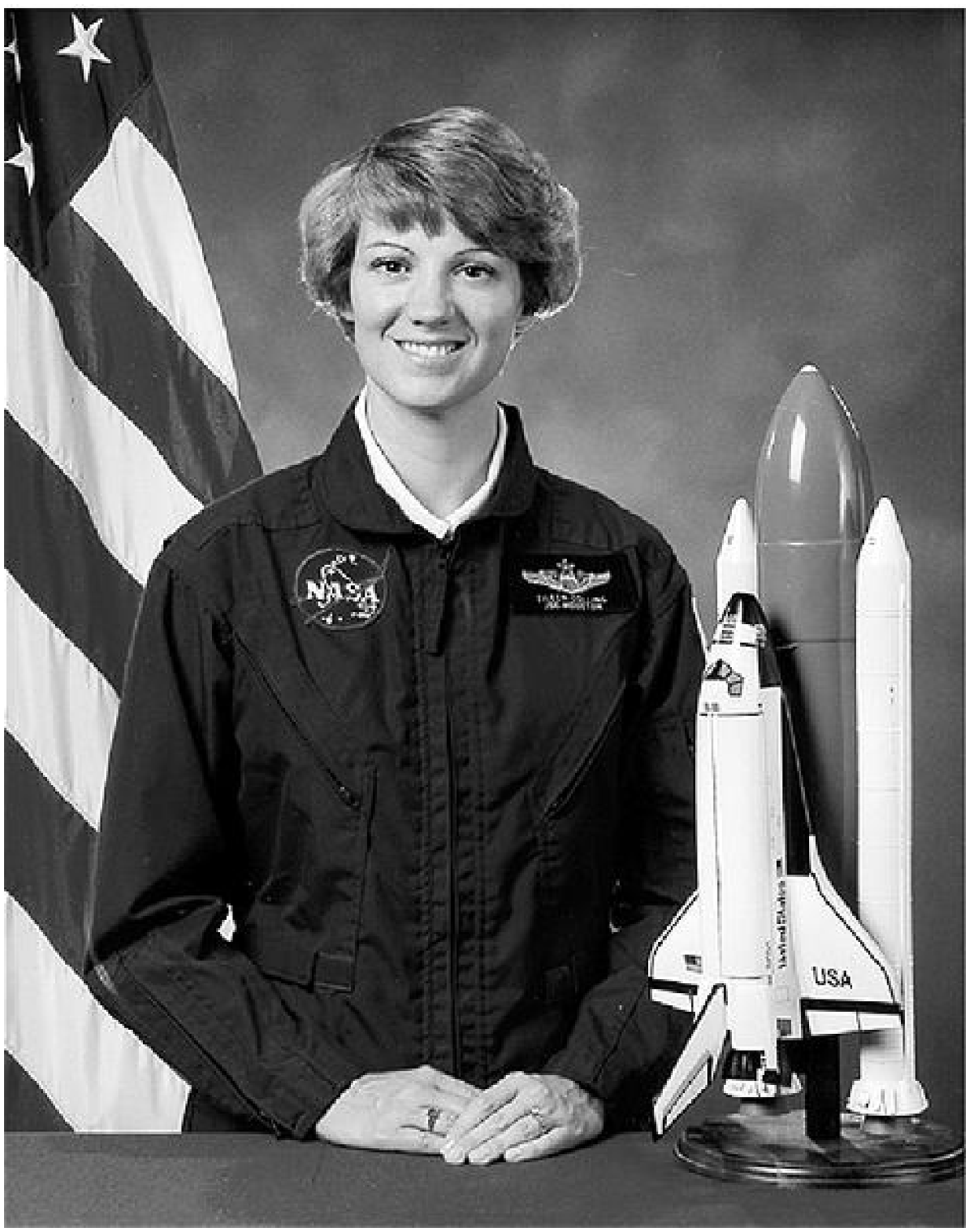} 
\caption{Col. Eileen Collins.
\label{f:EC}}
\end{center}
\end{figure}

\begin{figure}
\begin{center} 
\epsfysize=8cm
\epsfbox{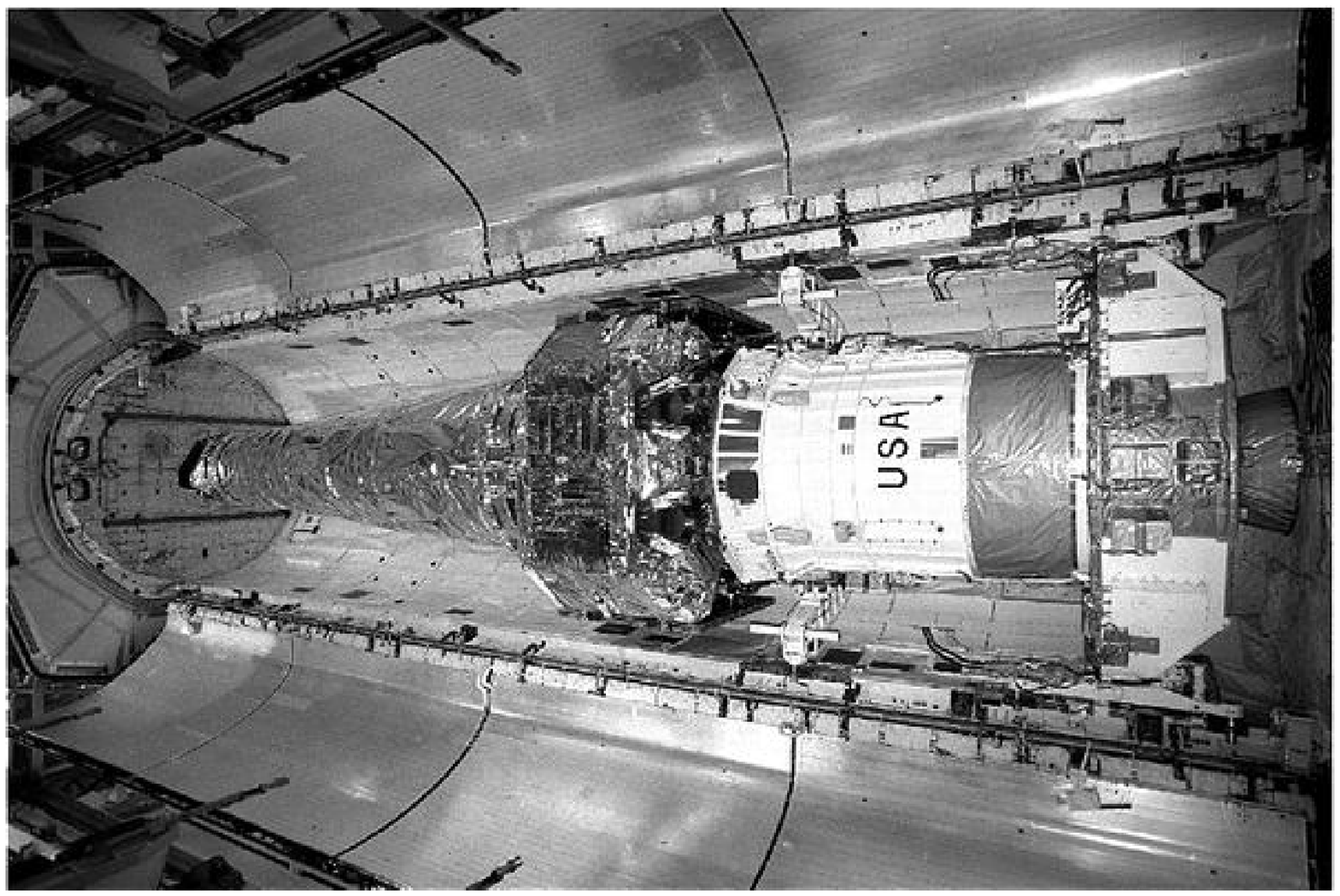} 
\caption{The \chandra\ X-ray Observatory with the IUS attached and mounted in
the
cargo bay of the Space Shuttle Columbia.
\label{f:cxo_ius_cbay}}
\end{center}
\end{figure}

After two attempts on the evenings of July 19, and July 21 the Observatory 
was launched on July 23, 1999 using the Space Shuttle Columbia. 
The Commander was Col. Eileen Collins (Figure~\ref{f:EC}), the first female
commander of a Shuttle flight. 
The rest of the crew were: Jeffrey Ashby the pilot; and mission specialists
Catherine Cady Coleman, Steven Hawley, and Michel Tognini.
With a second rocket system, the Inertial Upper Stage (IUS) attached, the
Observatory was both the largest and the heaviest payload ever launched by, and
deployed from, NASA's Space Shuttle. 
Figure~\ref{f:cxo_ius_cbay} shows the IUS mated to the Observatory and both
mounted in Columbia's cargo bay prior to launch.

\begin{figure}
\begin{center} 
\epsfysize=8cm
\epsfbox{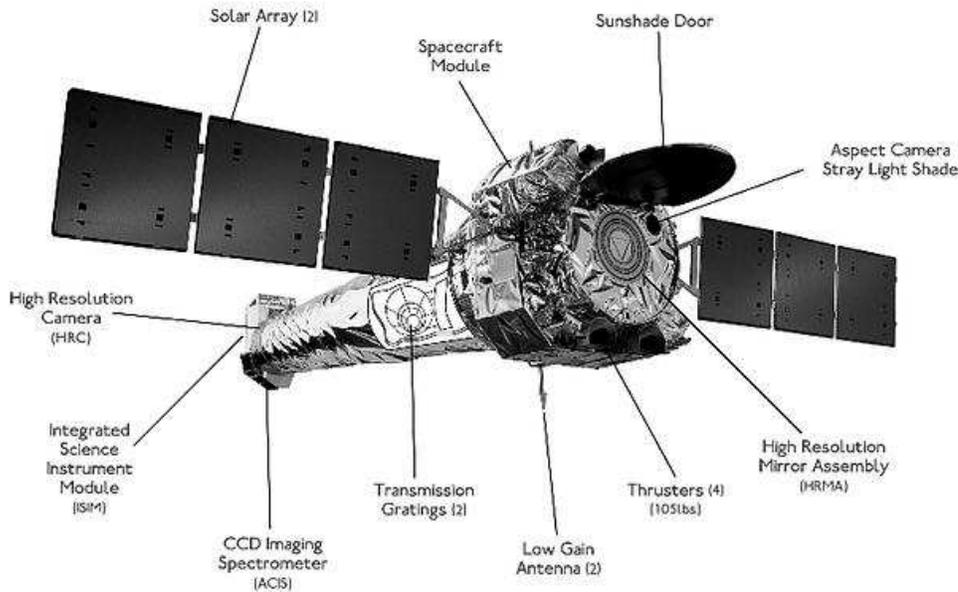} 
\caption{Artist's drawing of the \chandra\ X-ray Observatory with major
components
labeled.
\label{f:artist_concept_labeled}}
\end{center}
\end{figure}

After separation from the orbiter, the IUS performed two firings and then
separated from the Observatory.
The flight system, illustrated in Figure \ref{f:artist_concept_labeled}, is
13.8-m long by 4.2-m diameter, with a 19.5-m solar-panel wingspan. 
With extensive use of graphite-epoxy structures, the mass is only 4,800 kg,
including almost 1,000 kg of optics.
After five firings of an internal propulsion system - the last of which took
place 15 days after launch - the Observatory was placed in its highly
elliptical orbit.
This orbit has a nominal apogee of 140,000 km and a nominal perigee of 10,000
km. 
The inclination to the equator is 28.5$^o$.
With this orbit, the satellite is above the radiation belts for more than about
75\% of the 63.5-hour orbital period.
Uninterrupted observations lasting more than 2 days are thus possible.
The observing efficiency, which also depends on solar activity, varies from
65\% to more than 70\%.

The spacecraft provides pointing control, power, command and data management,
thermal control, and other such services to the scientific payload. 
Electrical power is obtained from two 3-panel silicon solar arrays that provide
over 2000 watts. 
Three 40-ampere-hour nickel-hydrogen batteries supply power during the rare
eclipses.
Two low-gain spiral antennas provide spherical communications coverage and the
transmission frequency is 2250 MHz.
The downlink provides selectable rates from 32 to 1024 kbps for communication
with NASA's Deep Space Network (DSN) of ground stations. 
Commands are sent at a frequency of 2071.8 MHz and the command rate is 2 kbps.
Data are obtained from the instruments at a rate of 24
kbs and are recorded using a solid-state recorder with 1.8 gigabits (16.8 hours)
of recording capacity.

Instrument data, together with 8 kbs of spacecraft data, primarily from the
aspect camera system, are downloaded to the DSN typically every 8 hours.
The ground stations then transmit the information to the \chandra\ Science
Center in Cambridge MA where the operations control center is located.
The Observatory is designed to operate autonomously, if necessary, for up to 72
hours and no ground intervention is required to place the Observatory in a safe
configuration after a fault is detected.
Safe mode has rarely been entered. 
The spacecraft systems and subsystems have no single-point failure-modes that
can threaten the mission. 

The principal elements of the payload are the pointing control and aspect system
(\S~\ref{s:aspect}) used to determine where the observatory was pointed, the
X-ray telescope (\S\ref{s:optics}), and the scientific instruments
(\S~\ref{s:cameras}, \S~\ref{s:ephin}, \& \S~\ref{s:gratings}).
The specified design life of the mission is 5 years; however, the only
perishable (gas for maneuvering) is sized to allow operation for more than 10
years. 
The orbit will be stable for decades. 

\section{Pointing Control and Aspect Determination System}
\label{s:aspect}

The system of sensors and control hardware that is used to point the
observatory, maintain the stability, and provide data for determining where the
observatory had been pointing is called the Pointing Control and Aspect
Determination (\pcad) system.  
Unlike HST, \chandra\ pointing requirements are not very stringent because
\chandra~detectors are essentially single-photon counters and
therefore an accurate post-facto history of the spacecraft pointing direction
is sufficient to reconstruct the X-ray image.

Here we discuss the \pcad~system, how it is used, and the flight performance. 

\subsection{Physical configuration}
\label{subss:physconfig}
The main components of the \pcad\ system are:

\begin{description}
\item{Aspect camera assembly (ACA) --} 11.2\,cm optical telescope, stray light
shade, two \ccd~ detectors (primary and redundant), and two sets of electronics.
\item{Inertial reference units (IRU) --} Two IRUs, each containing two 2-axis
gyroscopes. 
\item{Fiducial light assemblies (FLA) --} LEDs mounted near each X-ray detector
which are imaged in the ACA via the Fiducial Transfer System.
\item{Fiducial transfer system (FTS) --} directs light from the fiducial
lights to the ACA, via a retroreflector collimator (RRC) mounted at the X-ray
telescope~
center, and a periscope.
\item{Coarse sun sensor (CSS) --} Provides all-sky coverage of the sun.
\item{Fine sun sensor (FSS) --}  Has a 50$^o$ FOV and 0.02$^o$ accuracy.
\item{Earth sensor assembly (ESA) --} Conical scanning sensor, used
during the orbital insertion phase of the mission.
\item{Reaction wheel assembly (RWA) --} 6 momentum wheels which change
spacecraft attitude. 
\item{Momentum unloading propulsion system (MUPS) --} Liquid fuel thrusters
which allow RWA momentum unloading.
\item{Reaction control system (RCS) --} Thrusters which change spacecraft
attitude.
\end{description}

\subsubsection{ACA}
\begin{figure}
\begin{center} 
\epsfxsize=13cm
\epsfbox{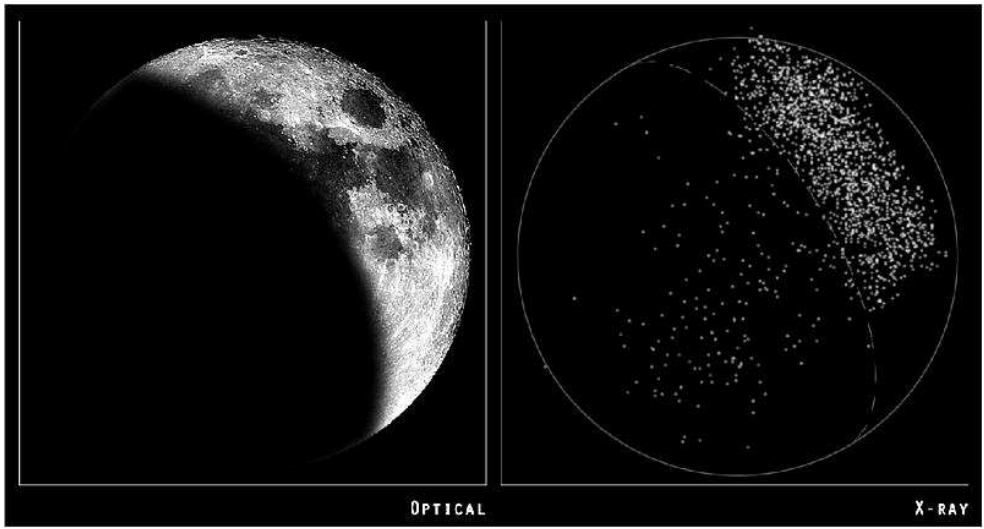} 
\caption{Optical and X-ray images of the Moon. 
Optical: Courtesy Robert Gendler; X-ray: NASA/CXC/SAO/J.Drake et
al. 
\label{f:moon}}
\end{center}
\end{figure}

The aspect camera assembly (Figure \ref{fg:aspect-fta}) includes a sunshade
($\sim$2.5 m long, $\sim$55 cm in diameter), a 11.2 cm, F/9 Ritchey-Chretien
optical telescope, and \ccd~ detector(s). 
This assembly and its related components are mounted on the side of the X-ray
telescope.
The camera's field of view is $1.4^o\times1.4^o$ and the sunshade is designed
to protect the instrument from the light from the Sun, Earth and Moon, with
protection angles of 47$^o$, 20$^o$ and 6$^o$, respectively.
Only light from the sun can damage the system.
Having either the Moon or the Earth in the field-of view only saturates
the detector output without incurring damage and therefore only limits the
aspects camera's utility. 
The Moon (Figure~\ref{f:moon})\footnote{Pictures that are publicly available at
the \chandra\ web site at
http://chandra.harvard.edu have credits labeled "Courtesy ... NASA/". The
acronyms may be found at this site.} has been viewed with \chandra, in part to
study the background signal, and in part to learn about the Moon's chemical
composition.

The aspect camera focal plane detector is a $1024\times1024$ Scientific Imaging
Technologies \ccd, with $24\times24$ micron ($5\times5$\arcsec) pixels, covering
the
spectral band between 4000 and 9000 \AA. 
The \ccd~ is deliberately placed out of focus (point source FWHM = 9~arcsec) to
spread the star images over several pixels in order to increase the accuracy of
the
centering algorithm, and to reduce variation in the point response function
over the field of view.  
There is a spare \ccd , which can be illuminated by rotating a mirror.

The ACA electronics track a small pixel region (either $4\times 4, 6\times 6,$
or $8\times 8$ pixels) around the fiducial light and star images.  
There are a total of eight regions available for tracking.  
Typically five guide stars and three fiducial lights are tracked.
The average background is subtracted on-board, and image centroids are
calculated by a weighted-mean algorithm. 
The image centroids and fluxes are used on-board by the \pcad, and are also
telemetered to the ground along with the raw pixel data.

\subsubsection{Fiducial lights and Fiducial Transfer System}
\label{ss:flights}
\begin{figure}
\centering
%\resizebox{3.7in}{!}{\includegraphics[1in,3in][7.5in,8in]{fig6.eps}}
\epsfxsize=13cm
\epsfbox{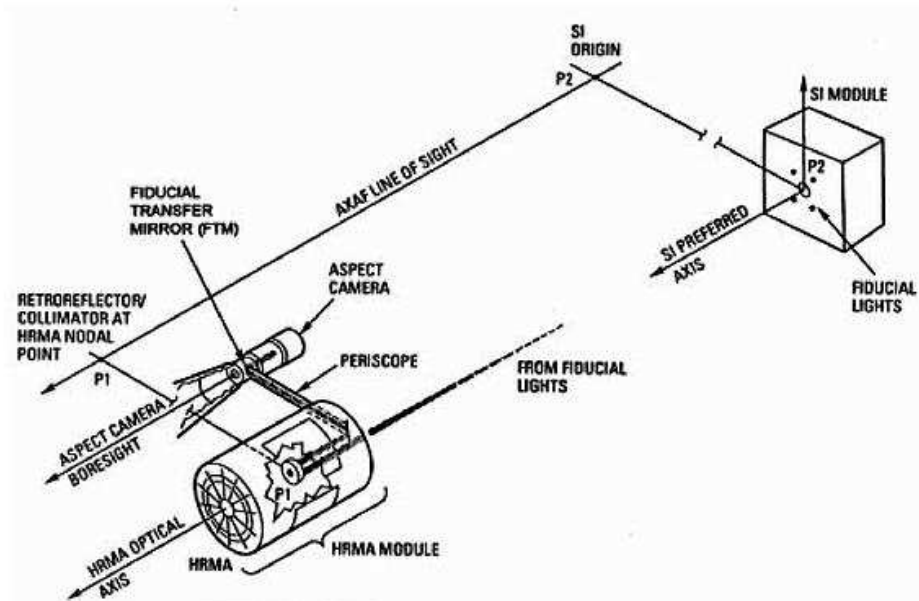} 
\caption{Schematic diagram of aspect camera assembly and the fiducial transfer
system.  
This illustrates the optical path for imaging the science instrument (SI)
fiducial lights onto the ACA.
The X-ray telescope is referred to as the High Resolution Mirror Assembly or
HRMA.}
\label{fg:aspect-fta}
\end{figure}

Surrounding each of the focal-plane detectors is a set of light emitting diodes,
or ``fiducial lights'', which serve to register the detector focal plane
laterally with respect to the \aca\ boresight. 
Each fiducial light produces a collimated beam at 635\,nm which is imaged onto
the ACA via a collimating lens, corner-cube retro-reflector and periscope
(Figure~\ref{fg:aspect-fta}).

\subsubsection{Inertial Reference Units}
Two Inertial Reference Units (IRU) are located in the front of the observatory
on the side of the X-ray telescope.  
Each IRU contains two gyroscopes, each of which measures an angular rate about 2
axes.  
This gives a total of eight channels. 
Data from four of the eight channels can be read out at one time. 
The gyros are arranged within the IRUs, and the IRUs are oriented, such that the
8 axes are in different directions and no three axes lie in the same plane. 
The gyros output pulses represent incremental rotation angles. 
In ``high-rate'' mode, each pulse nominally represents 0.75\arcsec, while in
``low-rate mode'' (used during all normal spacecraft operations) each pulse
represents nominally 0.02\arcsec.  

\subsubsection{Momentum control}
Control of the spacecraft momentum is required both for maneuvers and to
maintain stable attitude during observations. 
Momentum control is primarily accomplished using 6 Teldix RDR-68 reaction wheel
units mounted in a pyramidal configuration. 
During observing, with the spacecraft attitude constant apart from dither
(introduced to avoid having the flux from a point source illuminate only a
single focal plane detector pixel), external torques on the spacecraft
(e.g. gravity gradient, magnetic) will cause a buildup of momentum in the
reaction wheel assembly. 
Momentum is shed from the reaction wheels by firing small thrusters in the MUPS
and simultaneously spinning down the reaction wheels.

\subsection{Operating principles}

The aspect system serves two primary purposes: on-board spacecraft pointing
control and aspect determination, and post-facto ground aspect determination,
used in X-ray image reconstruction and celestial location.

The \pcad~ system has 9 operational modes (6 normal and 3 safe modes) which use
different combinations of sensor inputs and control mechanisms to control the
spacecraft and ensure its safety.   
In the normal pointing mode, the \pcad~ system uses sensor data from the ACA and
IRUs, and control torques from the RWAs, to keep the X-ray target well within
$\sim$30\arcsec\ of the desired location. 
This is done by smoothing (filtering) the data that have been taken during the
preceding time intervals using aspect camera star centroids (typically 5) and
angular displacement data from two of the 2-axis gyroscopes. 
On short time scales ($\sim$seconds) the spacecraft motion solution is dominated
by the gyroscope data, while on longer timescales it is the star centroids that
dominate.

The post-facto aspect determination is done on the ground and uses more
sophisticated processing and better calibration data to produce a more accurate
solution. 
The key improvements over the in-flight aspect come from better image
centroiding and
smoothing all available data over the observation period -- as opposed to only
a limited set of historical data. 
In addition, the aspect solution also accounts for the position of the
focal-plane instrument as determined by the images of the fiducial lights. 

\subsection{Performance}\label{ss:performance-asp}

The important \pcad~ system performance parameters and a comparison to the
original requirements are shown in Table~\ref{tb:aspect-reqs}.  
In each case the actual performance far exceeds the requirements.

\begin{figure}
\centering
\resizebox{2.7in}{!}{\rotatebox{-90}{\includegraphics{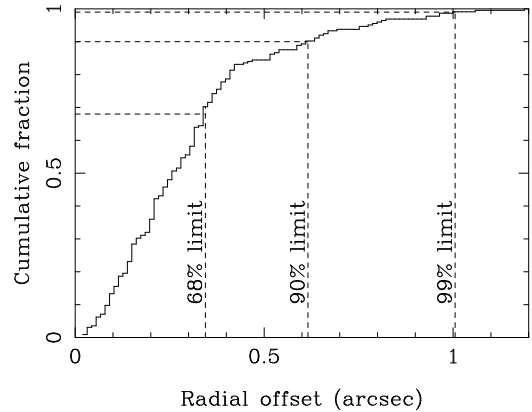}}}
\caption{
Cumulative histogram of celestial accuracy for \chandra\ X-ray source locations.
Radial offset is the distance in arcsec between the optical coordinate,
typically from the Tycho-2 catalog (see http://www.astro.ku.dk/\~erik/Tycho-2/),
and the \chandra\ position.
}
\label{fig:cel_loc}
\end{figure}

Celestial location accuracy measures the absolute accuracy of \chandra\ X-ray
source locations.  
Based on observations of 225 point sources detected within 2\arcmin\ of the
boresight and having accurately known coordinates, the 90\% source location
error circle has a radius of 0.64\arcsec\ (Figure~\ref{fig:cel_loc}).  
Fewer than 1\% of sources are outside a 1\arcsec\ radius.  

The image reconstruction performance measures the effective blurring of the
X-ray point spread function (\psf) due to aspect reconstruction.  
A direct measure of this parameter can be made by determining the time-dependent
jitter in the centroid coordinates of a fixed celestial source.  
Any error in the aspect solution will appear as an apparent wobble in the source
location.  
Unfortunately this method has limitations. 
When an Advanced CCD Imaging Spectrometer (ACIS - \S~\ref{s:cameras})
detector is at the focus, data are count-rate limited and we find only an upper
limit: aspect reconstruction effectively convolves the X-ray
telescope\ \psf\ with a Gaussian having FWHM of less than 0.25\arcsec.  
With an High Resolution Camera (HRC -  \S~\ref{s:cameras}) detector at the
focus, observations can produce acceptably high
count rates, but the current HRC photon positions have systematic errors due to
uncertainties in the HRC calibration.
These errors exactly mimic the expected dither-dependent signature of aspect
reconstruction errors, so no such analysis with HRC data has been done.  
An indirect method of estimating aspect reconstruction blurring is to use the
aspect solution to de-dither the ACA star images and measure the residual
jitter.  
We have done this for 350 observations and find that 99\% of the time the
effective blurring is less than 0.20\arcsec\ (FWHM).

Absolute celestial pointing is defined as the accuracy with which the
\chandra~line of sight (the line connecting the nominal aimpoint on the
detector and the X-ray telescope node) can be pointed toward a particular target
location on the sky, and is about 3\arcsec\ in radius. 
This result is based on the spread of apparent fiducial light locations for
$\sim 1000$ observations in the year 2002. 
It should be noted that the 3\arcsec\ value represents the repeatability of
absolute pointing on timescales of less than approximately one year.  
During the first 4 years of the mission, there was an exponentially decaying
drift in the nominal aimpoint of about 10\arcsec, most likely due to the
expected long-term relaxation in spacecraft structures.  
 
The \pcad~ 10-second pointing stability performance is measured by calculating
the RMS attitude control error (1-axis) over successive 10~second intervals.
The attitude control error is simply the difference between the ideal
(commanded) dither pattern and the actual measured attitude.  
Flight data show that after removing known systematic effects,  95\% of the RMS
error measurements are less than 0.04\arcsec\ (pitch) and 0.03\arcsec\ (yaw).  

\begin{table}
\centering
\caption{Aspect System Requirements and Performance}
\label{tb:aspect-reqs}
\begin{tabular}[t]{|p{14em}|c|c|}
%      Table title
\hline
%      Column labels
\hfil \bf Description \hfil & \hfil \bf Requirement \hfil  &  \bf \hfil Actual
\hfil \\
\hline
%      Table entries
Celestial location & 1.0\arcsec\  (RMS radius) & 0.6\arcsec\  \\ \hline
Image reconstruction & 0.5\arcsec\  (RMS diameter) & 0.3\arcsec\  \\ \hline
Absolute celestial pointing & 30.0\arcsec\  (99.0\%, radial) & 3.0\arcsec\  \\
\hline
\pcad~ 10\,sec pointing stability & 0.25\arcsec\ (RMS, 1-axis) & 0.043\arcsec\
(pitch) \\
& & 0.030\arcsec\ (yaw) \\ \hline
\end{tabular}
\end{table}

\subsection{ACA CCD Dark Current} \label{ss:dark_cal}

\begin{figure}
\centering
\resizebox{3.5in}{!}{\includegraphics{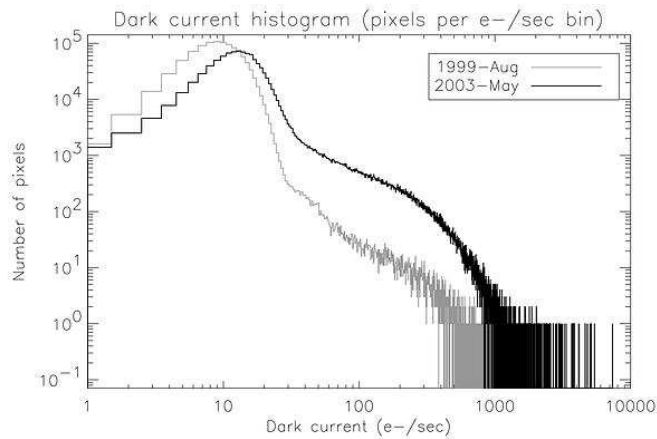}}
\caption{Differential histogram of dark current distribution for the ACA CCD
in 1999-Aug (gray) and 2003-Apr (black)}
\label{fig:dark}
\end{figure}

Damage caused by exposure to cosmic rays produces an increase in both the mean
\ccd~ dark current and the non-Gaussian tail of ``warm'' (damaged) pixels in
the aspect
camera CCD.
Figure~\ref{fig:dark} shows the distribution of dark current shortly after
launch (gray) and in 2003-Apr (black).  
The background non-uniformity caused by warm pixels (dark current $>
100$~e$^{-}$/sec) is the main contributor to the star centroiding error, though
the
effect is substantially reduced by code within the aspect data analysis software
which detects and removes most warm pixels from further consideration.

Prior to May-2003, the number of warm pixels was increasing at a rate of 2\% of
the CCD pixels per year.  
At that time the operating temperature of the CCD was lowered from -10~C to
-15~C which had the effect of decreasing the dark current by a factor of almost
2, and thereby reducing the number of warm pixels by 40\%.  
At the current operating temperature one expects no further degradation of X-ray
image quality (due to aspect) even after 15 years on
orbit\footnote{http://cxc.harvard.edu/mta/ASPECT/aca\_15yr\_perform}.

\section{The X-ray Optics \label{s:optics}}

The heart of the Observatory is the X-ray telescope made of four concentric,
precision-figured, superpolished Wolter-1 telescopes, similar in design to those
used for both the Einstein and Rosat observatories, but of much higher quality,
larger diameter, and longer focal length. 
Hughes Danbury Optical Systems (HDOS, Danbury, Connecticut) precision figured
and superpolished the 4-mirror-pair grazing-incidence X-ray optics out of
Zerodur blanks from Schott Glaswerke (Mainz, Germany).
Zerodur, a glassy ceramic, was chosen for its high thermal stability. 
Optical Coating Laboratory Incorporated (OCLI, Santa Rosa, California) coated
the optics with iridium, chosen for high x-ray reflectivity and chemical
stability.
The Eastman Kodak Company (Rochester, New York) aligned and assembled the
mirrors into the 10-m focal length telescope assembly.
Recently (2002), the \chandra\ Telescope Scientist, Leon Van Speybroeck,
received
the Rossi Prize of the High Energy Astrophysics Division of the American
Astronomical Society in recognition of his contributions to X-ray astronomy
especially for his role in the design and development of the \chandra\ optics.  

\begin{figure}
\begin{center} 
\epsfysize=8cm
\epsfbox{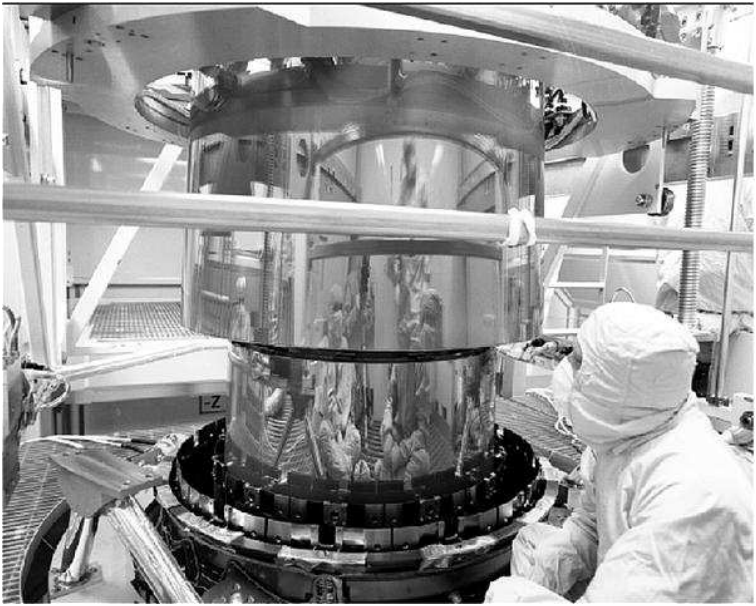} 
\caption{Photograph of the X-ray telescope during assembly and alignment. Here
one of the hyperboloids is being lowered into place prior to alignment and
bonding.
\label{f:optics}}
\end{center}
\end{figure}

The photograph shown in Figure~\ref{f:optics} was taken during the telescope
assembly and alignment process. 
The telescope is inverted in Figure~\ref{f:optics} and one sees one of the
hyperboloids being lowered into place.
Each individual element is 83.3 cm long and polished to better than a few
angstroms root-mean-square (RMS) microrougness. 
The inner surfaces of revolution are coated with 600 angstroms of iridium.
The Set of 8 mirrors weighs 956 kg. 
The focal Length is 10 meters, the outer diameter is 1.2 meters, the
field of view is 1$^o$ (FWHM) in diameter, and the clear aperture is 1136
cm$^2$.

\subsection{Point Spread Function}

\begin{figure}
\begin{center} 
\epsfysize=8cm
\epsfbox{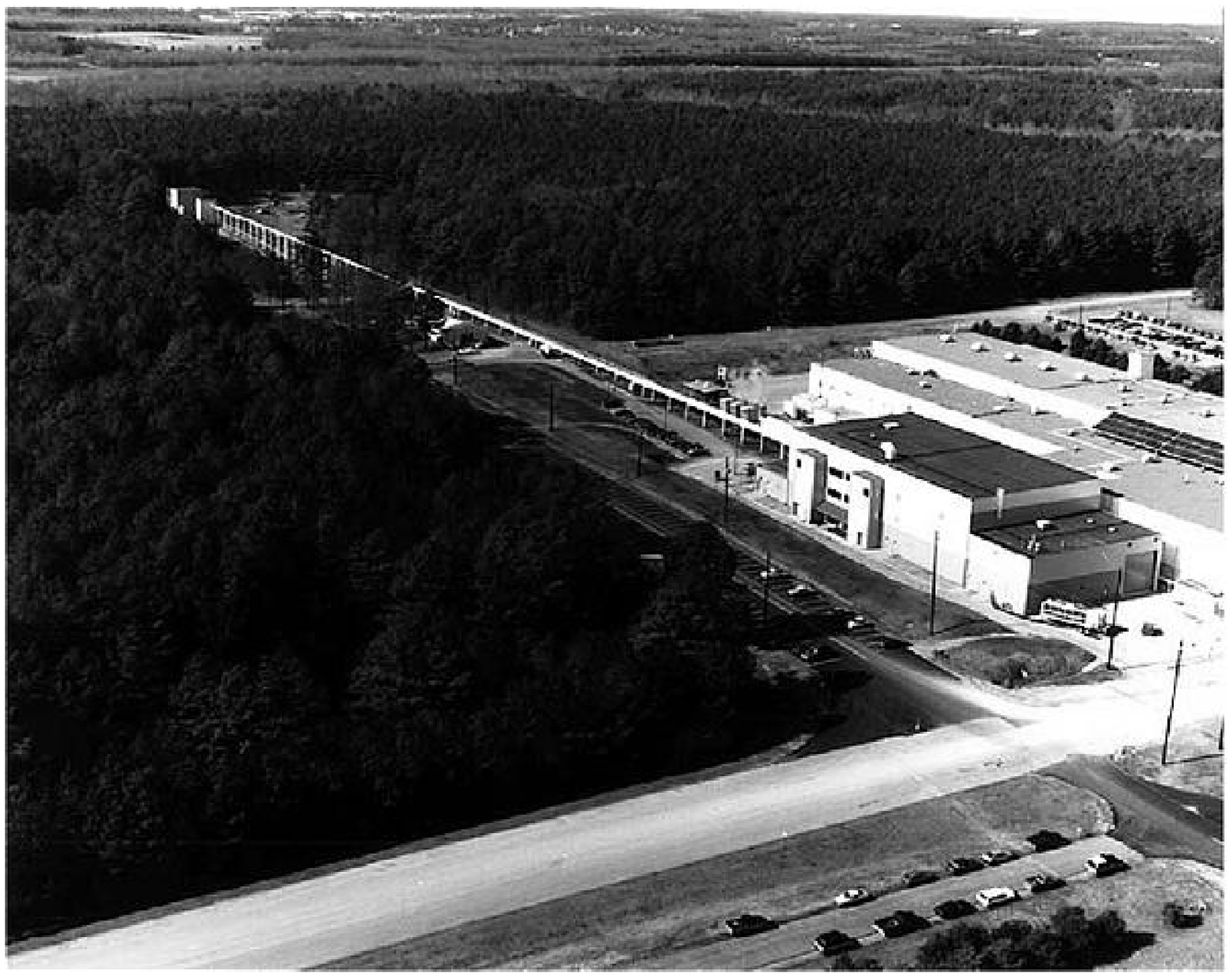} 
\caption{Aerial view of the X-ray Calibration Facility at NASA's Marshall Space
Flight Center. 
The large building to the right houses a thermal-vacuum chamber. 
A 525-m evacuated tube connects the chamber to various X-ray sources located in
the building to the far left.
\label{f:xrcf_air}}
\end{center}
\end{figure}

\begin{figure} 
\begin{center} 
\epsfysize=8cm
\epsfbox{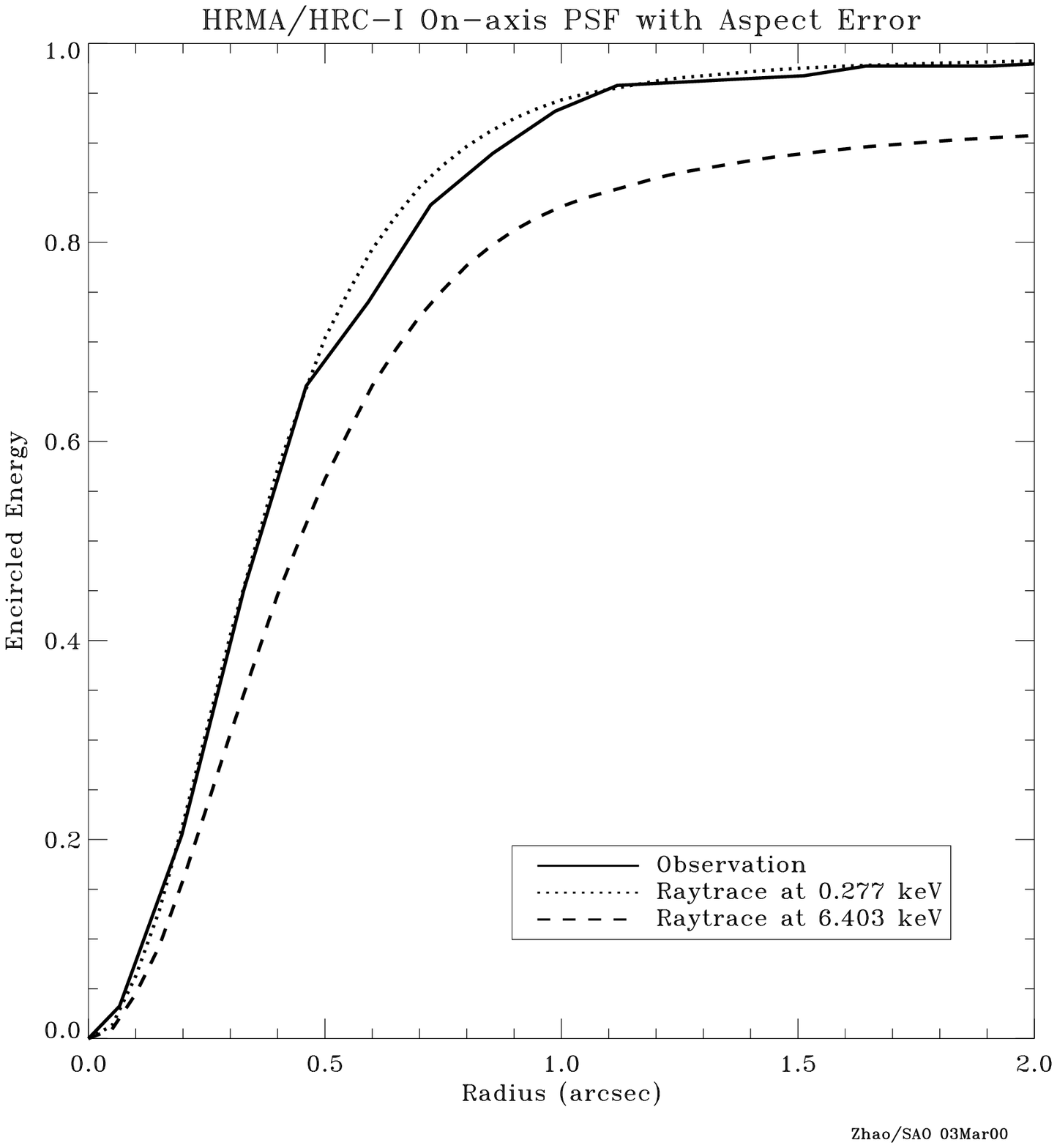} 
\caption{The predicted and observed fractional encircled energy as a function of
radius for an on-axis point source observed with the HRC-I at the focus of the
telescope. 
The calculations at two energies, 0.277 keV and 6.40 keV, include an aspect
solution with a rms uncertainty of 0.2 arcsec. 
Flight data from an observation of AR Lac are also shown.
\label{f:hrc_I_ee}
}  
\end{center}
\end{figure}

The telescope's point spread function, as measured during ground calibration,
had a full width at half-maximum less than 0.5 arcsec and a half-power
diameter less than 1 arcsec. 
The pre-launch prediction for the on-orbit encircled-energy fraction was that a
1-arcsec-diameter circle would enclose at least half the flux from a point
source. 
A relatively mild dependence on energy, resulting from diffractive scattering by
surface microroughness, attests to the better than 3-angstroms-rms surface
roughness measured with optical metrology during fabrication and confirmed by
the ground X-ray testing.
The ground measurements were performed at the X-ray Calibration Facility (XRCF)
at NASA's Marshall Space Flight Center (Figure~\ref{f:xrcf_air}).  
The effects of gravity, the finite source distance, and the size of the various
X-ray sources used to calibrate the observatory were unique to the ground
calibration and had to be accounted for to determine the on-orbit performance. 
On-orbit the performance also includes the spatial resolution of the detectors
and uncertainties in the aspect solution, although as discussed in
\S~\ref{ss:performance-asp} this latter is very small. 
The on-orbit performance met expectations as illustrated in
Figure~\ref{f:hrc_I_ee}.

\section{The Focal Plane Cameras \label{s:cameras}}
\subsection{ACIS} \label{ss:acis}

\begin{figure}
\begin{center} 
\epsfysize=8cm
\epsfbox{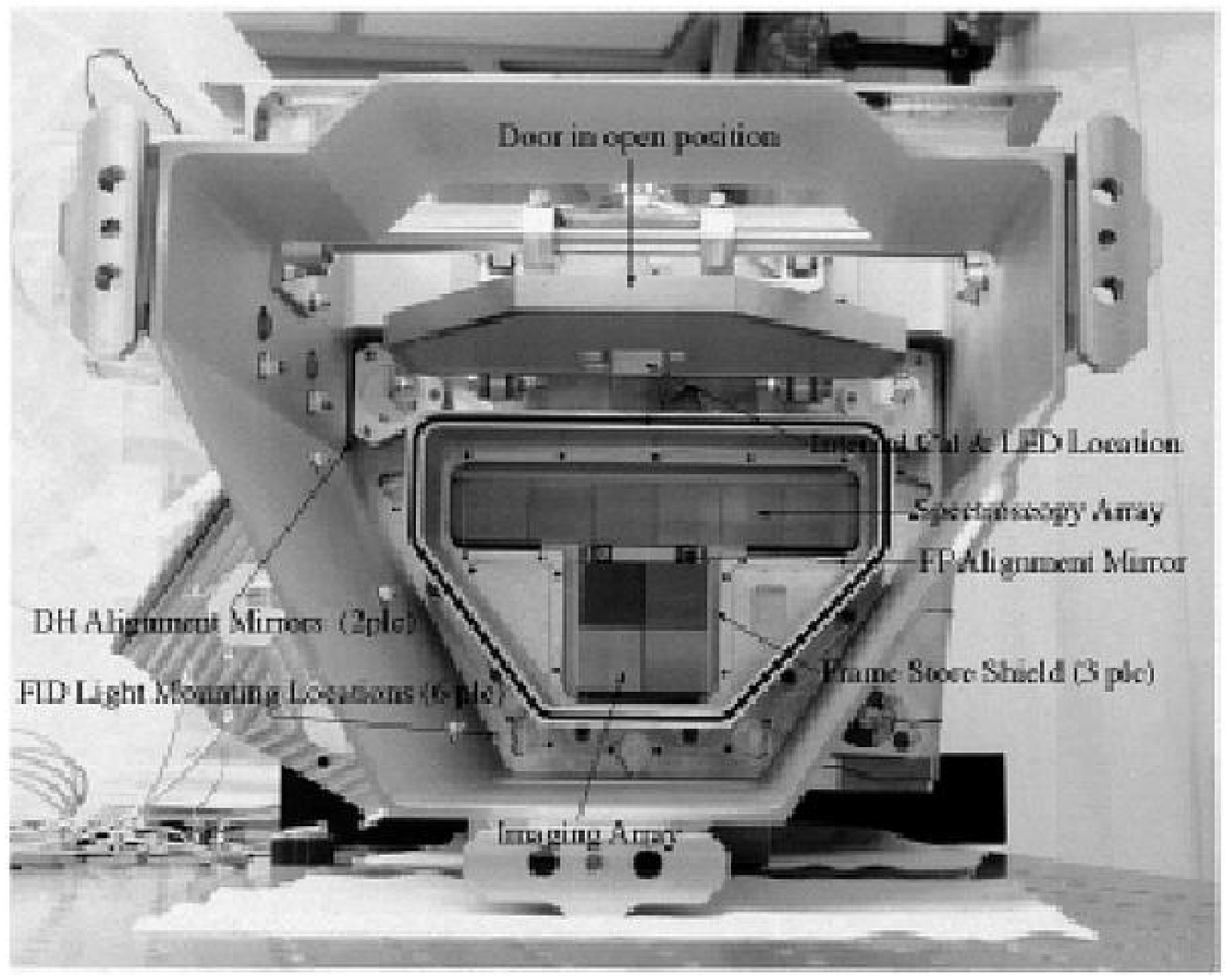} 
\caption{Photograph of the focal plane of ACIS, prior to installation of the
optical blocking filters.
The ACIS-I is at the bottom; the ACIS-S (the readout for the HETG), at the top.
\label{fig:acis}
}
\end{center}
\end{figure}

\begin{figure}
\begin{center} 
\epsfysize=8cm
\epsfbox{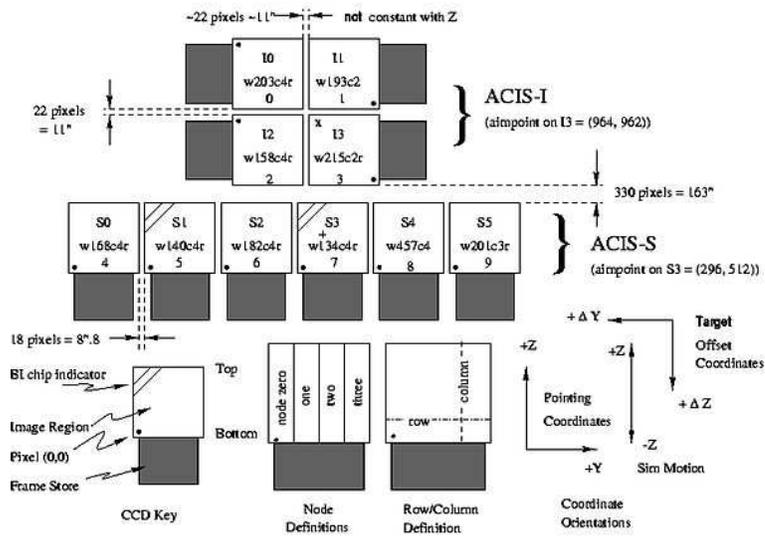} 
\caption{A schematic of the \acis~focal plane. 
The view is along the optical axis, from the source toward the detectors.
The nominal aimpoints are indicated by the `+' on S3 and the `x' on I3.   
The numerous ways to refer to a particular \ccd~are also shown: chip
letter+number, chip serial number, and \acis~chip number. 
The node numbering scheme is illustrated at the lower center of the figure.
\label{fig:acis_focal_plane}
}
\end{center}
\end{figure}

The Pennsylvania State University (PSU, University Park, Pennsylvania) and MIT
designed and fabricated the ACIS with CCDs produced by
MIT Lincoln Laboratory (Lexington, Massachusetts). 
Some subsystems and systems integration were provided by Lockheed-Martin
Astronautics (Littleton, Colorado).
Made of a 2-by-2 array of large-format, front-illuminated (FI), 2.5-cm-square,
CCDs, ACIS-I provides high-resolution spectrometric imaging over a
17-arcmin-square field of view.
ACIS-S, a 6-by-1 array of 4 FI CCDs and two back-illuminated (BI) CCDs mounted
along the transmission grating (\S~\ref{s:gratings}) dispersion direction,
serves both as the primary read-out
detector for the High Energy Transmission Grating (HETG - \S~\ref{ss:hetg}),
and, using the one BI CCD which can be placed at the aimpoint of the telescope,
also provides high-resolution spectrometric imaging extending to lower energies
but over a smaller (8-arcmin-square) field than ACIS-I.
Both ACIS detectors are covered with aluminized-polyimide filters, designed to
block visible light.
A picture of the instrument is shown in Figure~\ref{fig:acis} and a block
diagram with additional details is shown in Figure~\ref{fig:acis_focal_plane}
Many of the characteristics of the \acis~instrument are summarized in
Table~\ref{tb:ACIS_CHAR}.

{\footnotesize
  \begin{table}
\caption[Table of \acis Parameters]{\acis~
      Parameters} \label{tb:ACIS_CHAR}
  \centering
    \begin{tabular}{|lp{2.75in}|} \hline
       I-array& 4 CCDs placed tangent to the focal surface\\
       S-array& 6 CCDs placed tangent to the grating Rowland
circle\\
      \ccd~format& 1024 by 1024 pixels \\
      Pixel size& 24.0 microns (0.492$\pm$0.0001 arcsec)\\
      Array size& 16.9 by 16.9 arcmin ACIS-I\\
        & 8.3 by 50.6 arcmin \aciss\\
      On-axis effective Area&   $110\rm\,cm^2$ @ 0.5 keV ) \\
      ~~~(integrated over the \psf     &   $600\rm\,cm^2$ @
1.5 keV \\
      ~~~to $>$99\% encircled energy)     &   $40\rm\,cm^2$ @ 8.0
keV \\
      Quantum efficiency& $>80\%$ between 3.0 and 5.0 keV\\
       (frontside illumination)& $>30\%$ between 0.8 and 8.0 keV\\
      Quantum efficiency& $>80\%$ between 0.8 and 6.5 keV\\
       (backside illumination)& $>30\%$ between 0.3 and 8.0 keV\\
      Charge transfer inefficiency& : $\sim$2$\times$10$^{-4}$; :
$\sim$1$\times$10$^{-5}$ \\
      System noise& $<\sim2$ electrons (rms) per pixel \\
      Max readout-rate per channel& $\sim100$ kpix/sec\\
      Number of parallel signal channels& 4 nodes per \ccd~\\
      Pulse-height encoding& 12 bits/pixel \\
      Event threshold         &  : 38 ADU ($\sim$140 eV) \\
                              & : 20 ADU ($\sim$70 eV) \\
      Split threshold         &  13 ADU  \\
      Max internal data-rate&  6.4 Mbs ($100$ kbs $\times 4 \times 16$)\\
      Output data-rate& 24 kb per sec\\
      Minimum row readout time& 2.8 ms \\
      Nominal frame time & 3.2 sec (full frame) \\
      Allowable frame times &0.2 to 10.0 s \\
      Frame transfer time   & 40 ${\mu}$sec (per row) \\
      Point-source sensitivity&$4 \times 10^{-15}\rm ergs\,cm^{-2}\,s^{-1}\ in
      \ 10^4\, s$ \\
      &(0.4-6.0 keV)\\
      Detector operating temperature& $-90$ to $-120^\circ$C\\ \hline
    \end{tabular}
    
\end{table}
}% end footnotesize
\bigskip

\subsubsection{Spatial Resolution}\label{sss:acis_spat_res}

The spatial resolution for on-axis imaging with \acis~is limited by the
physical size of the \ccd~pixels (24.0 ${\mu}$m square $\sim$0.492
arcsec) and not the X-ray telescope. 
The spacecraft dither moves the image through a Lissajous pattern with an
amplitude of 16 ACIS-pixels which allows sub-sampling of the image.  
In addition, subpixel resolution can be achieved by using the distribution of
charge in the event island to modify the event position (e.g. Li et al. 2003).

Approximately 90\% of the encircled energy lies within 4 pixels ($\sim$ 2
arcsec) of the center pixel at 1.49 keV and within 5 pixels ($\sim$ 2.5 arcsec)
at 6.4 keV.  
Off-axis, the departure of the CCD layout from the ideal focal surface and the
increase of the telescope point-spread-function with off-axis angle become the
dominating factors.

\begin{figure}
\begin{center} 
\epsfysize=8cm
\epsfbox{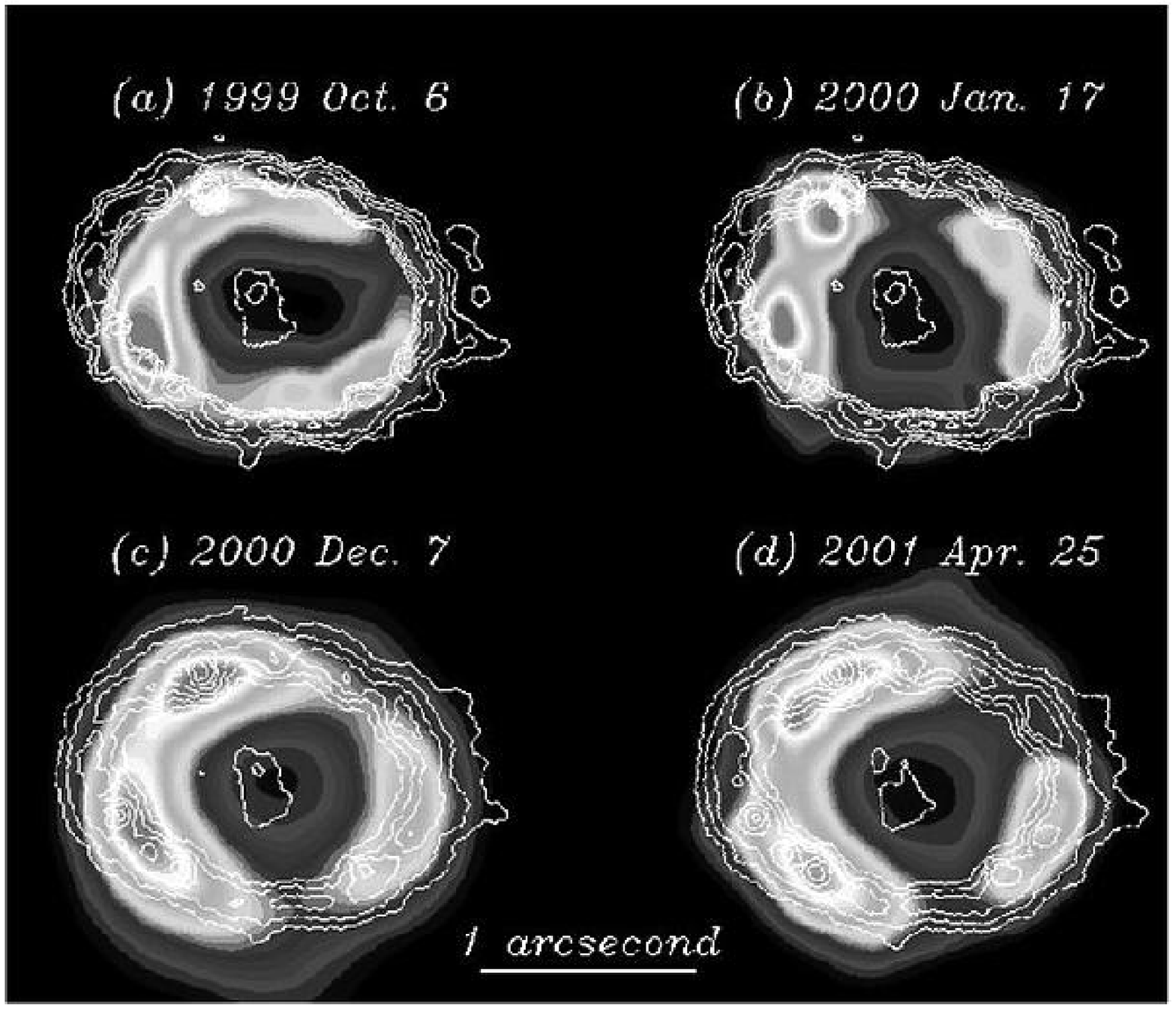} 
\caption{ACIS images and HST contours of the emission from SNR1987A.
Courtesy Dave Burrows.
\label{f:sn1987a}}
\end{center}
\end{figure}

Figure~\ref{f:sn1987a} demonstrates the spatial resolution using ACIS and shows
a
time history and relative positioning of the optical emission of SNR 1987A as
seen with HST together with the X-ray emission observed with \chandra\ (Burrows
et al. 2000).
The reverse, X-ray-emitting, shock, inside of the cooler, optically-emitting,
gas is a textbook example of the shock-heating of the interstellar medium
following the stellar explosion.  
The X-ray emitting ring
is only an arcsec in diameter, demonstrating the exciting new regime
of spatial scales that are being explored with the Observatory.
 
\subsubsection{Hot Pixels and Columns}

Radiation damage or manufacturing defects can cause individual pixels or 
entire columns to have anomalously high dark current.  
If the dark current  is large enough, the pulse height in the pixel or in an
entire column can regularly exceed the event threshold and produce spurious
events.  
These features, known as hot pixels and columns, are generally removed in 
data analysis.  
Because of the low operating temperature of $-120^\circ$C, which reduces dark
current, there are few hot pixel and columns on the ACIS CCDs.  
Currently less than 1\% of the entire 10 CCD focal plane are impacted and the
rate of increase in these features is very small, of
order a few pixels per year.

\subsubsection{Energy Resolution}\label{sss:acis_gain_enres}

Good spectral resolution depends upon the accurate determination of the total
charge deposited by a single photon. 
This in turn depends upon the fraction of charge collected, the fraction of
charge lost in transfer from pixel to pixel during read-out typically expressed
as the charge-transfer-inefficiency (CTI), and the ability of the readout
amplifiers to measure the charge. 
Spectral resolution also depends on read noise and the off-chip analog
processing electronics. 
The ACIS CCDs have readout noise less than 2 electrons RMS. 
Total system noise for the 40 \acis~ signal chains (4 nodes/CCD) ranges from 2
to 3 electrons (rms) and is dominated by off-chip analog processing electronics.

\begin{figure}
\begin{center} 
\epsfysize=8cm
\epsfbox{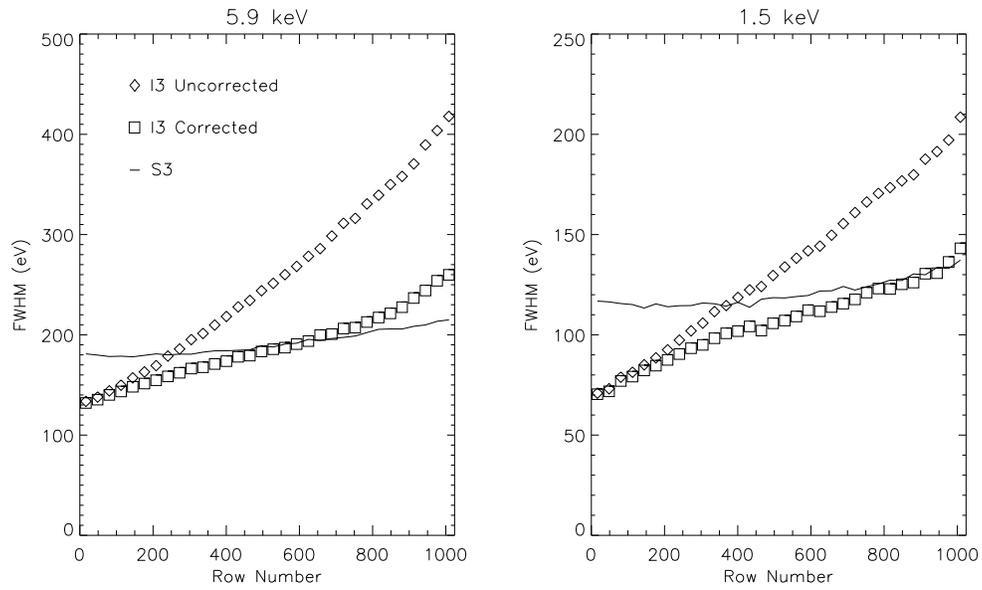}
\caption { 
The energy resolution at 2 energies of two of the CCDs (S3 a BI CCD and I3 a FI
CCD) as a function of row number for an operating temperature of $-120^\circ$C. 
The diamonds and squares are for the FI data before and after applying
a post-facto correction.
The solid curves are for the BI data.
\label{f:acis_rows} 
}
\end{center}
\end {figure}

The ACIS FI CCDs originally approached the theoretical limit for energy
resolution at almost all energies, while the BI were of somewhat lesser quality
due to imperfections induced in the manufacturing process (Bautz et al. 1998).
Subsequent to launch and orbital activation, the FI CCDs have developed
much larger CTI and the energy resolution of the FI CCDs has become a function
of the row number, being nearer pre-launch values close to the frame store
region and progressively degraded toward the farthest row (Prigozhin et al.
2000) as shown in Figure~\ref{f:acis_rows}.

The damage to the FI CCDs was caused by low energy protons, encountered during
radiation belt passages, and which Rutherford-scattered from the X-ray telescope
optical surfaces onto the focal plane.
Subsequent to the discovery of the degradation, operational procedures were
changed so that the ACIS instrument is not left at the focal position during
radiation belt passages.
Since then, degradation in performance has been limited
to the small, gradual increase due to cosmic rays that was predicted before
launch.
The BI CCDs were not impacted, consistent with the proton scenario, as it
is far more difficult for low energy protons to deposit their energy in the
buried channels where damage is most detrimental to performance.
These channels are near the CCD gates and the BI gates face in the direction
opposite to the telescope.

\begin{figure}
\begin{center} 
\epsfysize=8cm
\epsfbox{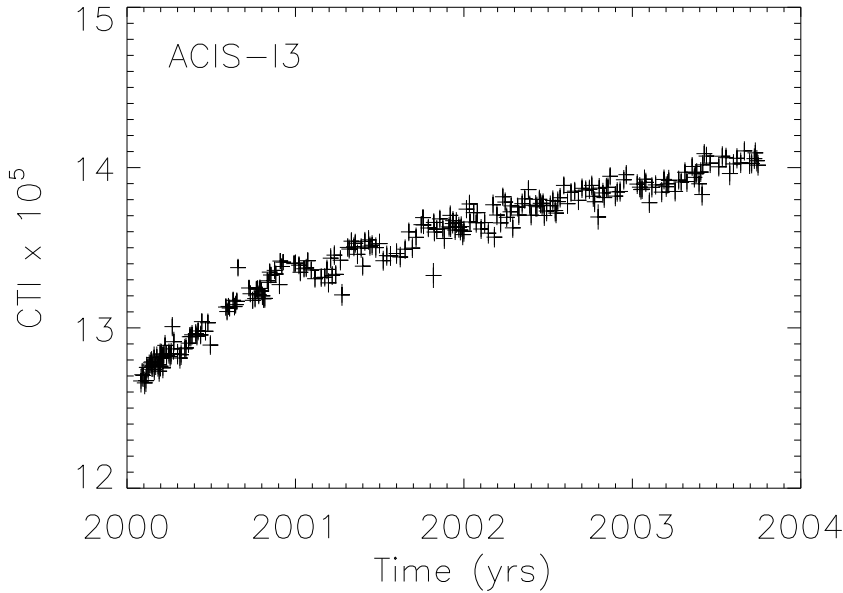}
\caption {\label{f:acis_fi_ctichange}  
The time-dependent change in parallel CTI at 5.9 keV for the FI CCD I3
subsequent to the proton damage early in the mission.
}
\end{center}
\end {figure}

\begin{figure}
\begin{center} 
\epsfysize=8cm
\epsfbox{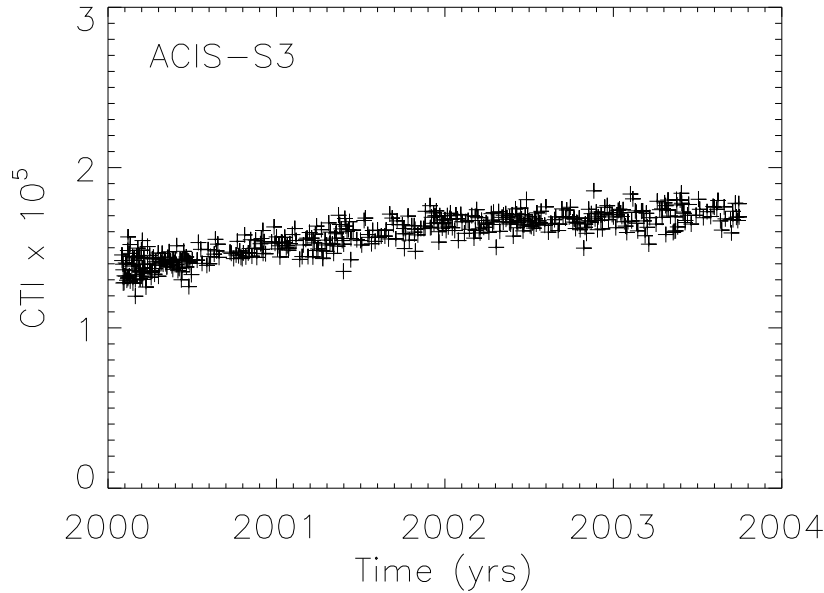}
\caption {\label{f:acis_bi_ctichange}  
The time-dependent change in parallel CTI at 5.9 keV for the BI CCD S3.
}
\end{center}
\end {figure}

Figures~\ref{f:acis_fi_ctichange} and \ref{f:acis_bi_ctichange} show the
time-dependent change in parallel (row-dependent) CTI on both FI and BI CCDs. 
The CTI is increasing at a rate of $3 \times 10^{-6}$ yr$^{-1}$ on the FI CCDs
and $1 \times 10^{-6}$ yr$^{-1}$ on the BI CCDs.  
CTI in the serial transfer direction (column-dependent) has not increased since
launch.

A post-facto software correction has been developed which can effectively
recover much of the resolution lost due to the increased, row-dependent CTI
(Townsley et al. 2002).  
For example, for the I3 FI CCD furthest from the framestore where resolution is
most degraded, the FWHM is improved from 420~eV to 260~eV at 5.9~keV and from
210~eV to 150~eV at 1.5~keV, as shown in Figure~\ref{f:acis_rows} 

\begin{figure}
\begin{center} 
\epsfysize=8cm
\epsfbox{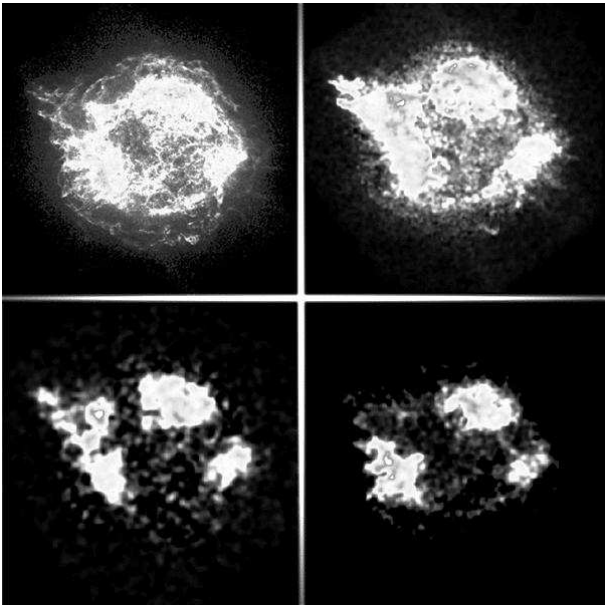} 
\caption{Images of the supernova remnant Cas A.  
Upper left: Broadband X-ray image. Upper right: Image made by X-rays from
silicon ions. Lower left: Image
made by X-rays from calcium ions. Lower right: Image made by X-rays from iron
ions. All images are 8.5 arc minutes on a side. (Credit: NASA/GSFC/U.Hwang et
al. )
\label{f:casa}}
\end{center}
\end{figure}

Many \chandra\ observing programs make use of ACIS spectral resolution to
achieve
their scientific goals.  
An example is shown in Figure~\ref{f:casa} (Hwang, Holt \& Petre 2000).  
This set of images of the supernova remnant Cassiopeia A (Cas A) was made by
using the CCD energy resolution to differentiate between emission from various
ions.  
The distinctive morphology of each image was used to study the evolving ejecta
distribution and its interaction with the surrounding medium.

\subsubsection{ACIS Time Resolution}

ACIS has a multitude of operating modes which offer different time resolution. 
In what is referred to as the standard timed exposure mode, the CCDs collect
data for a set period of time, then the imaging region of the CCD is quickly
transferred (40 $\mu$sec per transfer) to the framestore to be read out.  
For most efficient operation, the exposure time for a CCD frame is set equal to
the time required to read out the active region of the CCD.  
For six CCDs on and reading out the full frame, the frame time is 3.2
seconds and this is the mode which is typically used.
It is possible to configure the instrument to read out smaller subarrays and/or
operate fewer CCDs to reduce the frame time to a minimum of 0.2 seconds.

Higher time resolution can be obtained at the expense of one dimension of
spatial resolution.  
In what is referred to as the continuous clocking mode, data are moved through
the CCD at a rate of 3 msec per transfer.  
In this mode, the precise location of an event along a column is lost. 
On-orbit, the timing with ACIS has performed as expected. 

\subsubsection{Quantum Efficiency (QE)}

\begin{figure}
\begin{center} 
\epsfysize=8cm
\epsfbox{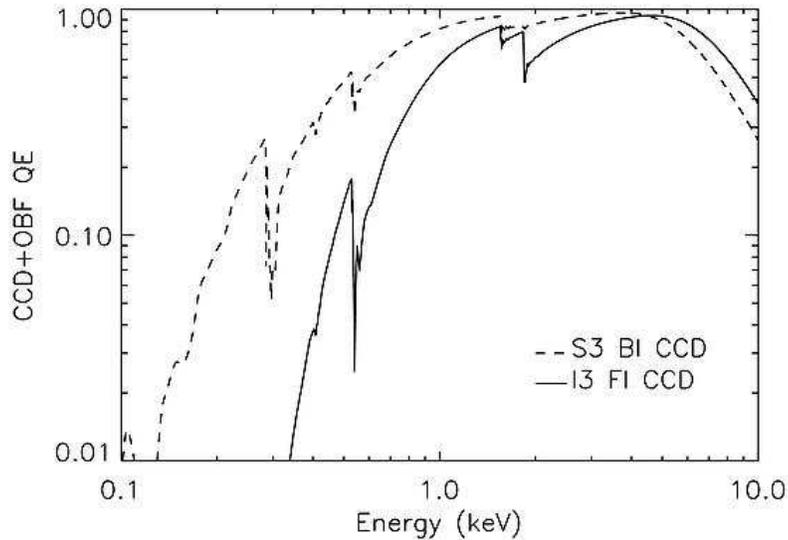}
\caption {\label{f:acis_qe}  
Nominal ACIS quantum efficiency convolved with the optical blocking filter for
the I3 FI CCD and the S3 BI CCD.
}
\end{center}
\end {figure}

The CCD quantum efficiencies for the I3 FI CCD and S3 BI CCD, convolved with the
appropriate optical blocking filter, are shown in Figure~\ref{f:acis_qe}. 
Characteristic absorption edges from materials in the CCD gate structures and
the filter are seen at low energies. 
At high energies the QE is bounded by the thickness of the sensitive depletion
region.  
The BI CCDs, whose gate structures face away from the telescope, have much
higher QE at low energies, but have lower QE at high energies because they are
thinner.

Charge transfer inefficiency degrades somewhat the uniformity of the nominal QE
by  redistributing the charge in an X-ray event island (an ``event island''
consists of an array of 3x3 or 5x5 pixels, depending on the selected mode of
operation) - so that it appears more like a cosmic ray and is consequently
rejected in analysis.  
Therefore,  regions of the CCD far from the readout can have slightly lower
effective QE than regions close to the readout.  
To date, and at the current operating temperature of $-120^\circ$C, the effect
of CTI on quantum efficiency uniformity is small.  
At energies above a few keV, there is a maximum range of $\sim$ 5\% - 10\% in QE
for the FI CCDs and a much smaller range for the BI CCDs. 

\subsubsection{Contamination}

The ACIS instrument has the coldest surfaces on the Observatory. 
Thus, naturally, there has been a gradual accumulation of a contaminating layer
since soon after launch which decreases the low energy detection efficiency and
introduces distinctive absorption edges.
The contaminant is believed to be deposited on the optical blocking filters and
not on the CCDs themselves as there is no clear ballistic path.
These filters are nominally at a temperature of -60 degrees C.   
Low Energy Transmission Grating (LETG - \S~\ref{ss:letg}) observations of bright
astrophysical continuum sources have been used to construct a detailed
empirical model of the contaminant (Marshall et al. 2004).  
The dominant element in the contaminant is carbon ($>$ 80\% by number and
excluding hydrogen), but small edges due to oxygen (7\%) and fluorine (7\%) are
also detectable. 

\begin{figure}
\begin{center} 
\epsfysize=8cm
\epsfbox{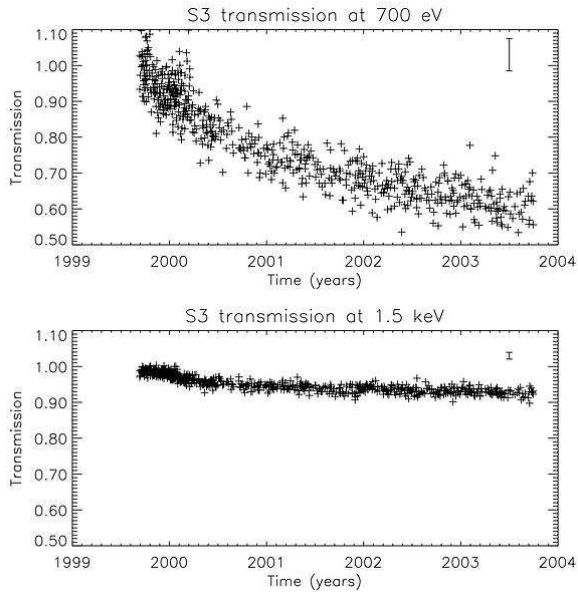}
\caption {\label{f:acis_contam_time}  
The time-dependent change in transmission at 700 eV and 1.5 keV due to the
deposition of a contaminant on the ACIS optical blocking filters.  
A typical error bar is shown in the upper right.
}
\end{center}
\end {figure}

Figure~\ref{f:acis_contam_time} shows the time dependent decrease in
transmission from the contaminant as measured using the on-board calibration
sources at 700 eV and 1.5 keV.  
After four years the transmission at 700~eV is about 60\% of the pre-launch
value and at 1.5~keV is about 93\% of the pre-launch value.  

An obvious response to contaminant deposition is to temporarily raise the
temperature of the focal plane.  
This procedure has not been immediately undertaken because of the experience,
early in the mission, in which a bakeout further degraded the CTI of the
already radiation damaged FI CCDs.  
Further lab studies of radiation damaged CCDs have confirmed this effect. 
Currently the risks and benefits of different bakeout scenarios are under
study.

\subsubsection{Background}

Cosmic-ray-induced events experienced by the ACIS CCDs are very effectively
minimized by on-board processing ($\sim$ 99\% and $\sim$ 79\% for FI and BI
CCDs).  
The remaining background can be separated into two parts; a slowly varying
quiescent component and flares in which the count rate can increase
dramatically over time scales of minutes to hours (Plucinsky \&Virani 2000).  
The quiescent background is due to high energy cosmic rays and is
anti-correlated with the solar cycle.  
The flares show some correlation with the orientation of the Observatory's orbit
with respect to the Earth's magnetosphere and with the Observatory's altitude
(Grant et al. 2002).
The frequency and magnitude of the flares are much more pronounced in the BI
CCDs than in the FI CCDs and less pronounced when either transmission grating is
inserted, consistent with the suggestion that the flaring events are caused by
weakly penetrating low energy protons that reach the focal plane after
reflection off the telescope.
The gate structure of the FI CCDs absorbs most of the protons for the majority
of the flares.  
The spectral shape of both the quiescent and flaring background is described in
more detail by Markevitch et al. (2003).

\begin{figure}
\begin{center} 
\epsfysize=8cm
\epsfbox{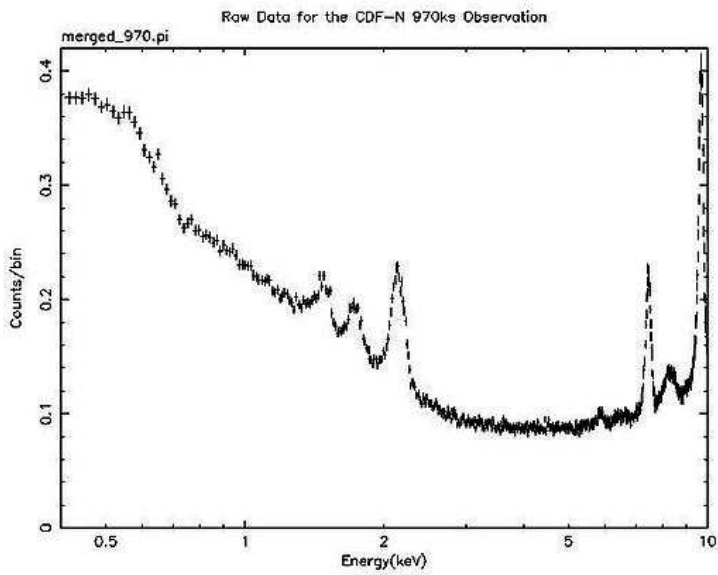}
\caption { 
A background spectrum deduced from  a 970 ksec exposure to the \chandra\ Deep
Field-North region after removing sources. The four prominent lines in order of
increasing energy are Al, Si, Au, and Ni. 
\label{f:acis_bk_spectrum} 
}
\end{center}
\end {figure}

The region of the Hubble Deep Field-North was observed using the ACIS-I array,
for 970 ksec in a series of twelve pointings spanning a period of 15 months and
provided an excellent representation of the background and its spectrum.
This spectrum, after removing the contribution from point sources, is shown
in Figure~\ref{f:acis_bk_spectrum}.
The prominent spectral lines in Figure~\ref{f:acis_bk_spectrum} are from cosmic
ray induced fluorescence of the gold-coated collimator, the nickel-coated
substrate of the collimator, the silicon in the CCDs and from aluminum used in
various places in the housing and the filter coating.
In general, the background produced by the fluorescent lines is only about 2.6\%
of the background not found in the lines in the soft (0.5-2.0 keV) band and
13.5\% of the flux in the hard (2.0-10.0 keV) band.
A number of on-ground data filtering techniques have been developed which
significantly suppress background.  
Brandt et al. (2001) have noted an approach wherein the background can be
reduced by up to 36\%, while only reducing source counts by 12\%. 
Another technique, which requires a particular ACIS mode of operation, can
reduce the background by 30-80\% depending on the energy of interest with a
minimal loss ($\sim$ 2\%) of source events (Vikhlinin 2002).

\subsection{HRC}\label{ss:hrc}

\begin {figure} 
\begin{center} 
\epsfysize=8cm
\epsfbox{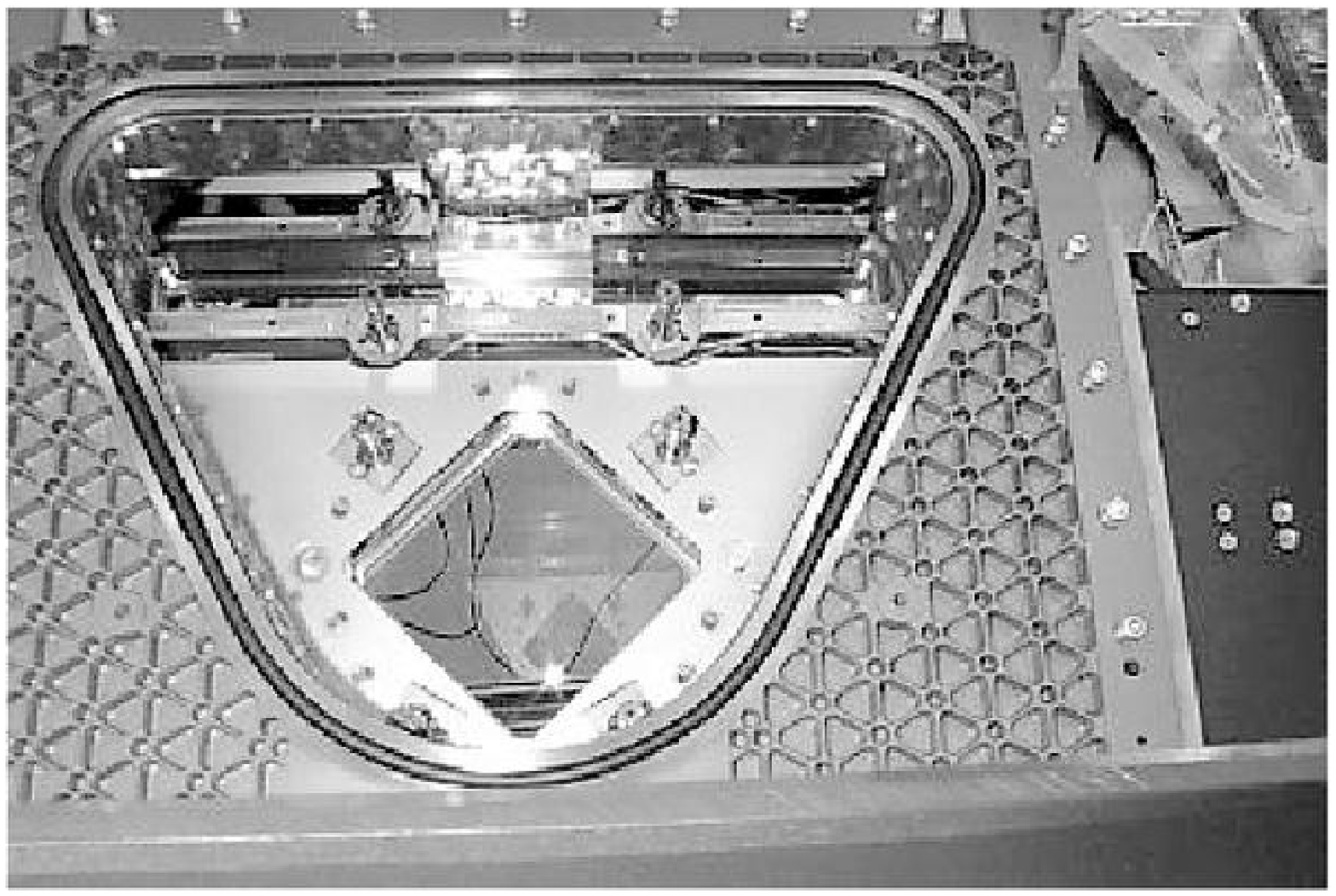}
\caption {
Photograph of the HRC.
The HRC-I (imager) is at the bottom; the HRC-S (the readout for the LETG), at
the top. 
\label{f:hrc_picture} 
}
\end{center}
\end {figure}

\begin{figure} 
\begin{center} 
\epsfysize=8cm
\epsfbox{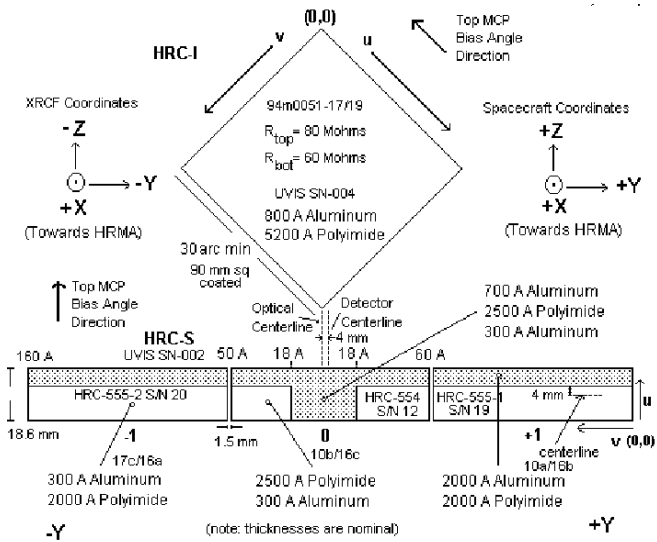} 
\caption{A schematic of the HRC focal plane geometry as viewed along the
optical axis from the telescope toward the focal plane.
\label{f:hrc_layout}
}  
\end{center}
\end{figure}

The Smithsonian Astophyscial Observatory (SAO, Cambridge MA),  designed and
fabricated the HRC (Murray et al. 2000) shown
in Figure~\ref{f:hrc_picture}.
Made of a single 10-cm-square microchannel plate (MCP), the HRC-I provides
high-resolution imaging over a 30-arcmin-square field of view.
Comprising 3 rectangular MCP segments (3-cm $\times$ 10-cm each) mounted
end-to-end along the grating dispersion direction, a second detector, the
HRC-S, serves as the primary read-out detector for the LETG.
Both detectors have cesium-iodide-coated photocathodes and are covered with
aluminized-polyimide UV/ion shields.
A schematic of the HRC layout is shown in Figure~\ref{f:hrc_layout}, and a
summary of the characteristics is given in Table \ref{t:hrc_par}.

\begin{table}
  \centering
  \scriptsize{
  \caption{HRC Parameters}\label{t:hrc_par}
  \smallskip
  \begin{tabular}{|lll|} \hline
	 HRC-I:
			& CsI-coated MCP pair
			& $90\times90$ mm coated \\
                        & & ($93\times93$ mm open) \\
	 HRC-S:
		        & CsI-coated MCP pairs
			& 3-$100\times20$ mm \\[1mm]
    Field of view        & HRC-I: & $\sim 30\times30$ arcmin \\
   		         & HRC-S:& $6\times99$ arcmin \\
     MCP Bias angle:&&	$6^\circ$\\[1mm]
    UV/Ion Shields: && \\
	& HRC-I: & 5520 \AA\ Polyimide, 763 \AA\ Al\\
	& HRC-S: &\\

	&\hspace*{0.5in}Inner segment & 2750 \AA\ Polyimide, 307 
\AA\ Al\\
	&\hspace*{0.5in}Inner segment ``T'' & 2750 \AA\ Polyimide, 793 
\AA\ Al\\
	&\hspace*{0.5in}Outer segment  & 2090 \AA\ 
Polyimide, 304 \AA\ Al\\
	&\hspace*{0.5in}Outer segment (LESF)  & 2125 \AA\ 
Polyimide, 1966 \AA\ Al\\
%	& HRC-S: Low-Energy Suppression Filter  & \\
%	& (LESF) & \\
%	& \hspace*{0.5in}Outer segment: 170--55 \AA\ & 2000 \AA\ Polyimide, 2000 
%\AA\ Al\\
%	&\hspace*{0.5in}Inner segment: $\la$55 \AA\ & 2500 \AA\ Polyimide, 1000 
%\AA\ Al\\[1mm]
    Spatial resolution   
		& FWHM &  $\sim 20\mu$m, $\sim 0.4$ arcsec \\
                & & \\
		& HRC-I: pore size & 10$\mu$m \\
                & HRC-S: pore size & 12.5$\mu$m \\
		& HRC-I: pore spacing & 12.5$\mu$m \\
                & HRC-S: pore spacing & 15$\mu$m \\
		& pixel size (electronic readout)& $6.429\mu\rm m$\\
		& & [0.13175 arcsec pixel$^{-1}$]\\[1mm]
 Energy range:& &	$0.08-10.0$ keV\\
    Spectral resolution &  
		$\Delta E/E$ & $\sim 1$ @1keV \\
		    MCP\ Quantum efficiency  & & 30\% @ 1.0 keV \\
	& & 10\% @ 8.0 keV\\[1mm]
      On-Axis Effective Area: & HRC-I, @ 0.1 keV&	$5\,\rm cm^2$ \\
		    & HRC-I, @ 1 keV&	$227\,\rm cm^2$ \\

    Time resolution      & & 16 $\mu$sec \\[1mm]
    Limiting Sensitivity  & point source, 5 count detection in $3\times10^5$ s
			   & $7 \times 10^{-16}\rm erg\,cm^{-2}\,s^{-1}$\\
			& (power law spectrum: $\alpha$ = 1.4, & \\
			&$\rm N_H = 3 \times 10^{20}$ cm$^{-2}$)& \\[1mm]
On-orbit & HRC-I~& 9$\times 10^{-6}$\,cps arcsec$^{-2}$ \\ %HD's Oct99 memo
quiescent background & HRC-S~& 1.8$\times 10^{-4}$\,cps ${\rm (res.\
elm.)^{-1}}$\\
%     &       & ${\rm (0.06\AA\times 0.1 mm)}$ \\
(prior to ground processing)       &       & (0.07\AA$\times$ 0.1 mm) \\
       Intrinsic dead time      & & 50 $\mu$s \\[1mm]
    Constraints:   & telemetry limit & 184\,cps \\
			& maximum  counts/observation/aimpoint & 450000 cts \\
     & linearity limit (on-axis point source) & \\ 
     & HRC-I~& $\sim$ 5\,cps (2 cps\ $\rm pore^{-1}$) \\ 
     &HRC-S~& $\sim$ 25\,cps (10 cps\ $\rm pore^{-1}$) \\ 

\hline

  \end{tabular}
  }
\end{table}

\subsubsection{Spatial Resolution \& Encircled Energy}

\begin{figure} 
\begin{center} 
\epsfysize=8cm
\epsfbox{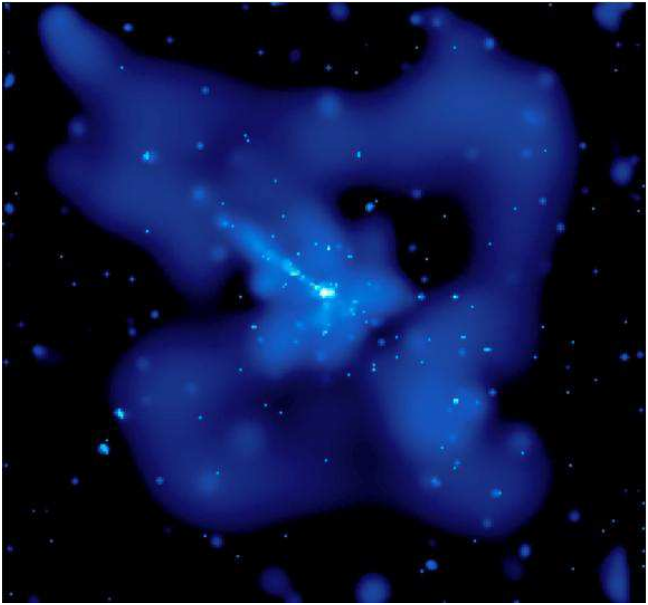} 
\caption{HRC-I image of Centaurus A covers 30' by 30' and highlights the large
field of view. 
\label{f:cenA}
}  
\end{center}
\end{figure}

The intrinsic PSF of the HRC is well modeled by a Gaussian with a FWHM of
$\sim20\mu$m ($\sim$ 0.4 arcsec). 
The \hrc\ pixels are $6.429 ~\mu$m (0.13175 arcsec). 
Approximately 90\% of the encircled energy lies within a 7-pixel radius
region (0.9 arcsec) from the center pixel.
The image resolution with the HRC at the focus degrades off-axis as
the telescope PSF increases with increasing off-axis angle and with increasing
deviation between the various HRC detection surfaces and the curved telescope
focal surface. Figure~\ref{f:cenA} is a spectacular example of an image made
with the HRC. 

\subsubsection{Energy Resolution}
The pulse-height amplitude of each event is telemetered, however, the energy
resolution is very poor and is not used for data analysis.  

\subsubsection{Spatial Variations}

\begin{figure} 
\begin{center} 
\epsfysize=8cm
\epsfbox{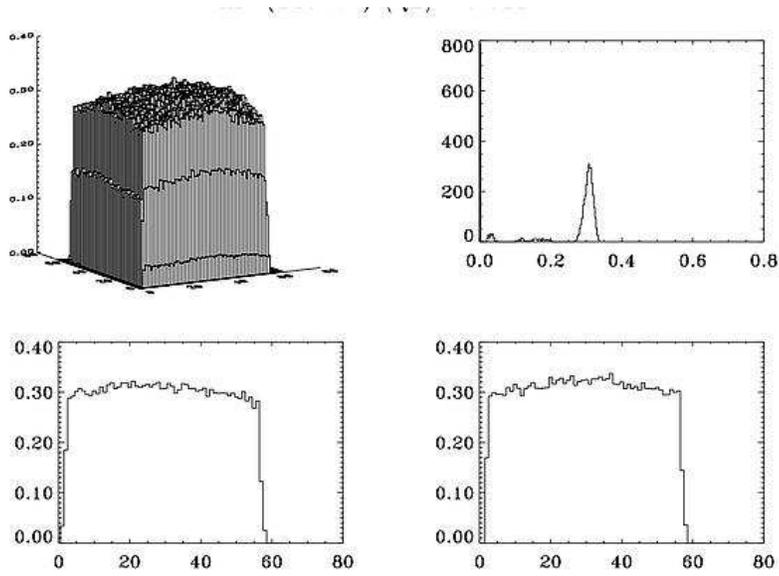} 
\caption{HRC-I quantum efficiency as a function of position at 1.5 keV.
Top left - as a function of position. Top right - distribution of quantum
efficiency values. Bottom left - efficiency as a function of one spatial
coordinate. Bottom right - efficiency as a function of the other spatial coordinate.
\label{f:hrc_i_flat}}  
\end{center}
\end{figure}

There is spatial variation in the MCP electron gain across both
instruments.  
However, the detector quantum efficiency is unaffected by this variation as
illustrated in Figure~\ref{f:hrc_i_flat}. 

\subsubsection{UV/Ion Shields}\label{sss:hrc_uv_ion}

\begin{figure} 
\begin{center} 
\epsfysize=8cm
\epsfbox{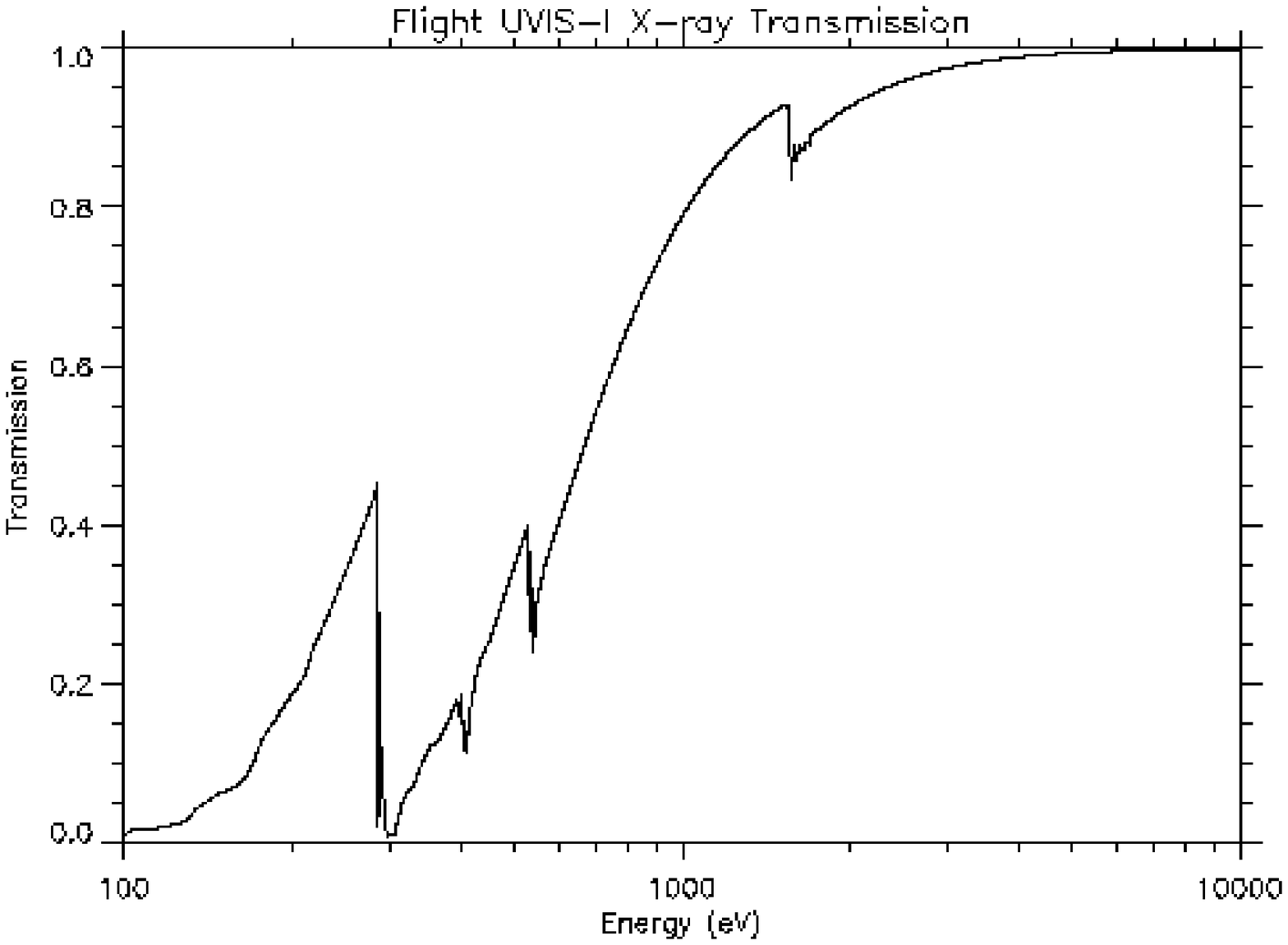} 
\caption{
The transmission of the HRC-I UV/Ion shield as a function of energy. 
\label{f:hrc_i_filter}
}
\end{center}
\end{figure}

\begin{figure} 
\begin{center} 
\epsfysize=8cm
\epsfbox{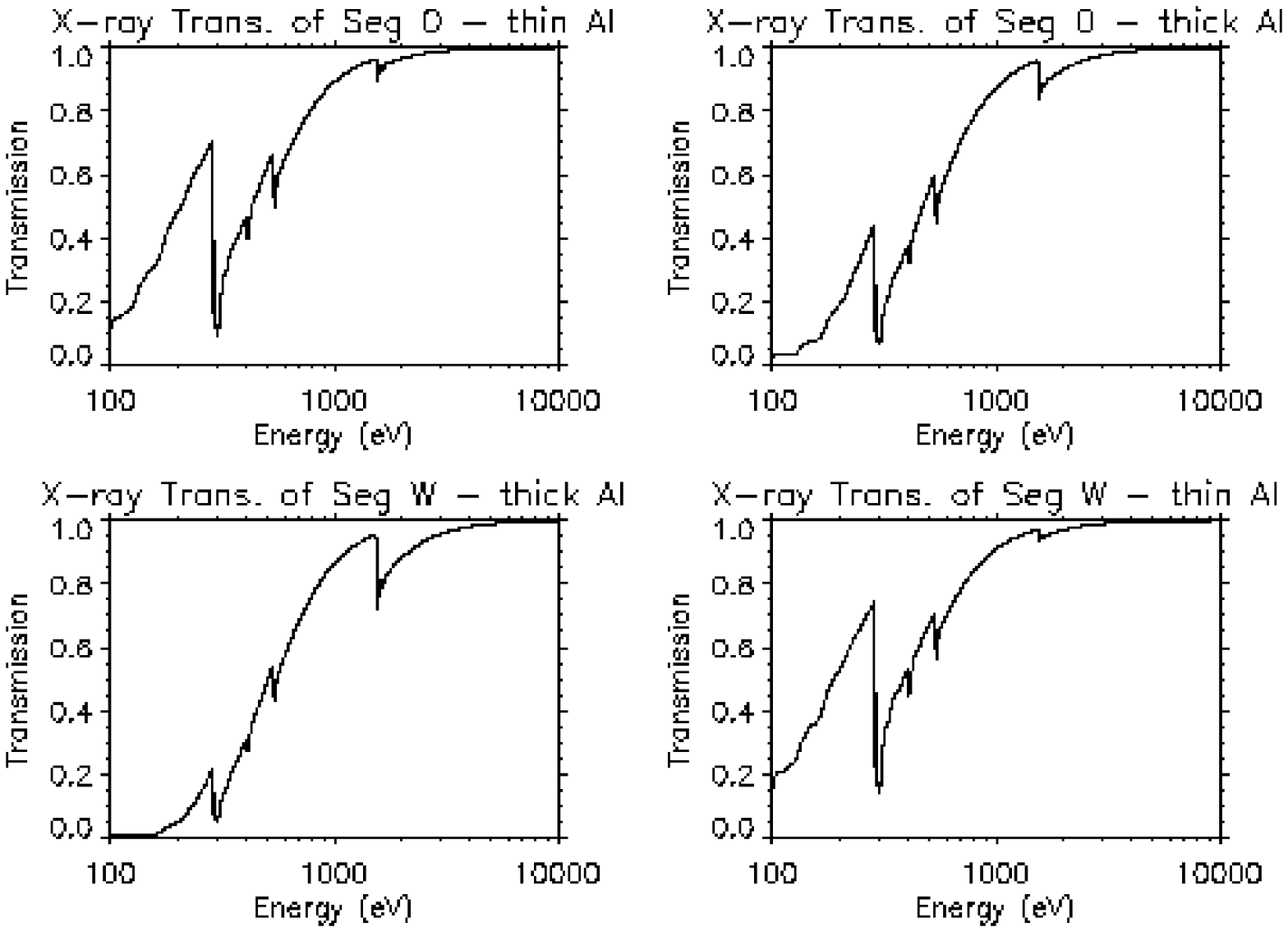} 
\caption{
The transmissions of the HRC-S UV/Ion shields for (upper left) the thin Al inner
segment (seg 0); (upper right) the thick Al inner segment (seg 0, ``T'');
(lower left) the thick Al outer segments (``wings'' or seg +1,-1); (lower
right) the thin Al outer segments (``wings'' or seg +1,-1).
\label{f:hrc_s_filter}
}
\end{center}
\end{figure}

The placement, composition, and thickness of various UV/ion shields (filters)
were shown in Figure~\ref{f:hrc_layout}. 
The shields are present to suppress out-of-band radiation from the ultraviolet
through the visible. 
This suppression is particularly important for observing sources which have
bright XUV and UV fluxes. 
The transmission of the UV/ion-shields are shown in Figures
~\ref{f:hrc_i_filter} and  ~\ref{f:hrc_s_filter}.
The HRC/S UV/ion shields are designed to allow the center segment to be used for
imaging and to allow the outer segments to detect the longer wavelengths
dispersed by the LETG, up to 170\AA.

As part of the in-flight calibration program the bright A-star Vega (A0 V,
U=0.02, B=0.03, V=0.03) was observed with both HRC detectors. 
The predicted HRC-I count rate (assuming no X-ray emission from the star) was
$7\times 10^{-4}$cps and an upper limit 
of $1~\times~10^{-3}$ cps was observed. 
The image of Vega was also placed on three regions of the HRC-S - the inner
segment ``T'', the thin aluminum inner segment, and on one of the thin aluminum
outer segments (Figure~\ref{f:hrc_layout}). 
The predicted count rates were 1, 400, and 2000 cps, respectively. 
The corresponding observed rates were 0.2, 240, and 475 cps, well within the
allowed uncertainty. 
The star Sirius was also observed with the HRC-S/LETG in order to obtain a soft
X-ray spectrum of Sirius B (a white dwarf).
Sirius A ( A1 V, V=-1.46, B-V=0.01) was seen in 0-th order at about the expected
count rate.
Thus, the filter is working within the accuracy of the pre-launch calibration 
and the performance has been stable over the first four years on-orbit.

Scattered UV, far-UV, and extreme-UV (XUV) light from the Sun or bright
Earth may cause a background strongly dependent on viewing geometry. 
The spacecraft was designed to limit the contribution from stray scattered
radiation to 0.01 ${\rm cts~cm^{-2}~s^{-1}}$ on the HRC. 
The imaged components of scattered radiation are dependent on the solar cycle,
but are at most $\sim$ 0.01 ${\rm cts~cm^{-2}~s^{-1}}$ for most lines of sight.

\subsubsection{Quantum Efficiency and Effective Area}

\begin{figure} 
\begin{center} 
\epsfysize=8cm
\epsfbox{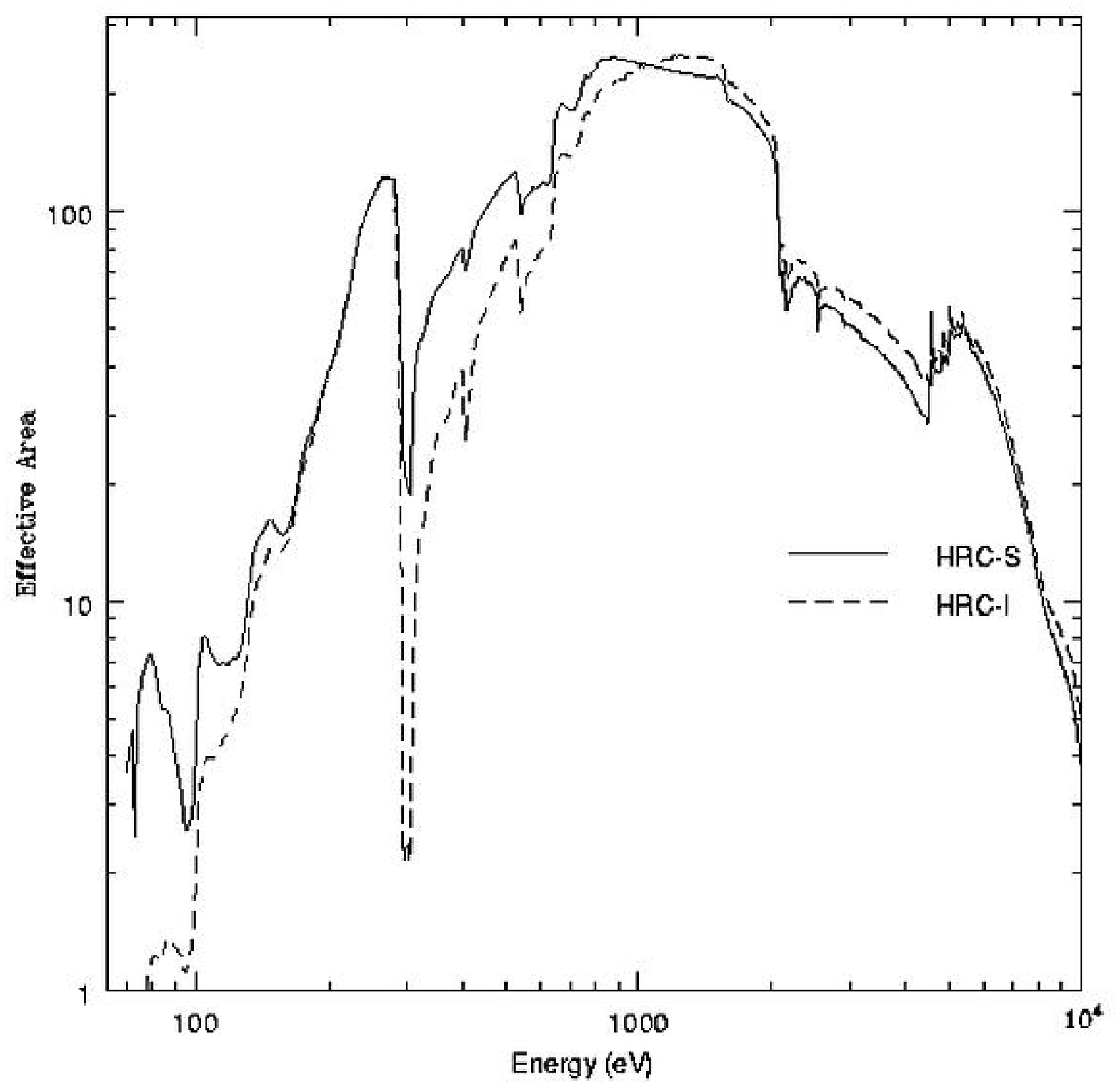}
\caption{
The telescope/HRC-I (dashed) and the center section of the telescope/HRC-S
(solid) effective area as a function of energy, integrated over the PSF.
Absorption edges are due to the iridium on the mirror, the CsI coating, and the
polyimide/Al UV/Ion shield.
\label{f:hrc_effa}
}
\end{center}
\end{figure}

The combined HRC-I and HRC-S effective areas are the product of the telescope
effective area, the quantum efficiency of the HRC detectors and the
transmission of the appropriate UV/Ion shield.
The current best estimates of these effective areas are shown, 
integrated over the point spread function, in Figure~\ref{f:hrc_effa}. 

\subsubsection{On-Orbit Background}\label{sss:hrc_bckgrd}

\paragraph{HRC-I}

The raw HRC-I counting rate on orbit is about 250 cps total. 
These are mostly cosmic ray events which are also detected in the
anti-coincidence shield. 
After anticoincidence, the valid event rate is about 50 c/s over the field
yielding a background rate of 10${\rm ^{-5}}$ ${\rm cts~s^{-1}~arcsec^{-2}}$. 
The background is generally flat, or at least smoothly varying over the field
with no more than a 20\% difference  between the center (higher rate) and edges
(lower rate) of the detector. 

\paragraph{HRC-S}

The anti-coincidence shield of the HRC-S is not working because of a timing
error in the electronics.  
As a result the event rate is very high and would exceed the total telemetry
rate limit, thus impacting observing efficiency. 
To cope with this problem, on-board data collection is limited to a region which
is about 1/2 of the full width and extends along the full length of the
detector. 
With this change, the quiescent background rate is about 85 cps.

Both HRC detectors experience occasional fluctuations in the background
due to variations in the charged particle background. 
These times of enhanced background are typically short (a few minutes to a few
tens of minutes) and are anywhere from a factor a 2 to a factor of 10 over the
quiescent rate. 
The increased background appears to be uniformly distributed over the detector
and introduces no apparent image artifacts.

\subsubsection{Ghost Images}\label{sss:hrc_ghost_images}

There is a very faint ``ghost'' image displaced 10 arcsec on one side of every
source in the field of view in the HRC-I.
These ghost images are due to events where the electronic signals saturate one
or more amplifiers in the event processing chain.
An event processing algorithm has been developed that eliminates the bulk of
this spurious image.
The intensity of the ghost is $< 3$\% of the source signal without filtering and
$< 0.1$\% after filtering.
The same feature is present in the HRC-S, but at a much reduced intensity.

\subsubsection {HRC Timing} \label{sss:hrc_timing}

The HRC time resolution of 16 $\mu$sec offers the highest precision timing of
the
two imaging cameras.
However, the HRC was mis-wired so that the time of the event associated with the
j-th trigger is that of the previous (j-th -1) trigger. 
If the data from all triggers were routinely telemetered, the mis-wiring would
not be problematical and could be dealt with by simply reassigning the time tag. 
Since the problem has been discovered, new operating modes have been defined
which allow one to telemeter all data whenever the total counting rate is
moderate to low, albeit at the price of higher background. 
For very bright sources the counting rate is so high that information associated
with certain triggers are never telemetered. 
In this case, the principal reason for dropping events is that the on-board,
first-in-first-out (FIFO) buffer fills as the source is introducing events at a
rate faster than the telemetry readout. 
Events are dropped until readout commences freeing one or more slots in the
FIFO. 
This situation can also be dealt with (Tennant et al. 2001) and time resolution
of the order of a millisecond can be achieved.

\begin{figure} 
\begin{center} 
\epsfysize=8cm
\epsfbox{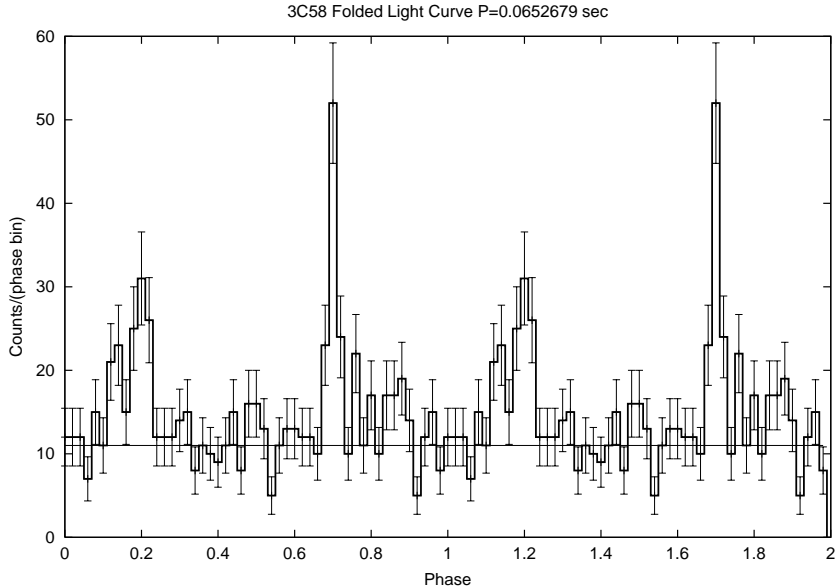}
\caption{
Light curve of the pulsing X-ray source in 3C58. 
\label{f:hrc_3C58}
}
\end{center}
\end{figure}

An example of the excellent timing performance of the HRC-S (Imaging) mode is
seen in the light curve for 3C58 (Murray et al. 2002) shown in
Figure~\ref{f:hrc_3C58}. 
3C58 is a supernova remnant with a central compact source which was suspected to
be a pulsar, but for which no pulsations had been previously observed. 
A \chandra~-HRC-S observation showed the pulse period to be about 65 msec and
the pulse profile to consist of a very short main pulse (a few msec wide) and a
somewhat broader interpulse.

More details of the HRC and its performance may be found in Murray et al. 2000, 
Kenter et al. (2000), and Kraft et al. (2000).

\section {Electron Proton Helium Instrument (EPHIN) \label{s:ephin}} 

Mounted on the spacecraft and near the X-ray telescope is a particle detector:
the Electron, Proton, Helium INstrument (EPHIN).
The EPHIN instrument was built by the Institut f\"ur Experimentelle und 
Angewandte Physik, University of Kiel, Germany, and a forerunner 
was flown on the SOHO satellite. 

EPHIN consists of an array of 6 silicon detectors with anti-coincidence. 
The instrument is sensitive to electrons in the energy range 250 keV - 10 
MeV, and hydrogen and helium isotopes in the energy range 5 - 53 MeV/nucleon. 
Electrons above 10 MeV and nuclei above 53 MeV/nucleon are registered with
reduced  capability to separate species and to resolve energies. 
The field of view is 83$^o$  with a geometric factor of 5.1 cm$^2$ sr. 

EPHIN is used to monitor the local charged particle environment as part of the
scheme to protect the focal-plane instruments from particle radiation damage.
EPHIN is also a scientific experiment in its own right. 
A detailed instrument description is given in Mueller-Mellin et al. (1995). 

\section{Gratings \label{s:gratings}}

\begin{figure} 
\begin{center} 
\epsfysize=8cm
\epsfbox{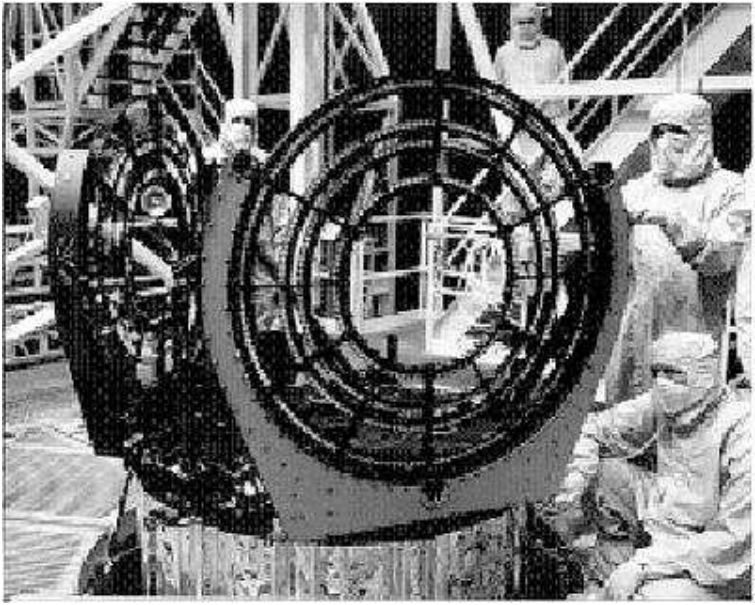}
\caption { 
Photograph of the LETG and HETG mounted to the spacecraft structure. 
Photograph is from TRW.
\label{f:gratings}
}
\end{center}
\end {figure}

Aft of the X-ray telescope are 2 objective transmission gratings (OTGs) -
the Low-Energy Transmission Grating (LETG) and the High-Energy Transmission
Grating (HETG).
Positioning mechanisms are used to insert either OTG into the converging beam
where they disperse the x-radiation onto the focal plane.
Figure~\ref{f:gratings} shows the gratings mounted behind the X-ray
telescope in their retracted position.

\subsection{LETG}\label{ss:letg}

The Space Research Institute of the Netherlands (SRON, Utrecht, Netherlands) and
the Max-Planck-Instit\"ut f\"ur extraterrestrische Physik (MPE, Garching,
Germany) designed and fabricated the LETG.  
The LETG is a slitless spectrometer employing a Rowland geometry.  
The diffraction grating itself is comprised of 540 individual gold grating
elements held in place by an aluminum structure designed such that the
individual elements form parts of the surface of a Rowland torus.  
The individual grating facets are circular with a diameter of approximately
1.6~cm. 
Each facet consists of bars with $0.5\mu \times 0.5\mu $ square cross-section,
held in place by a fine perpendicular grid of thicker bars and a triangular
lattice.  
The main characteristics of the LETG are summarized in Table~\ref{t:letg}.  

The LETG is put into operation by rotating the holding structure into its
place in the light path about 300~mm behind the exit aperture of the
telescope. 
The 540 grating facets, mounted 3 per module, then lie tangent to the Rowland
toroid which includes the focal plane.  
With free-standing gold bars of about 991-nm period, the LETG provides
high-resolution spectroscopy ($E/\Delta E$ $>1000$) between 80 and 175 \AA\/
(0.07 -- 0.15 keV) and moderate resolving power at shorter wavelengths.  
The nominal LETG~wavelength range accessible with the HRC-S as the detector is
1.2 -- 175 \AA~(0.07 -- 10 keV); ACIS-S coverage is 1.2 -- 65 \AA~(0.20 -- 10
keV).

  \begin{table}
  \caption{LETG~Parameters}
  \centering
\small{
  \label{t:letg}
\begin{tabular}{ll}\hline

Rowland diameter   &   $8637$ mm \\ %(effective value, subject to revision)\\
%      Grating material          &       gold\\
%      Facet frame material      &       stainless steel\\
%      Module material           &       aluminum \\
      LETG~grating parameters  &       \\
        \quad Period            &       0.99125 $\pm 0.000087$ $\mu$m \\
        \quad Thickness         &       0.474 $\pm 0.0305$ $\mu$m \\
        \quad Width             &       0.516 $\pm 0.0188$ $\mu$m \\
%        \quad Bar Side Slope    &       83.8 $\pm 2.27$ degrees \\
      Fine-support structure    &\\
        \quad Period            &       25.4$\mu$m\\
        \quad Thickness         &       2.5$\mu$m\\
%        \quad Obscuration       &       $<10\%$\\
%        \quad Dispersion        &       29.4~\AA/mm\\
%        \quad Material          &       gold\\
      Coarse-support structure  &\\
        \quad Triangular height &       2000$\mu$m\\
        \quad Width             &       68$\mu$m\\
        \quad Thickness         &       $<30\mu$m\\
%        \quad Obscuration       &       $<10\%$\\
%        \quad Dispersion        &       2320~\AA/mm\\
%        \quad Material          &       gold\\

Instrument Bandpass    & 1.2-175 \AA\ (70-10000 eV) (HRC-S) \\
                            & 1.2-65 \AA\ (200-10000 eV) (ACIS-S) \\
Resolution ($\Delta\lambda$, FWHM) &$0.05$ \AA\ \\
Resolving Power ($\lambda/\Delta \lambda$) &  $\geq 1000$ (50--160 \AA)\\
                                & $\approx20\times\lambda$ (3-50 \AA)\\
Dispersion                      & 1.148 \AA/mm\\

Detector angular size   & 3.37' $\times$ 101' (HRC-S) \\
                        & 8.3' $\times$ 50.6' (ACIS-S) \\
%Pixel size              & 6.43 $\times$ 6.43 $\mu$m (\hrcs) \\
%                        & 24.0 $\times$ 24.0 $\mu$m (ACIS-S) \\ 
Temporal resolution     & 16 $\mu$sec (HRC-S~in Imaging Mode, center segment
only) \\
                        & $\sim$10 msec (HRC-S~in default mode)\\
                        & 2.85 msec--3.24 sec (ACIS-S, depending on
                        mode) \\
\hline

    \end{tabular}
}
\end{table}

The on-orbit performance of the LETG is similar to pre-flight predictions
(e.g.\ Brinkman et al.\ 1997; Predehl et al.\ 1997; Dewey et al.\
1998),  though some minor problems with the readout detectors have provided
challenges for data analysis and the implementation of some observations.  
These are: (i) a high HRC-S background rate resulting from inoperability of
anti-coincidence veto of energetic particle events; (ii) small-scale spatial
non-linearity in HRC-S event position determination; (iii) the build-up of a
contamination layer on the ACIS-S filter that significantly reduces the
effective area of the LETG+ACIS-S combination for wavelengths longward of $\sim
18$~\AA\ ($\sim0.7$~kev); all of which were discussed previously in
\S~\ref{ss:acis} and ~\ref{ss:hrc}

The capabilities and usage of the LETG gratings are defined by their spectral
resolution and energy range.  
From an astrophysical perspective, salient attributes include: the ability to
study the very softest X-ray sources at high spectral resolution, such as hot
white dwarfs, novae, and isolated neutron stars, whose radiative output peaks
at wavelengths longward of 20~\AA\ or so ($\sim 0.6$~keV); coverage of two
decades in wavelength such that the precise shapes of the spectral energy
distributions of sources with strong non-thermal emission components can be
investigated, including intervening absorption systems; inclusion of the
resonance lines of all of the astrophysically important C, N and O trio for
study in emission and absorption; the highest spectral resolving power 
\chandra\ has to offer, reaching $\lambda/\Delta\lambda\sim 2000$ at $\ga
100$~\AA ; and simultaneous spectroscopy and high time resolution using the
HRC-S detector (\S~\ref{ss:hrc}). 

\subsubsection{Dispersion Relation}

The
dispersive properties of the LETG are calibrated and monitored in-flight using
the spectra of stellar coronae.  
The primary calibration target is the evolved binary Capella (G2~III+G8III).
While not ideal from the standpoint of being a binary - the projected
orbital speed of the two components is 36~km~s$^{-1}$ and can cause discernible
shifts and broadening of spectral lines at \chandra's highest resolution -
Capella is the brightest coronal X-ray source in the sky and provides
calibration information with the minimum of observing time.

Stellar coronal spectra are dominated by emission lines from highly ionized
astrophysically abundant elements---predominantly C, N, O Ne, Mg, Si, Ar and
Fe.  
Shortward of 40~\AA\ (0.3~keV) or so, the H-like and He-like ions of these
elements provide resonance lines with very precisely calculated wavelengths
that are ideal for examining and monitoring dispersion relations. 
In principle, these lines can be supplemented with the forest of Fe lines from
the ``L-shell'' complex (transitions among valence electrons with n=2 ground
states).
However, at the LETG resolution of up to $\lambda/\Delta\lambda\sim 2000$, many
of the wavelengths of these lines are known much less precisely and must be
treated with caution.  
Longward of 40~\AA\, laboratory wavelengths of light element L-shell lines and
transitions of the type $\Delta n=0$ in Fe L-shell ions provide the
wavelength benchmarks.  
Before use in calibration, lines are first screened based on spectrum
synthesis-type calculations for the presence of significant blends that might
skew the observed wavelengths from their true values.

\begin{figure}
\begin{center}
\epsfysize=8cm
\epsfbox{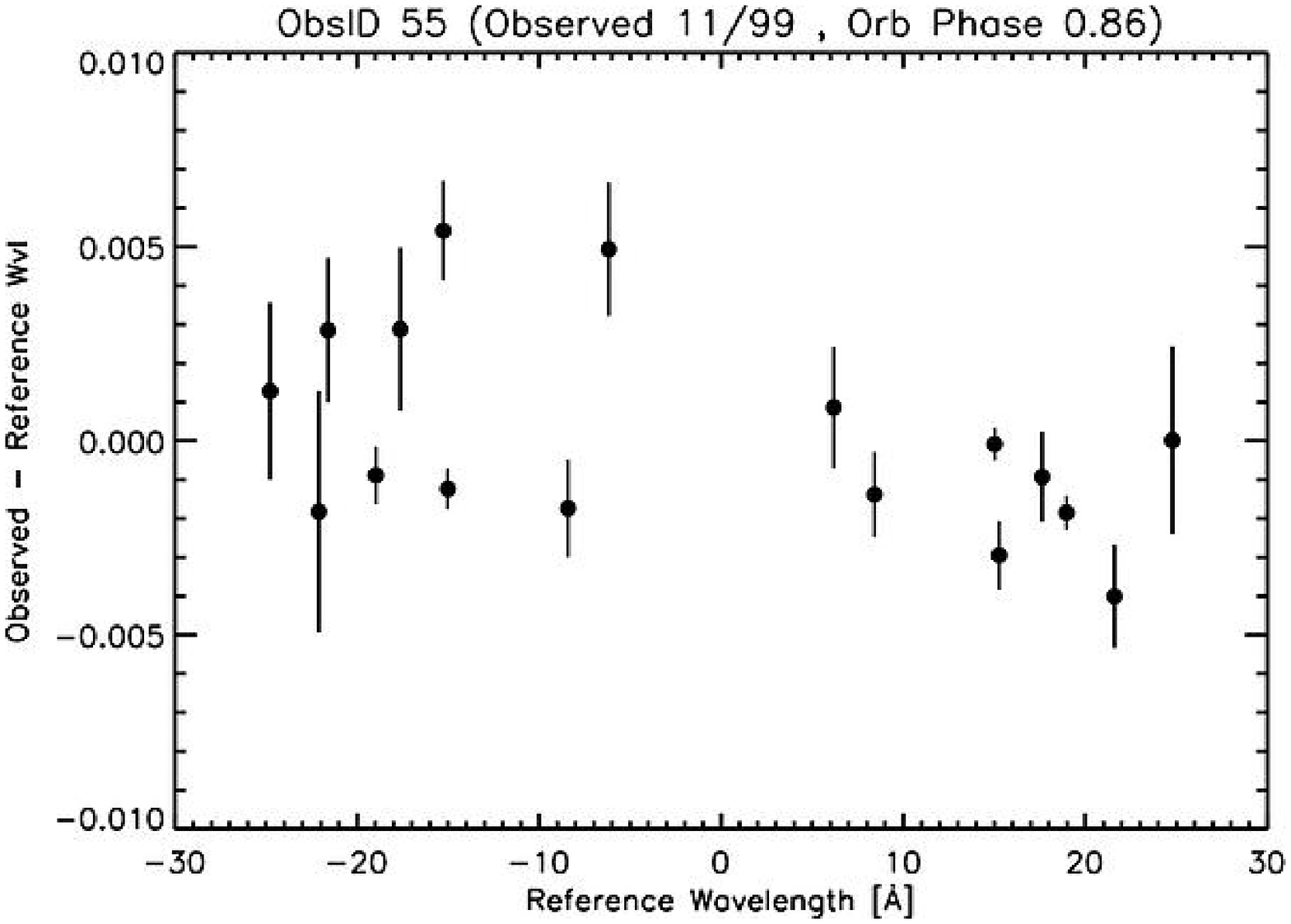}
\epsfysize=8cm
\epsfbox{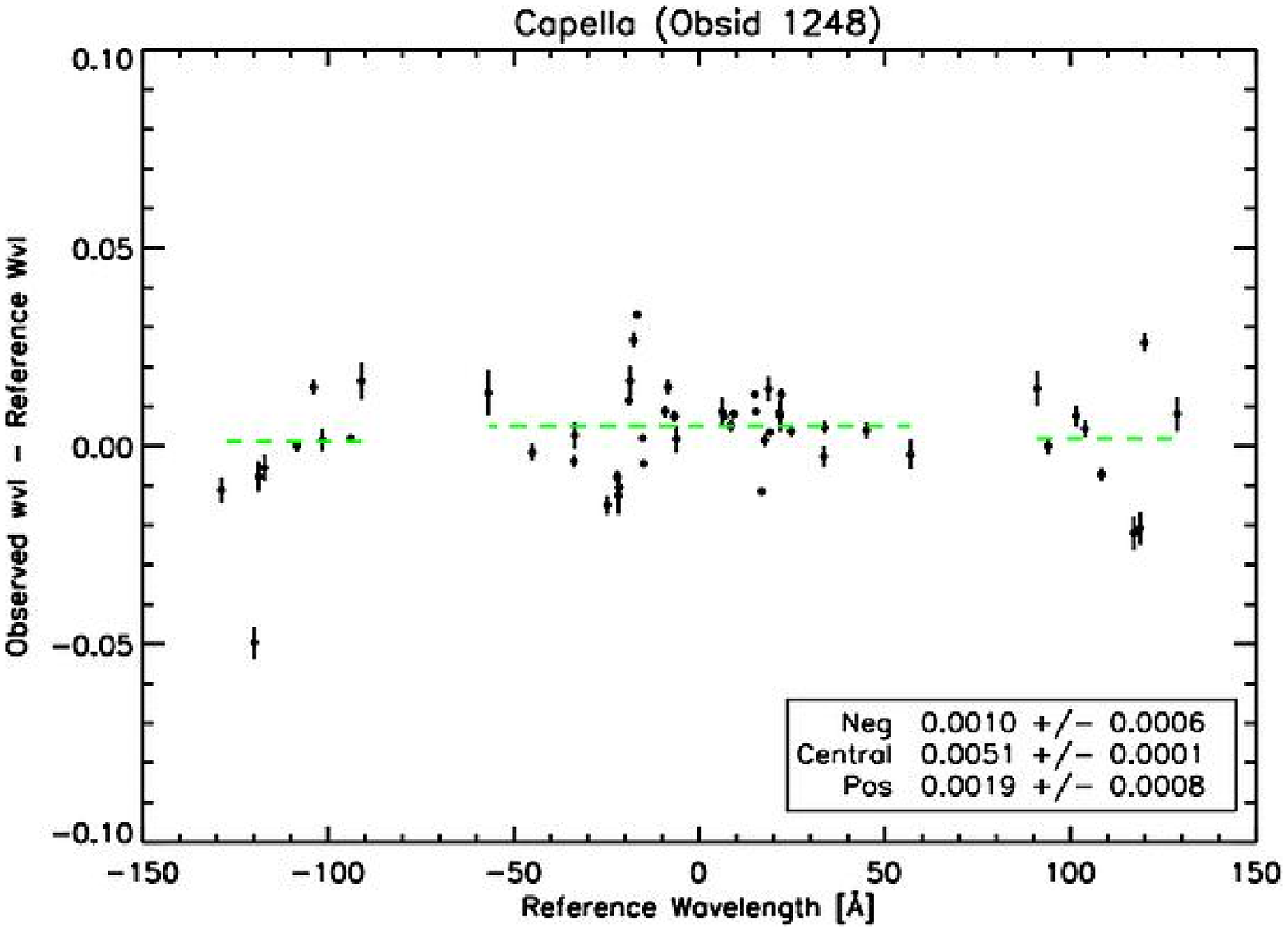}
\caption{The difference between observed and theoretical wavelengths as a
function of wavelength for relatively unblended spectral lines in the coronae
of Capella, seen by the LETG+ACIS-S (top), LETG+HRC-S (bottom).  
Small departures from zero are expected based on the residual orbital velocities
and on the presence of hidden blends.  
In the case of the HRC-S, however, the much larger departures are cause by
imaging non-linearities in the detector.
}
\label{f:letg_disprel}
\end{center}
\end{figure}

The current LETG dispersion relations with the HRC-S and ACIS-S detectors as
illuminated by Capella (after correction for the spacecraft and Capella
systemic radial velocity differences) are shown in Figure~\ref{f:letg_disprel}.  
In the case of ACIS-S, the RMS deviation in predicted vs.\ observed line
wavelength amounts to only $\sim 0.002$~\AA\ or 0.01~\%.  
The scatter seems larger than the errors bars suggest should be the case; this
is likely due primarily to hidden line blends.  
To put these results in perspective, the net orbital velocity of the Capella
components is of a similar magnitude and, ultimately, we do not know from which
binary component the lines are emitted.

In the case of the HRC-S the RMS deviation between observed and predicted
wavelengths for the set of relatively unblended lines observed amounts to
0.013~\AA---almost an order of magnitude higher than with ACIS-S, though at
100~\AA\ this again
amounts to an effect of order 0.01\% . 
Close examination of Figure~\ref{f:letg_disprel}, however, reveals especially
large departures, especially at wavelengths falling on the
outer HRC-S plates, and on the central plate near 20~\AA.  
These differences between predicted and observed wavelengths have been found
to be caused by small non-linearities in the imaging characteristics of the
detector.  
These are understood in terms of small errors in the positions of events, as
determined by the combination of detector electronics and
ground telemetry processing.
Correcting for these small position errors is not trivial and future
improvements to the dispersion relation will rely on empirical corrections
based on a large accumulation of data such as that in
Figure~\ref{f:letg_disprel}.

\subsubsection{Resolving Power}

\begin{figure}
\begin{center}
\epsfysize=8cm
\epsfbox{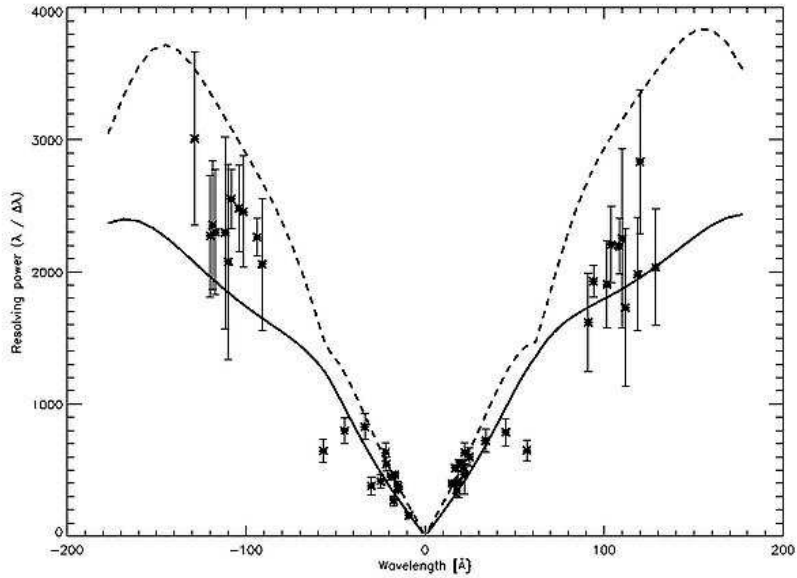}
\caption{LETG/HRC-S spectral resolving power derived from lines thought not to
be affected significantly by blending from several observations of Capella and
Procyon.  
Measured line widths were corrected for source orbital, rotational, and thermal
motions.
Optimistic (dashed line) and conservative (solid line) predictions
of pre-flight models are also shown.  
The deviations from approximate linearity near $\pm60$ \AA\ and at the longest
wavelengths arise from deviations of the HRC-S surface from the Rowland circle.
Deviations from a smooth curve are likely caused by hidden blends not predicted
by the radiative loss model, or by imaging non-linearities of the HRC-S along
the dispersion axis.}
\label{f:letg_res} 
\end{center}
\end{figure}

The dominant contribution to the LETG line response function (LRF) and
instrument resolving power is the telescope \psf, which is $\sim25~\mu$m FWHM,
depending on energy.  
When the LETG is used with the HRC-S, the intrinsic uncertainty in photon
position determination adds another small contribution of order 15-$20~\mu$m.
The LETG itself does not contribute any significant broadening.
Uncertainties in correcting photon event positions for the observatory
aspect, which occurs during ground data processing, also introduces an
additional, but small, blurring of order a few $\mu$m.
For spectral lines with $\la 1000$ counts, the combined LETG+HRC LRF can be
well-approximated by the function

\begin{equation} 
I(\lambda) = {[ 1 + \left({\lambda\over{\lambda_c}}\right)^2]^{-\beta}} 
\end{equation}
with FWHM$\sim40$~$\mu$m, or $\sim 0.05$~\AA.  
The LETG~resolving power determined by fitting this function to lines seen in
spectra of Capella and Procyon is illustrated in Figure~\ref{f:letg_res},
together with pre-flight model predictions.

\begin{figure}
\begin{center}
\epsfysize=8cm
\epsfbox{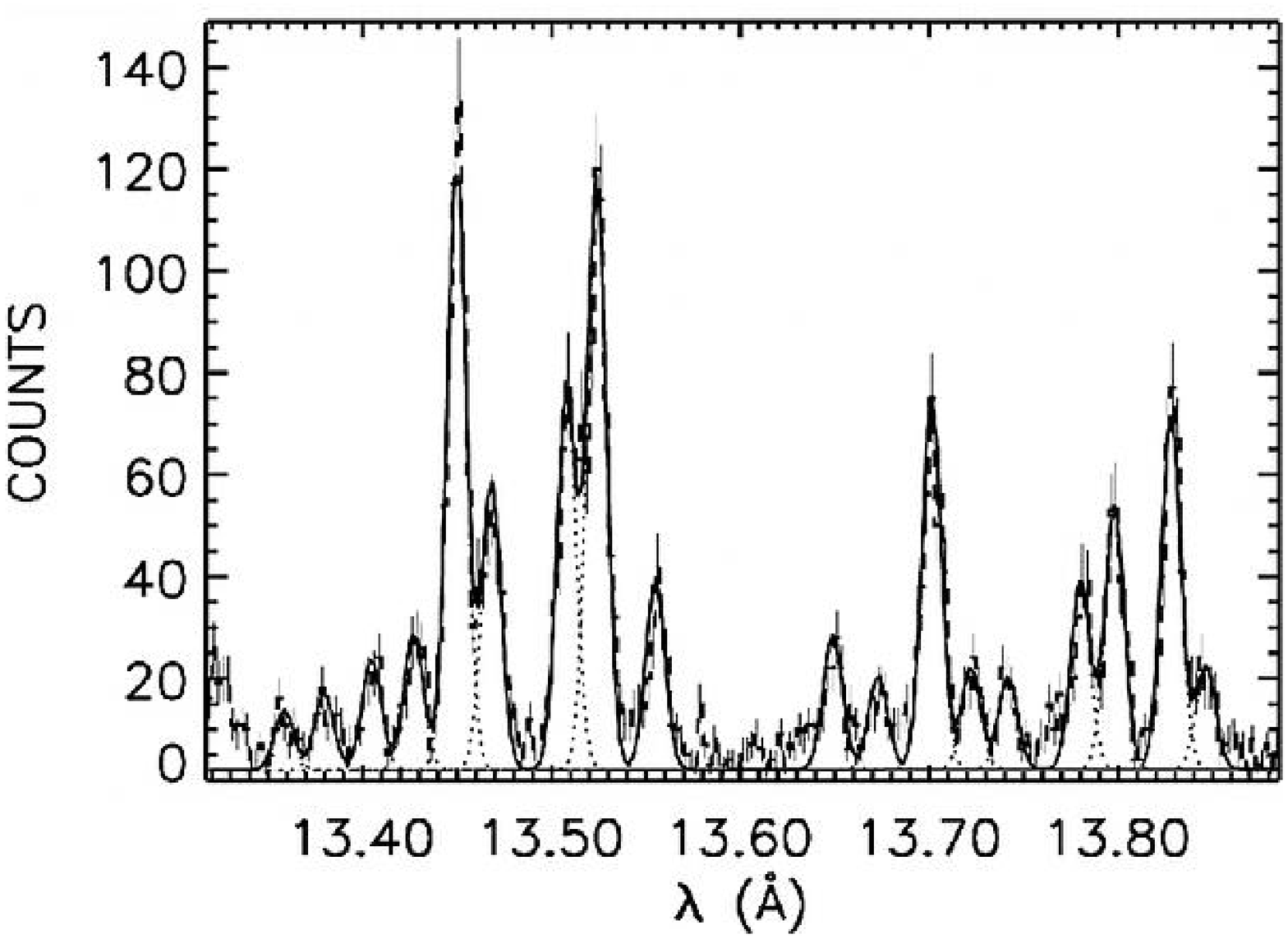}
\epsfysize=8cm
\epsfbox{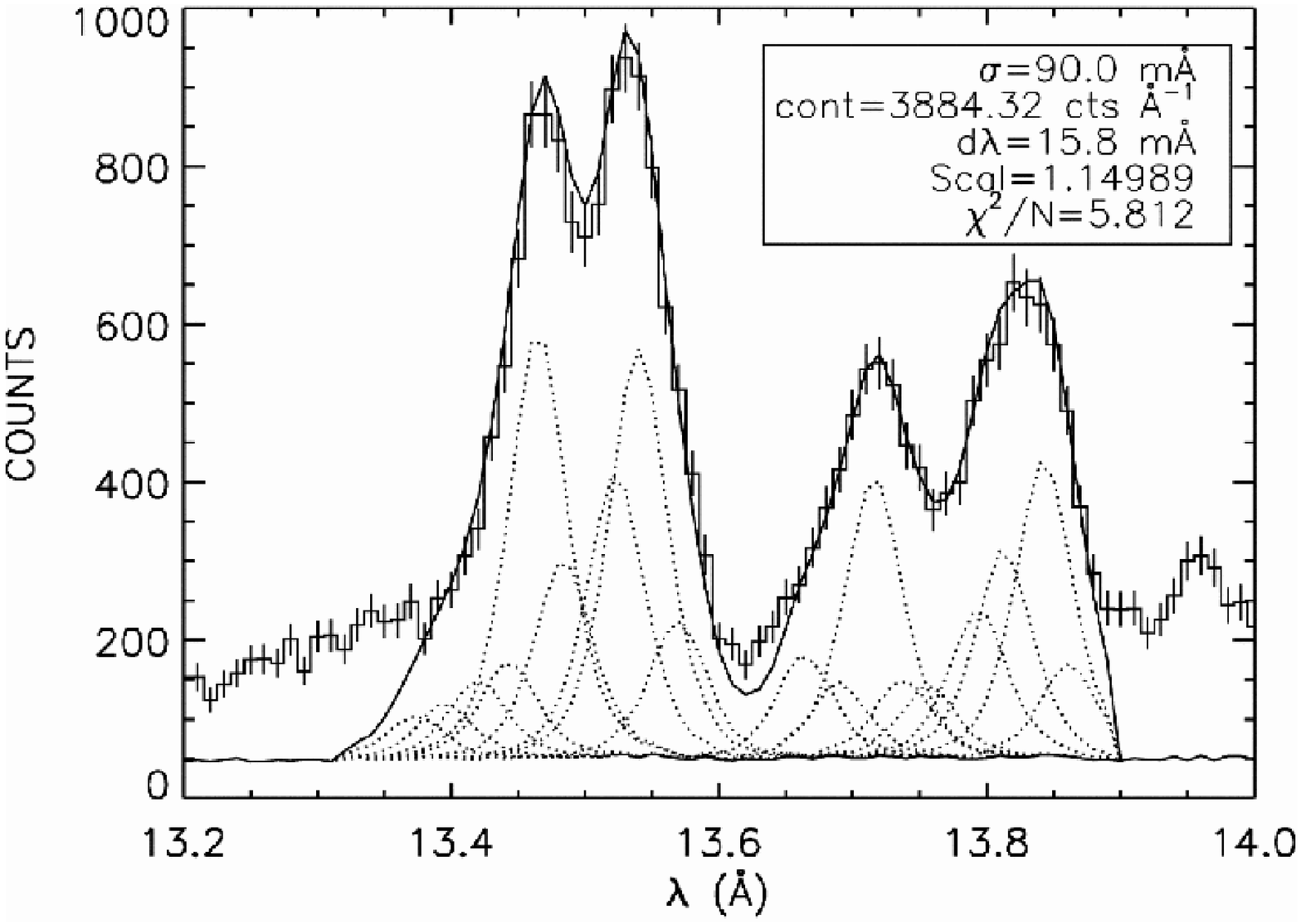}
\caption{Top: HEG summed plus and minus first order spectrum of Capella in the
13-14~\AA\ range, fitted with a model comprised of 18 emission lines and a
constant continuum. 
Bottom: The best-fit model obtained from the HEG spectrum, scaled and overlaid
on the LETG spectrum. 
The scaling parameters are listed in the top right inset.
From Ness et al.\ (2003).}
\label{f:letg_ne9}
\end{center}
\end{figure}

That the current LETG LRF is now quite well-described has recently been
demonstrated by Ness et al. (2003), who have succeeded in modeling the
spectrum of Capella in the extremely complex 13-14~\AA\ region that contains
the He-like Ne density-sensitive lines.  
At the LETG resolution, these lines are heavily blended with Fe lines -
predominantly Fe~XIX.  
This spectral region cannot simply be modeled based on optically thin radiative
loss models because the exact placement of the Fe lines is crucial, and model
wavelengths are not of sufficient accuracy. 
Ness et al.\ (2003) circumvented this problem by examining in detail the same
spectral region as seen using the HETG (\S~\ref{ss:hetg}), in which individual
blending lines could be empirically identified and isolated. 
Using the HETG-derived line list, and relative intensities, Ness et al.\ (2003)
reconstructed the LETG spectrum and were able to produce an impressive fit to
the observations as illustrated in Figure~\ref{f:letg_ne9}.

\begin{figure}
\begin{center}
\epsfysize=12cm
\epsfbox{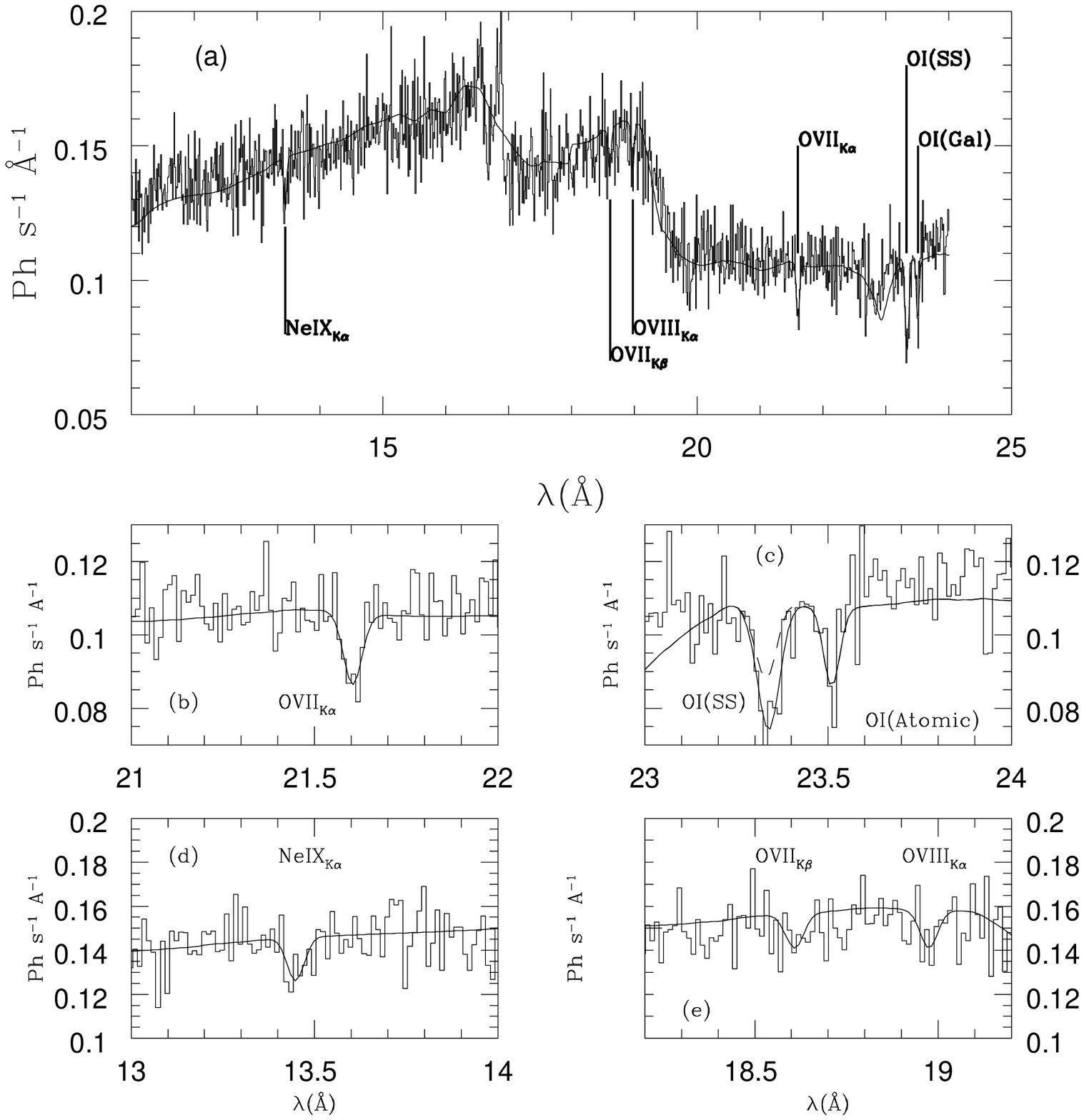}
\caption{The LETG/HRC-S spectrum of PKS~2155-304 in the 11-24~\AA\ range
illustrating the continuum (well-described by a power law with Galactic
absorption), together with six absorption lines (solid line). The
apparent continuum shape reflects the instrument response.
Four portions of the spectrum are magnified: (b) O VII K; (c) atomic and
solid-state (i.e., molecular) O I; (d) Ne IX K; (e) O VII K and O VIII
K. }
\label{f:letg_nicastro}
\end{center}
\end{figure}

While the LETG LRF is slightly broader than a Gaussian of the same FWHM away
from line center, a key characteristic is this rather modest extension of the
line wings. 
This greatly aides the sensitivity of the instrument: overlapping line wings are
a source of noise for understanding a true continuum level, or for extracting
line fluxes.
One example where this sensitivity was crucial is the detection of narrow line
absorption in the spectrum of the blazar PKS2155-304 by Nicastro et al.\
(2002), illustrated in Figure~\ref{f:letg_nicastro}. 
O VII K$_\alpha$ and Ne IX K$_\alpha$ resonant absorption lines were detected,
together with a more tentative identification of absorption due to O VIII
K$_\alpha$ and O VII K$_\beta$. 
In the same line-of-sight toward PKS~2155-304, the Far Ultraviolet Spectroscopic
Explorer spectrum shows complex O VI 2s2p absorption, including one
significantly blue-shifted component.
Nicastro et al.\ (2002) interpreted the combined X-ray and UV data in terms of
absorption in a low-density intergalactic plasma collapsing toward our Galaxy. 
The presence of such an absorber had been predicted by numerical simulations of
the warm-hot intergalactic medium (e.g.\ Hellsten et al.\ Kravtsov 1998; et
al.\ 2002).  
Detecting these shadows of our local Universe is extremely difficult even with
observatories such as  \chandra\ - the absorption features are relatively
weak, and narrow.  
A spectrometer such as the LETG is required: one with high spectral resolution,
a sharp, well-defined instrumental profile, and with sufficient sensitivity
throughout the soft X-ray range to accumulate high continuum signal from the
background source.

\subsubsection{Bandpass and Effective Area}

\begin{figure}
\begin{center}
\epsfysize=8cm
\epsfbox{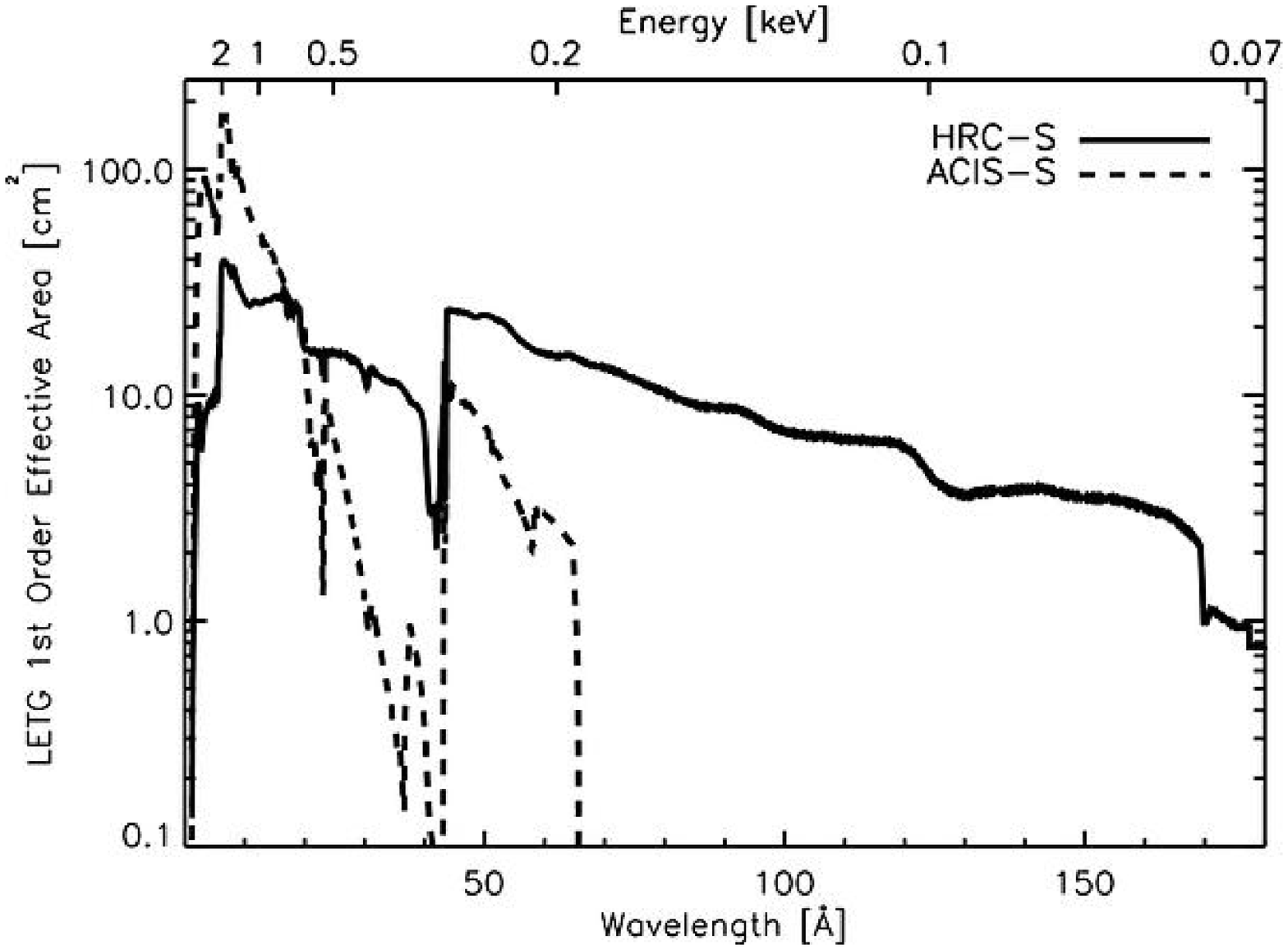}
\caption{  \chandra\ LETG/HRC-S and LETG/ACIS-S first-order effective areas
as a function of wavelength and energy.}
\label{f:letg_areas}
\end{center}
\end{figure}

The first order effective areas of the LETG with both HRC-S and ACIS-S detectors
is illustrated in Figure~\ref{f:letg_areas}.  
Toward the short wavelength limit of $\sim 1.2$~\AA\ ($\sim 10$~keV), the area
is limited by the rapidly falling reflectivity of the telescope combined with
the decreasing efficiency of first order diffraction --- the LETG grating bars
become effectively transparent to such short wavelength photons.  
Both LETG/ACIS-S and LETG/HRC-S effective areas are characterized by a step near
6~\AA, corresponding to the increase in reflectivity of the X-ray telescope
longward of the Ir M edges.  
This peak is accentuated by a peak in first order diffraction efficiency.  
Toward longer wavelengths, both the LETG diffraction efficiency and the X-ray
telescope reflectivity are essentially flat, and the shapes of the
effective area curves are dominated by the transmittance of the detector
filters and the detector quantum efficiencies, including any contamination.  
Easily visible in these curves are the absorption edges of C, N and O, that are
major constituents of the polyimide filters.

\begin{figure}
\begin{center}
\epsfysize=8cm
\epsfbox{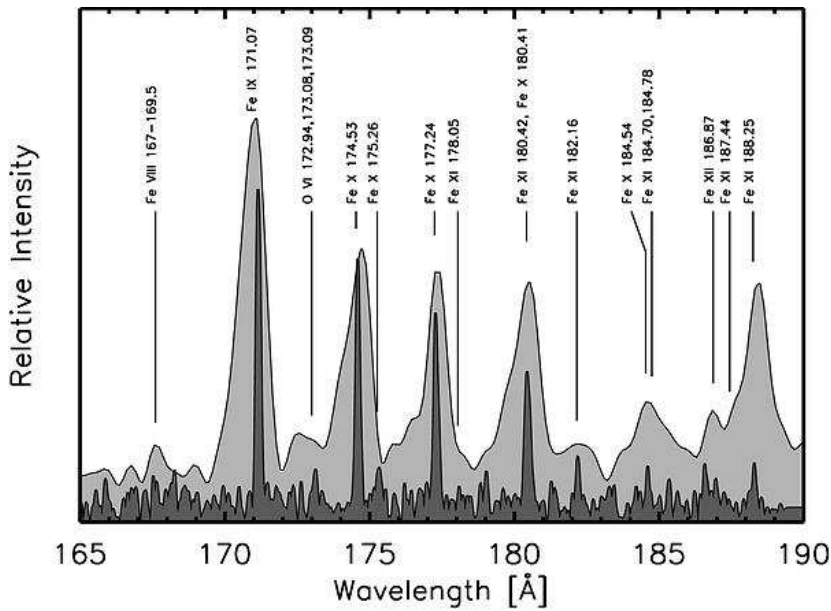}
\caption{\chandra\ LETG and EUVE Medium Wavelength Spectrometer spectra of the
165-190~\AA\ range in Procyon (F5~IV).  
The \chandra\ spectrum was obtained with the target
off-axis in the LETG dispersion direction so as to extend the wavelength range.  
The prominent lines are due to intermediate charge states of Fe.}
\label{f:letg_procyon}
\end{center}
\end{figure}

With the HRC-S detector, the nominal LETG long wavelength cut-off is
170~\AA\ (0.07~keV).  
This limit is simply a dictate of the physical extent of the detector.  
It is interesting that the Al coating of the HRC-S UVIS becomes more transparent
longward of the Al~L edges near 170 and 171~\AA, and, by acquiring a target at
an off-set position along the dispersion axis and at a small cost in
resolution, the wavelength coverage can actually be stretched beyond 170~\AA. 
A spectrum of the late-type subgiant Procyon (F5~IV) was obtained in this way
and is compared in Figure~\ref{f:letg_procyon} with the same spectral region
observed by the Extreme Ultraviolet Explorer (EUVE).

\begin{figure}
\begin{center} 
\epsfysize=8cm
\epsfbox{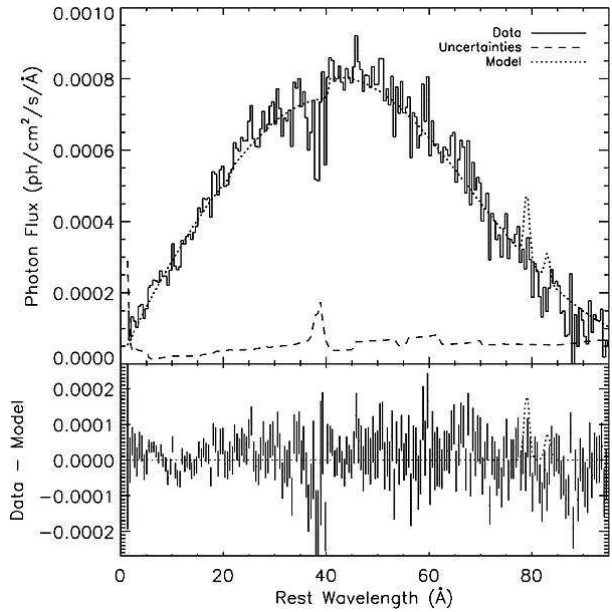} 
\caption{First order LETG+HRC-S spectrum of Mrk 478 binned at 0.5 \AA\
(top) and residuals from the best-fit power-law model.  
Residuals near the C-K edge in first order most likely result at least partly
from systematic errors in the effective area calibration.  
The continuum appears bereft of line absorption or emission features, contrary
to the expectations of Hwang \& Bowyer (1997) based on their analysis of EUVE
spectra. 
The particular lines suggested as being present in EUVE spectra by Hwang \&
Bowyer (1997) are included in the model in the top panel (but were not used to
calculate the residuals in the bottom panel).}
\label{f:letg_mk478}
\end{center}
\end{figure}

In order to interpret spectra correctly over the whole of the soft X-ray
range---covering two orders of magnitude in photon energy---instrumental
calibration becomes a vital concern.  
The LETG/HRC-S effective area calibration has been adjusted post-launch
based on the observed spectra of ``well understood'' cosmic sources
and on cross-calibration with the other \chandra\ instruments, and
is believed to be accurate to 15\%\ or so (absolute) in first order.
Marshall et al.\ (2003) were able to match the LETG+HRC-S spectrum of
Mkn~478 with a pure continuum model from 1.2-100~\AA, applying only
minor adjustments to the contributions from higher spectral orders
that are mixed in with the first order signal.  Mkn~478 lies in a
direction out of the galaxy that has a particularly low neutral
hydrogen column density, and so remains a strong source at these
longer wavelengths.  The first order spectrum is illustrated in
Figure~\ref{f:letg_mk478}.  Using these data Marshall et al.\ (2003) argued for
the absence of both a strong warm absorber, as expected based on other Seyfert
1 observations (see \S\ref{s:discoveries}) and a lack of emission lines at
longer wavelength that were predicted based on an analysis of earlier EUVE
spectra (Hwang \& Bowyer 1997).

\subsection{HETG\label{ss:hetg}}

The Massachusetts Institute of Technology (MIT, Cambridge, Massachusetts)
designed and fabricated the HETG.
The HETG employs 2 types of grating facets~--- the Medium-Energy Gratings (MEG),
mounted behind the X-ray telescope's 2 outermost shells, and the High-Energy
Gratings (HEG), mounted behind the X-ray telescope's 2 innermost shells.
With polyimide-supported gold bars of 400-nm and 200-nm periods, respectively,
the HETG provides high-resolution spectroscopy from 0.4 to 4 keV (MEG, 30 to 3
\AA) and from 0.8 to 8 keV (HEG, 15 to 1.5 \AA).

\begin{figure}
\begin{center} 
\epsfxsize=15cm
\epsfbox{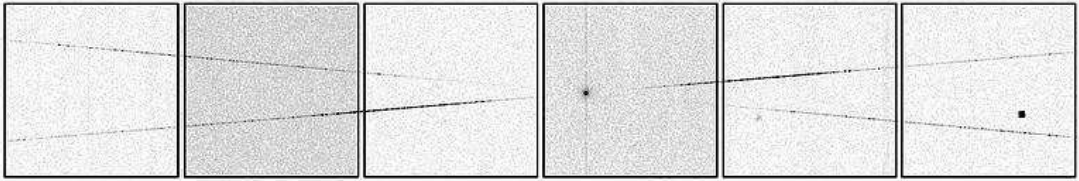} 
\caption{The X-ray image of the HETG spectrum of Capella. The image of the flux
dispersed by the MEG is in the line extending from the lower left to the upper
right. The bright dot in the center is the image of the zeroth order. The bright
square to the right is the effect of a bright pixel and the dither motion.}
\label{f:hetg_x}
\end{center}
\end{figure}

There are 192 medium energy gratings (MEGs), each about 25 mm square, that
intercept rays from the outer telescope shells and are optimized for medium
energies.
There are 144 high energy gratings (HEGs), also 25 mm square, that intercept
rays from the two inner shells and are optimized for high energies.  
Both gratings are mounted on a single support structure and therefore used
concurrently.  
The two sets of gratings are mounted with their grating bars at different angles
so that the dispersed images from the HEG and MEG will form a shallow $X$
centered on the undispersed (zeroth order) position; one leg of the $X$ is from
the HEG, and the other from the MEG (see Canizares et al. 2000 and
Figure~\ref{f:hetg_x}).  
The HETG is designed for use with the spectroscopic array of ACIS-S, although
other detectors may be used for particular applications.
Here we restrict the discussion to use with ACIS-S.

\begin{figure}
\begin{center} 
\epsfxsize=15cm
\epsfbox{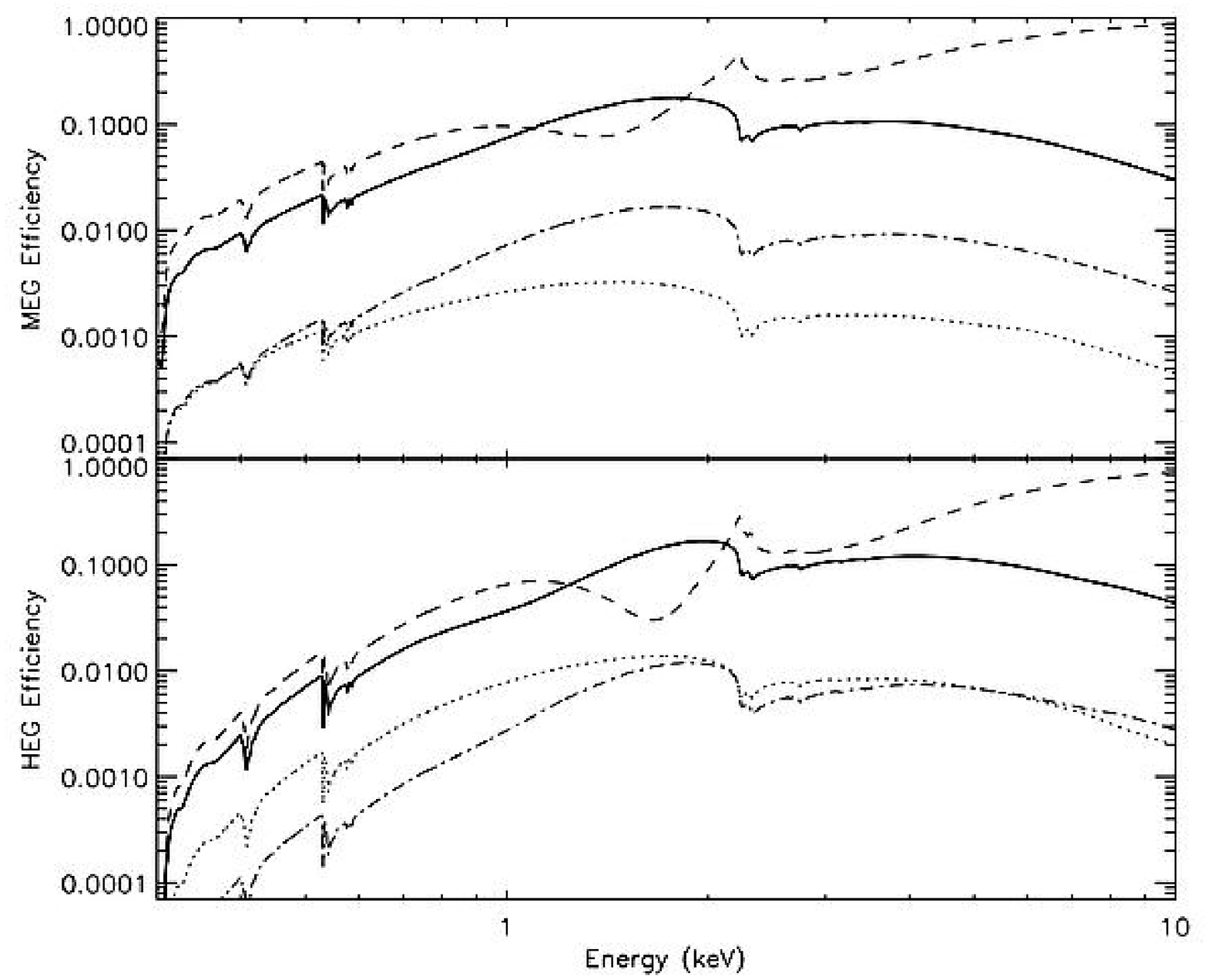} 
\caption{The MEG (top) and HEG (bottom) dispersion efficiencies as a function of 
energy for orders 0 (dashed), 1 (solid), 2 (dotted), and 3 (dash-dot).}
\label{f:hetg_efficiency}
\end{center}
\end{figure}

The gratings are electroplated gold bars supported on polyimide
membranes.  
The heights and widths of the bars were chosen to maximize the efficiency into
first order and reduce the zeroth order throughput.  
Setting the bar width to half of the grating period suppresses even orders and 
improves first order efficiency for rectangular bar profiles.  
This nulling of even orders is achieved in the MEGs.  
For the HEGs, the bar height was increased in order to improve the first order
efficiency above 1.2 keV where the gold becomes partially transparent. 
In this case the fabrication process produced bar widths that were 60\% of the
period.
A consequence is that the second order efficiency is comparable
to that in third order. 
The MEG and HEG efficiency for the zeroth, first, second, and third orders are
shown as a function of energy in Figure~\ref{f:hetg_efficiency}.
A summary of HETG characteristics is given in Table~\ref{t:hetg}.

  \begin{table}
    \caption{HETG Parameters}
    \label{t:hetg}
    \centering
    \begin{tabular}{|lll|}
	\hline
	 &	HEG	&	MEG	\\
	\hline
      HETG Rowland Spacing & \multicolumn{2}{c|}{$8632.65$ mm }\\
      HETG Effective Area			&
						\multicolumn{2}{c|}{$28\rm\,cm^2$ @ 6.5 keV}\\
	\quad (MEG+HEG first orders,	&
						\multicolumn{2}{c|}{$200\rm\,cm^2$ @ 1.5 keV}\\
	\quad\quad  with ACIS-S ) 		&
						\multicolumn{2}{c|}{$59\rm\,cm^2$ @ 1.0 keV}\\
      Energy Range	&		$0.8-10.0$ keV & $0.4-5.0$ keV \\
      Wavelength Range	&		$1.2-15$ \AA  & $2.5-31$\AA \\
      Resolving Power ($E/\Delta E$) &	$1070-65$ & $970-80$ \\
      Resolution (FWHM, \AA)	& 	$0.012$ \AA~FWHM	& 	$0.023$ \\

      Absolute Wavelength Accuracy (\AA) & $ \pm 0.006$ & 	$ \pm 0.011$ \\
      Relative Wavelength Accuracy (\AA) & $ \pm 0.0028$ & 	$ \pm 0.0055$ \\

      Angle on ACIS-S ($^\circ$)		&	$-5.235 \pm 0.01$ &	$4.725 \pm 0.01$ \\
      Wavelength Scale (\AA~/ ACIS pixel) & 0.0055595 & 0.0111185 \\
      Diffraction Efficiency, 0.5 keV &	1.4\% & 3.6\% \\
	\quad\quad ($\pm$1 orders), 1.5 keV	& 22\% & 32\%  \\
		\quad\quad\quad\quad	6.5 keV & 17\% & 13\%  \\
      HETG Zeroth-order Efficiency at 0.5 keV & \multicolumn{2}{c|}{4.5\%}\\
		\quad\quad\quad\quad	1.5 keV  & \multicolumn{2}{c|}{8\%}\\
		\quad\quad\quad\quad	6.5 keV  &	\multicolumn{2}{c|}{60\%}\\
	%%dd Facet parameters
      Grating Facet Parameters	&	& \\
      \quad Bar material  &		Gold	&	Gold\\
      \quad Period (\AA)  &			2000.81 & 4001.41 \\
      \quad Bar thickness (\AA)	&		5100 & 3600 \\
      \quad Bar width (\AA)   &         1200 & 2080 \\
      \quad Polyimide support thickness (\AA) & 9800 & 5500  \\
	\hline
    \end{tabular}
  \end{table}

\subsubsection{Imaging}

\begin{figure}
\begin{center} 
\epsfysize=8cm
\epsfbox{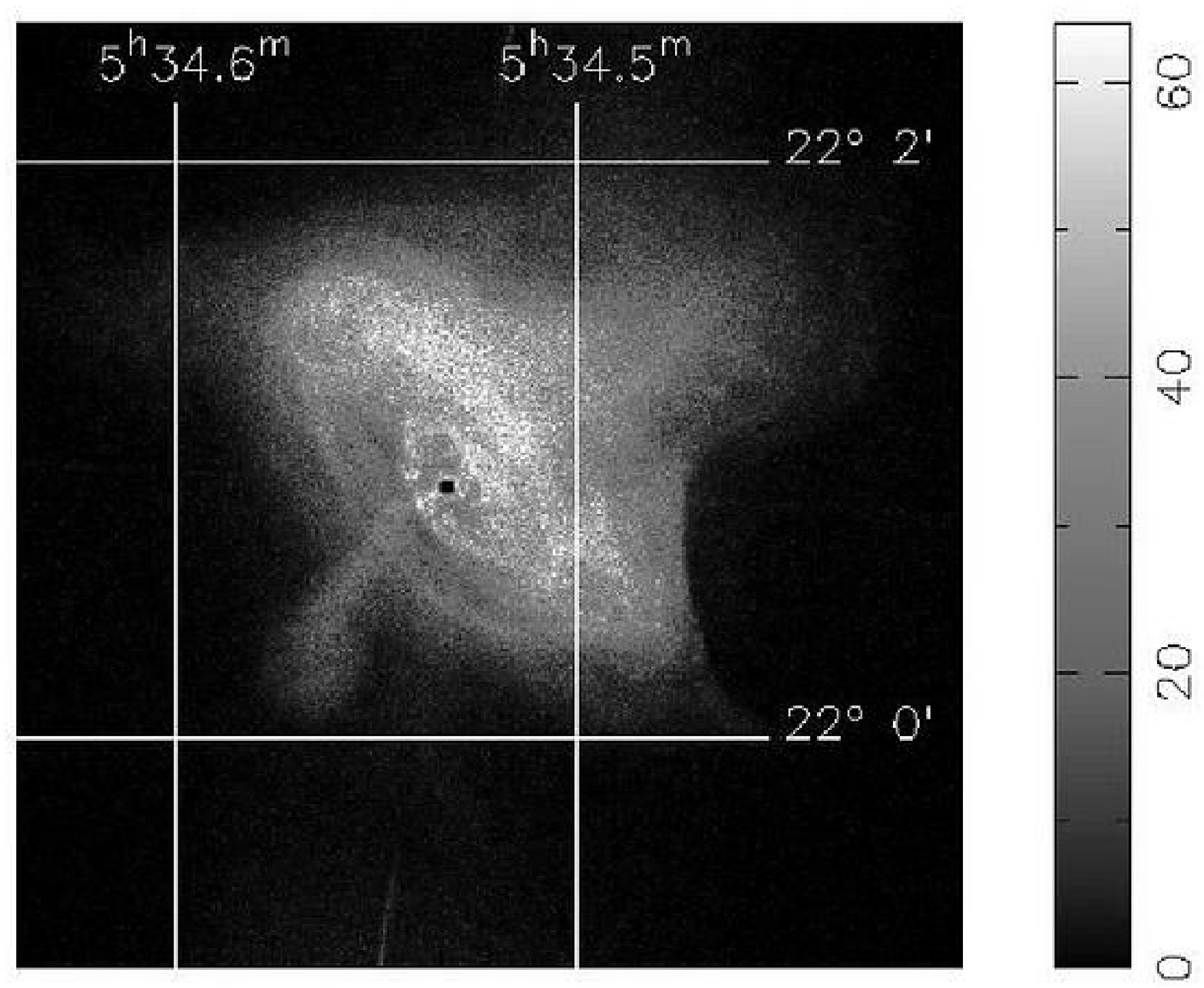} 
\caption{HETG/ACIS-S zero-order image of the central 200" x 200" of the Crab
Nebula.
\label{f:crab}}
\end{center}
\end{figure}

The zeroth order image is not significantly degraded from that without the HETG
in place. 
The HETG/ACIS-S in zeroth order was used to view and discover the fascinating
spatial structure in the X-Ray emission from the Crab Nebula as shown in the
now famous image - Figure~\ref{f:crab} (Weisskopf et al. 2000).  

\subsubsection{Wavelength Accuracy}

\begin{figure}
\begin{center} 
\epsfysize=8cm
\epsfbox{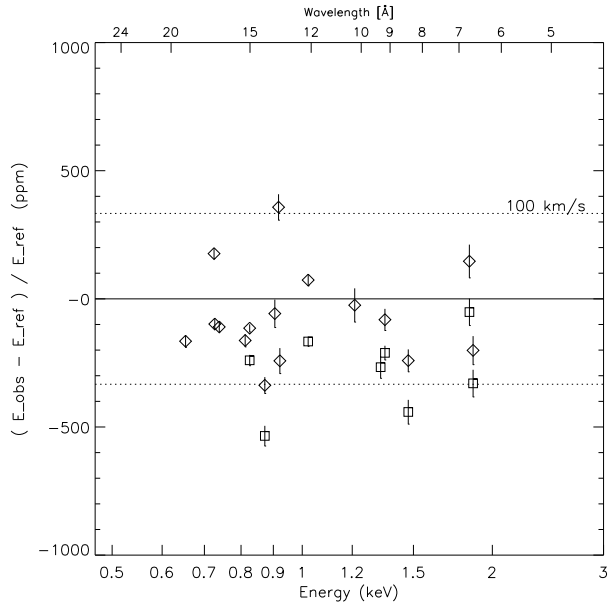} 
\caption{Measurements of lines in one observation of Capella (observation ID
2583).
{\em Open squares:} Lines from the HEG portion of the spectrum.  
{\em Open diamonds:} Lines from the MEG portion of the spectrum.
The average wavelength error is seen to correspond to velocity erros of less
than 100 km/s.  
\label{f:hetgs_dispersion_a}}
\end{center}
\end{figure}

The relative wavelength accuracy is determined by the placement of the CCDs and
the accuracy of the dispersion scale, which, in turn, is set by the HEG
and MEG grating periods and the distance from the grating assembly to the focal
plane.
The locations of the CCDs were precisely established in flight using emission
lines from the
calibration source Capella.
The location of the CCD gaps are accurate to 0.011 \AA\ (0.006 \AA) based on
measurements with the MEG (HEG) (Marshall, Dewey, \& Ishibashi, 2004).
The net result is that the wavelengths are typically accurate to better than 100
km/s as shown in Figure~\ref{f:hetgs_dispersion_a}

\subsubsection{Line Response Function and Spectral Resolution}

\begin{figure}
\begin{center}
\epsfysize=8cm
\epsfbox{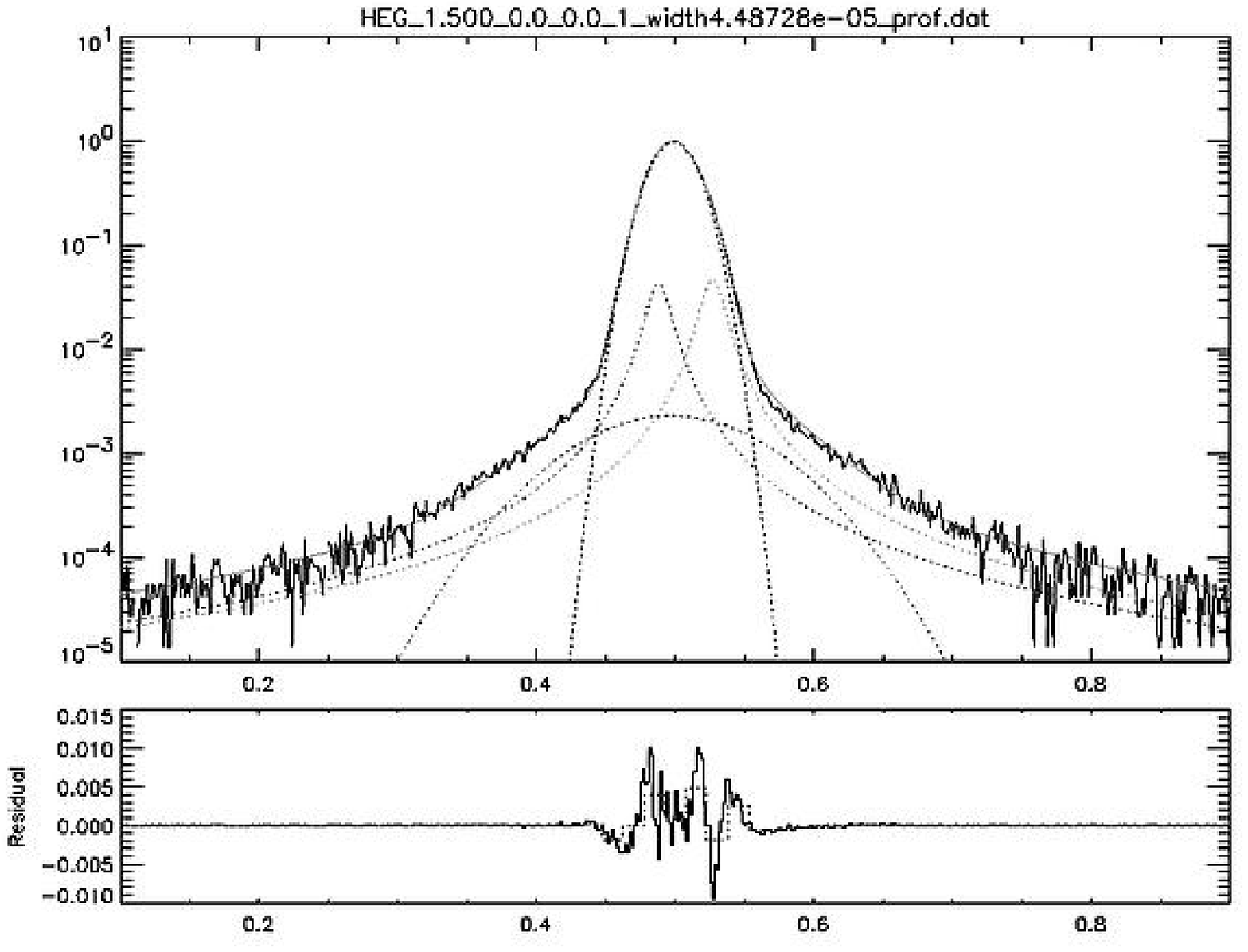} 
\caption{Decomposition of the HETG LRF and a comparison with simulated data. 
This example is for the HEG in positive first order.  
The LRF is modeled with narrow and broad Gaussians and two Lorentzian's, one for
each of the wings to allow for asymmetry.
The narrow Gaussian dominates the LRF and the Lorentzian wings are rarely
detectable in flight data.
The residuals (bottom panel) are largest in the core, but are not statistically
significant.
\label{fig:hetgs_lrf}}
\end{center}
\end{figure}

\begin{figure}
\begin{center}
\begin{tabular}{cc}
\epsfysize=4.5cm
\epsfbox{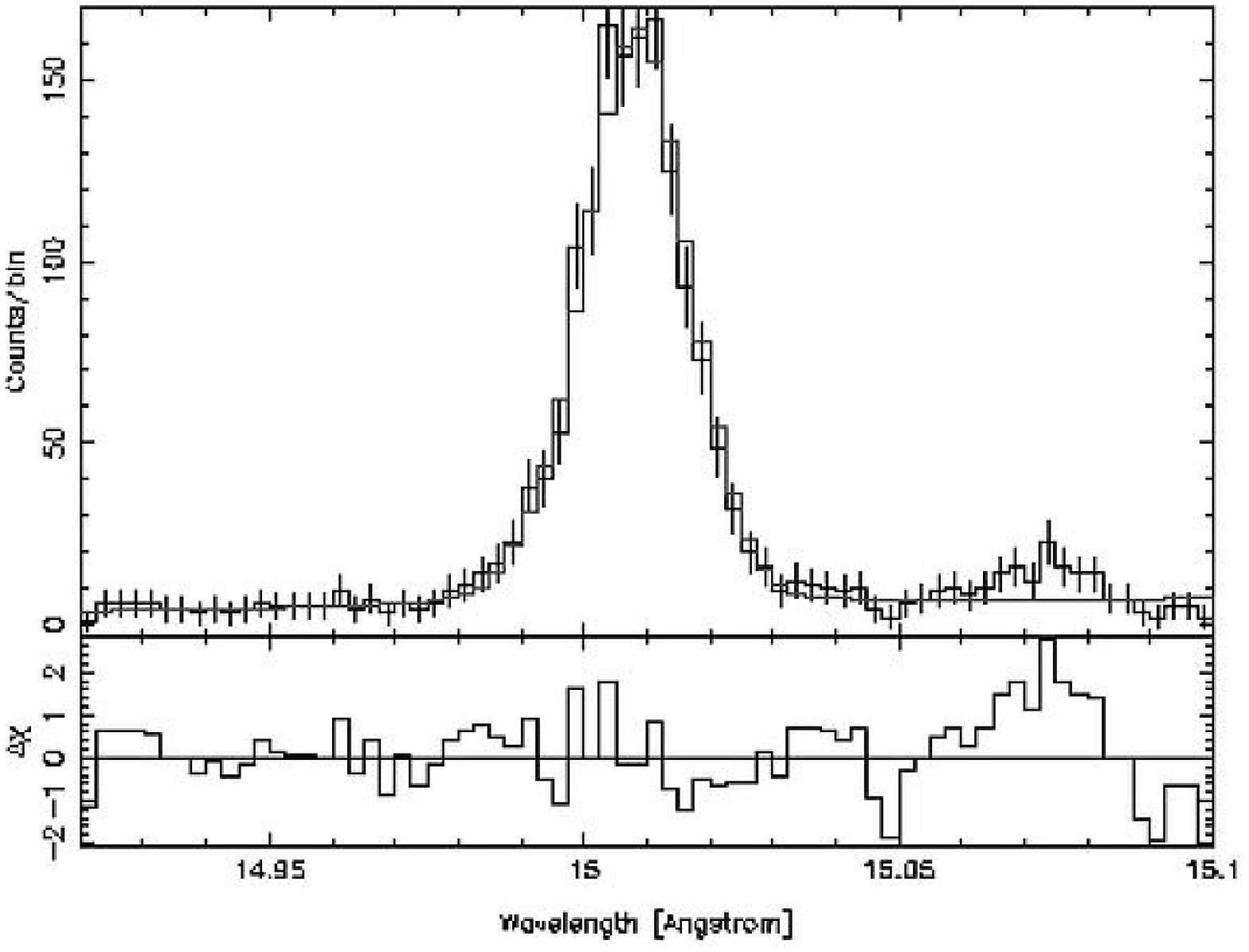}
\epsfysize=4.5cm
\epsfbox{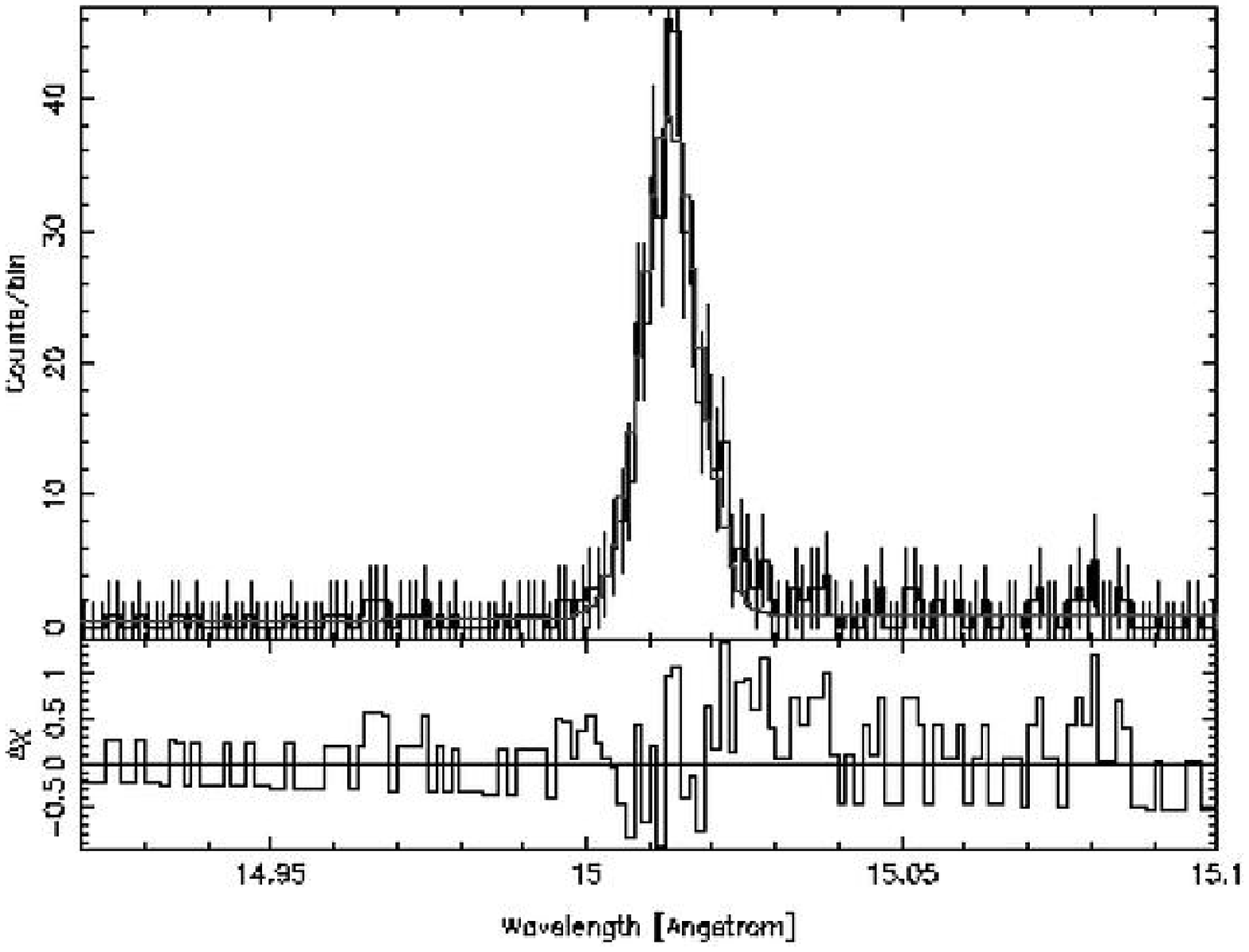}
   \end{tabular}
   \end{center}
   \caption{Comparison of data from Capella and the LRF model for the Fe {\sc
XVII} line at 15.013 \AA.  
The fit to the MEG (HEG) data is on the left (right).
\label{f:hetg_fe17}}
   \end{figure}

The HETGS LRF has a Gaussian-like core with extended wings. 
The model of the HETG LRF is comprised of two Gaussians and two Lorentzians with
the narrow Gaussian dominating as shown in Figure~\ref{fig:hetgs_lrf}.
The LRFs derived from the fits match in-flight data extremely well as shown in
Figure~\ref{f:hetg_fe17} (Marshall, Dewey, \& Ishibashi, 2004).
The spectral resolution is the FWHM of the LRF and is essentially independent of
wavelength and is 0.012\AA\ for the HEG and 0.023\AA\ for the MEG. 

\subsubsection{Effective Area}

\begin{figure}
\begin{center}
\epsfysize=8cm
\epsfbox{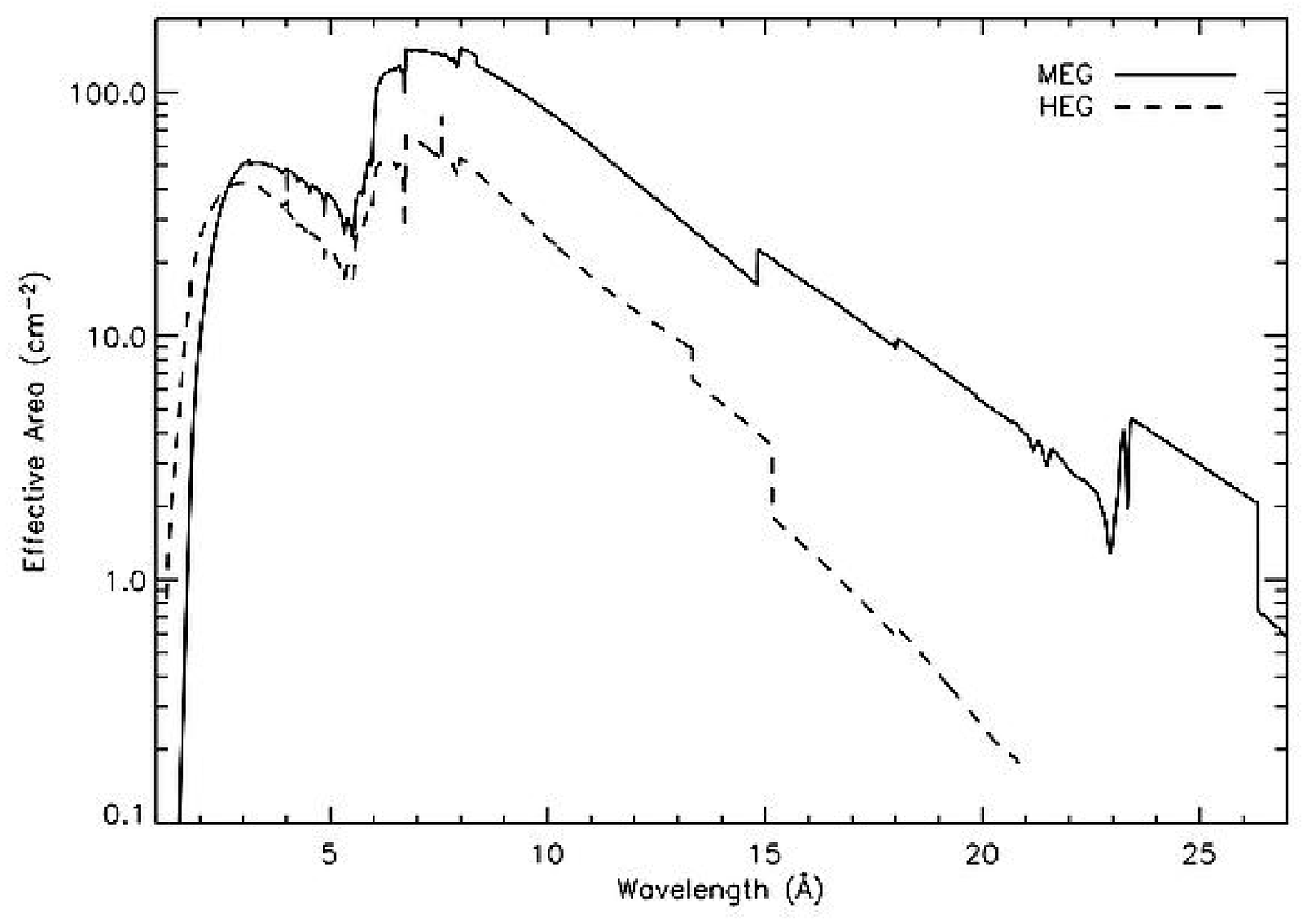}
\caption{First Order effective area of the HETG/ACIS-S combination accounting
for contamination at the end of 2003.
\label{f:hetgs_area}}
\end{center}
\end{figure}

\begin{figure}
\begin{center}
\epsfysize=8cm
\epsfbox{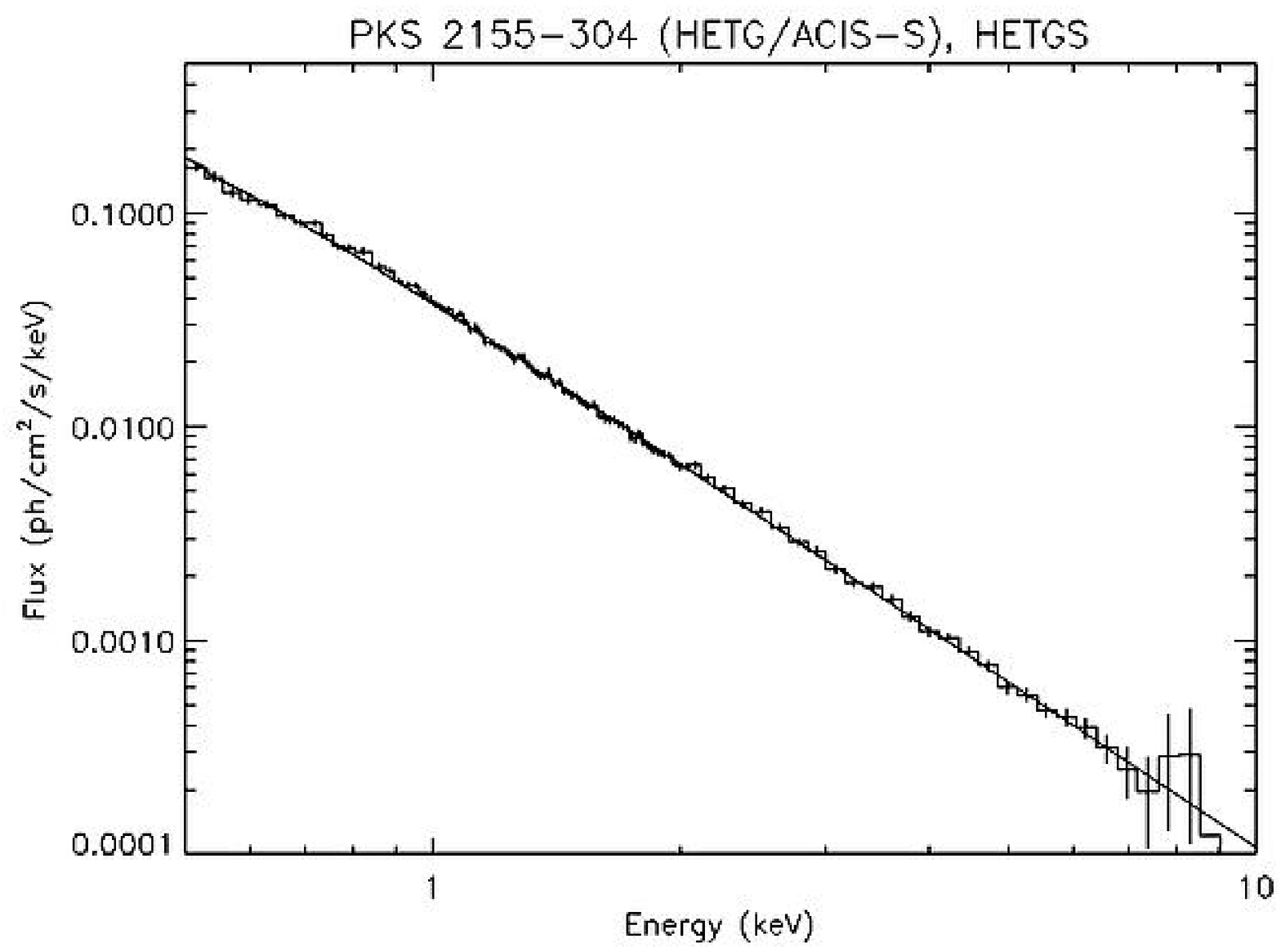}
\caption{HETGS spectrum of PKS 2155-304 taken in May of 2000.
The data were fit to a broken (double) power model; systematic deviations are
consistently less than 10\%.
\label{f:hetgs_pks2155}}
\end{center}
\end{figure}

There are many ingredients in the effective area of the HETG/ACIS-S combination. 
These include the telescope, the gratings and the various CCD detectors and
their filter.
The first order effective areas for the HEG and MEG, after adjusting the BI CCDs
quantum efficiency to agree with that
inferred for the FI CCDs, and accounting for the molecular
contamination of the ACIS filter as of the end of 2003,
is shown in Figure~\ref{f:hetgs_area}.  
This effective area was used to analyze the spectrum of the BL Lac object,
PKS 2155-304.  
This object is a \chandra\ calibration source and its spectrum is assumed to be
featureless. 
Figure~\ref{f:hetgs_pks2155} shows the results of the analysis of an observation
taken in May of 2000. 
There are no systematic residuals larger than 10\% and the spectrum appears
featureless at the 3\% level. 

\begin{figure}
\begin{center}
\epsfysize=8cm
\epsfbox{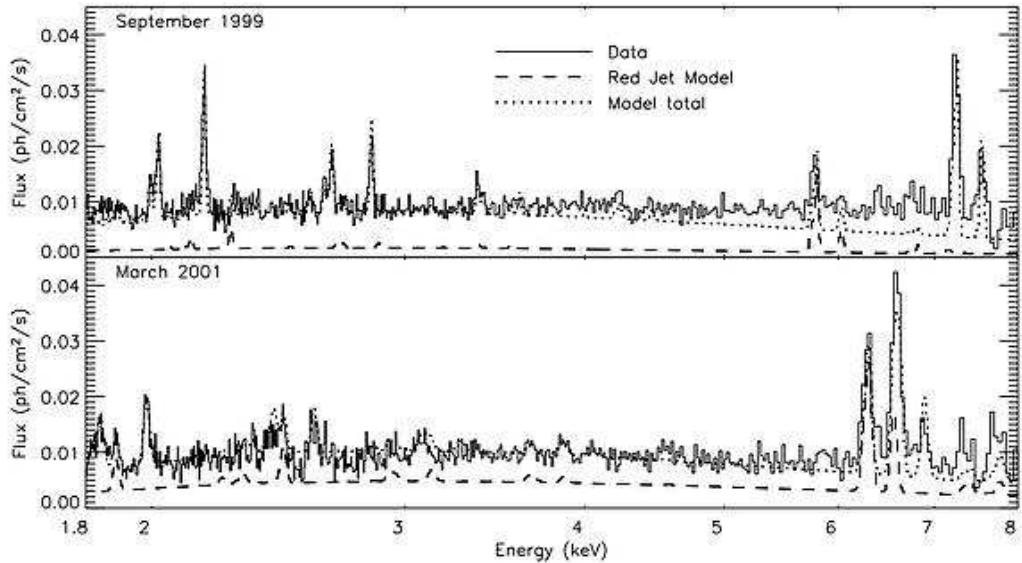}
\caption{HETG/ACIS-S spectra of SS 433. These spectra are rich in red- and
blue-shifted emission lines from highly ionized elements: Si, S, Ar, Ca, and Fe
near 2.1, 2.8, 3.5, 4.3, and 6-7 keV, respectively.  
The dashed lines are models of the red jet and the dotted lines are the sum
of the models for the blue and red jets.  
{\em Top:} The spectrum obtained in September 1999 shows strong and narrow lines
from the blue jet but the lines from the red jet are weak. 
{\em Bottom:} The spectrum obtained in March 2001 shows much stronger lines from
the red jet and all lines are broader than in the earlier observation.
\label{f:hetgs_ss433} }
\end{center}
\end{figure}

An example of spectroscopy with the HETG is shown in Figure~\ref{f:hetgs_ss433}
which shows two HETG/ACIS-S spectra (Marshall, Canizares, \& Schulz, 2002;
Lopez et al. 2004) of the binary source SS 433 taken at different times in the
binary period. 
The spectra are dominated by strong emission lines on a moderately strong
continuum. 
Here the lines are associated with the oppositely directed, mildly relativistic
jets that precess with a period of 164 days (Margon et al. 1977; Abell \&
Margon, 1979). 
Marshall, Canizares, \& Schulz (2002) showed that the emission lines are
collisionally excited and consistent with a model of an adiabatically expanding
outflow cooling from $10^8$K to $<10^7$K.  
Additionally, the electron density of the jet was estimated to be 10$^{14}$
cm$^{-3}$ by measuring the ratio of the intercombination and forbidden lines of
the Si {\sc xiii} triplet, which in turn provided an estimate for the size of
the jet.

The lines from the approaching (blue) jet in the first observation were strong
enough to show that the resolved lines had a common blueshift, indicating that
the jet was produced in a uniform conical outflow.
The Doppler broadening of the lines gave the opening angle of the jet, 1.2$^o$.
The March 2001 observation took place during orbital eclipse, and the modeling
of the red jet by Lopez et al. (2004) found that the visible portion of the jet
was $\sim 10^{11}$ cm long.  
Using the system geometry and some system parameters derived from optical
observations of the accretion disk, these authors estimated that the mass
of the compact object in SS 433 is about 16 $M_{\sun}$, confirming that it is
indeed a black hole.

\begin{figure}
\begin{center}
\epsfysize=8cm
\epsfbox{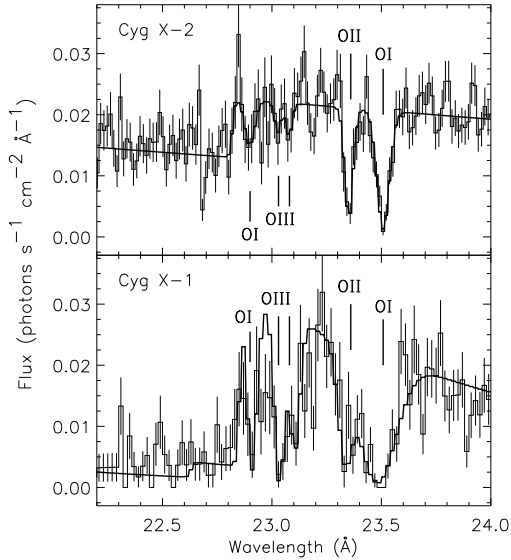}
\caption{Absorption spectra in the vicinity of the oxygen K edge for Cyg X-2
(top) and Cyg X-1 (bottom) with line identifications.
These data were used to measure the neutral and ionized oxygen abundances along
two lines of sight in the interstellar medium (Juett, Schulz, \& Chakrabarty
2004).}
\label{f:hetgs_ism_abs}
\end{center}
\end{figure}

Another example of spectroscopy with the HETG was its use to begin to determine
the abundances and ionization fractions of gas in the interstellar medium (ISM). 
Juett, Schulz, \& Chakrabarty (2004) presented highly resolved spectra of the
oxygen $K$-shell interstellar absorption edge using X-ray binaries as sources
(Figure~\ref{f:hetgs_ism_abs}).  
The $K\alpha$ and $K\beta$ absorption lines from neutral, singly, and doubly
ionized oxygen were identified as well as expected absorption edges.
The observed wavelength of O I $K\alpha$ absorption line was used to adjust the
wavelength of the theoretical predictions of the absorption cross sections.  
Juett, Schulz, \& Chakrabarty (2004) also placed a limit on the
velocity dispersion of the neutral lines of $\leq$200~km~s$^{-1}$, consistent
with measurements at other wavelengths.  
Finally, the HETG measurements determined the oxygen ionization fractions in the
ISM in these lines of sight.  
This work demonstrated the utility of X-ray spectroscopy for studies of the ISM  
and future work with the Observatory will provide measurements of the relative
abundances and ionization fractions for elements from carbon to iron for many
different lines of sight.

\section{Additional Discoveries \label{s:discoveries}} 

The first X-rays focused by the telescope were observed on August 12, 1999. 
Figure~\ref{f:crab} showed one of the early images. 
This image of the Crab Nebula and its pulsar included a major new discovery
(Weisskopf et al. 2000) -
the bright inner elliptical ring showing the first direct observation of the
shock front where the wind of particles from the pulsar begins to radiate in
X-rays via the synchrotron process.
Discoveries of new astronomical features in \chandra\ images have been the rule,
not the exception.

The Observatory's capability for high-resolution imaging enables detailed
studies of the structure of extended X-ray sources, including
supernova remnants, astrophysical jets, and hot gas in galaxies and clusters of
galaxies. 
Equally important are \chandra's unique contributions to high-resolution
dispersive spectroscopy. 
As the capability for visible-light spectroscopy initiated the field of
astrophysics about a century ago, high-resolution X-ray spectroscopy now
contributes profoundly to the understanding of the physical processes in cosmic
X-ray sources and is the essential tool for diagnosing conditions in hot
plasmas. 
The high spectral resolution of the \chandra\ gratings isolates individual lines
from the myriad of spectral lines, which would overlap at lower resolution.
The additional capability for spectrometric imaging allows studies of
structure, not only in X-ray intensity, but also in temperature and in chemical
composition. 
Through these observations, users are addressing several of the most exciting
topics in contemporary astrophysics.

\begin{figure}
\begin{center} 
\epsfysize=8cm
\epsfbox{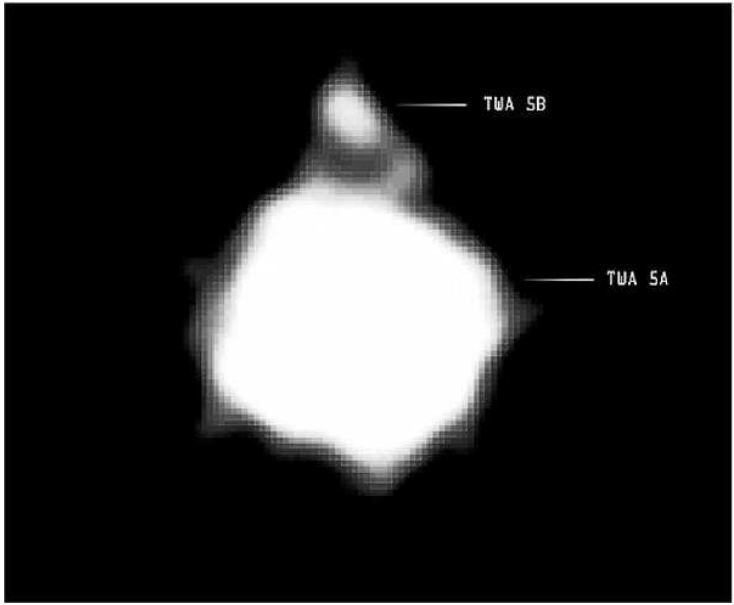} 
\caption{X-rays produced by TWA 5B, a brown dwarf orbiting a young 
binary star system known as TWA 5A. The image is 6-arcsec on a side. Courtesy of
NASA/CXC/Chuo U.
\label{f:browndwarf}}
\end{center}
\end{figure}

In addition to mapping the structure of extended sources, the high angular 
resolution permits studies of discrete sources, which would otherwise be
impossible. 
An example is shown in Figure~\ref{f:browndwarf} where one sees X-rays
produced by TWA 5B, a brown dwarf orbiting a young binary star system known as
TWA 5A6. 
This observation not only demonstrates the importance of the \chandra\
angular resolution but also addresses the question as to how do brown dwarfs
heat their upper atmospheres (coronas) to X-ray-emitting temperatures of a few
million degrees.

From planetary systems to deep surveys of the faintest and most distant objects,
the scientific results from the first years of \chandra\ operations have
been exciting and outstanding. 
We conclude this paper with an overview of some of these results. 

\begin{figure}
\begin{center} 
\epsfysize=8cm
\epsfbox{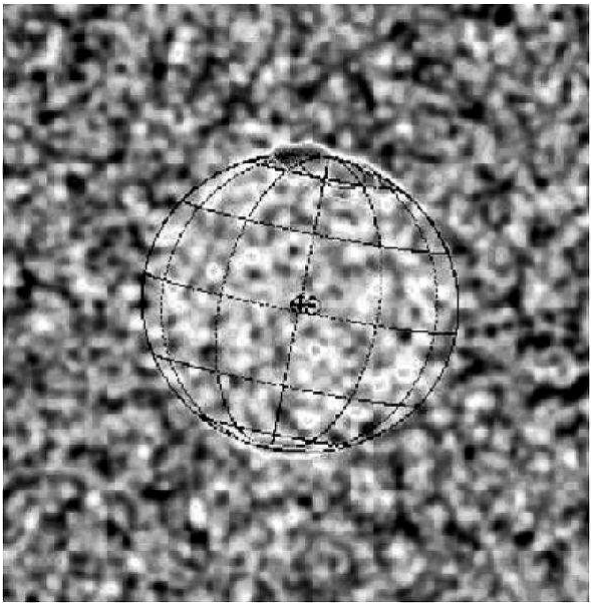} 
\caption{\chandra\ image of Jupiter showing the hot spots at high latitudes. The
image is 50--arcsec on a side. Courtesy R. Elsner.
\label{f:jupiter}}
\end{center}
\end{figure}

\begin{figure}
\begin{center} 
\epsfysize=8cm
\epsfbox{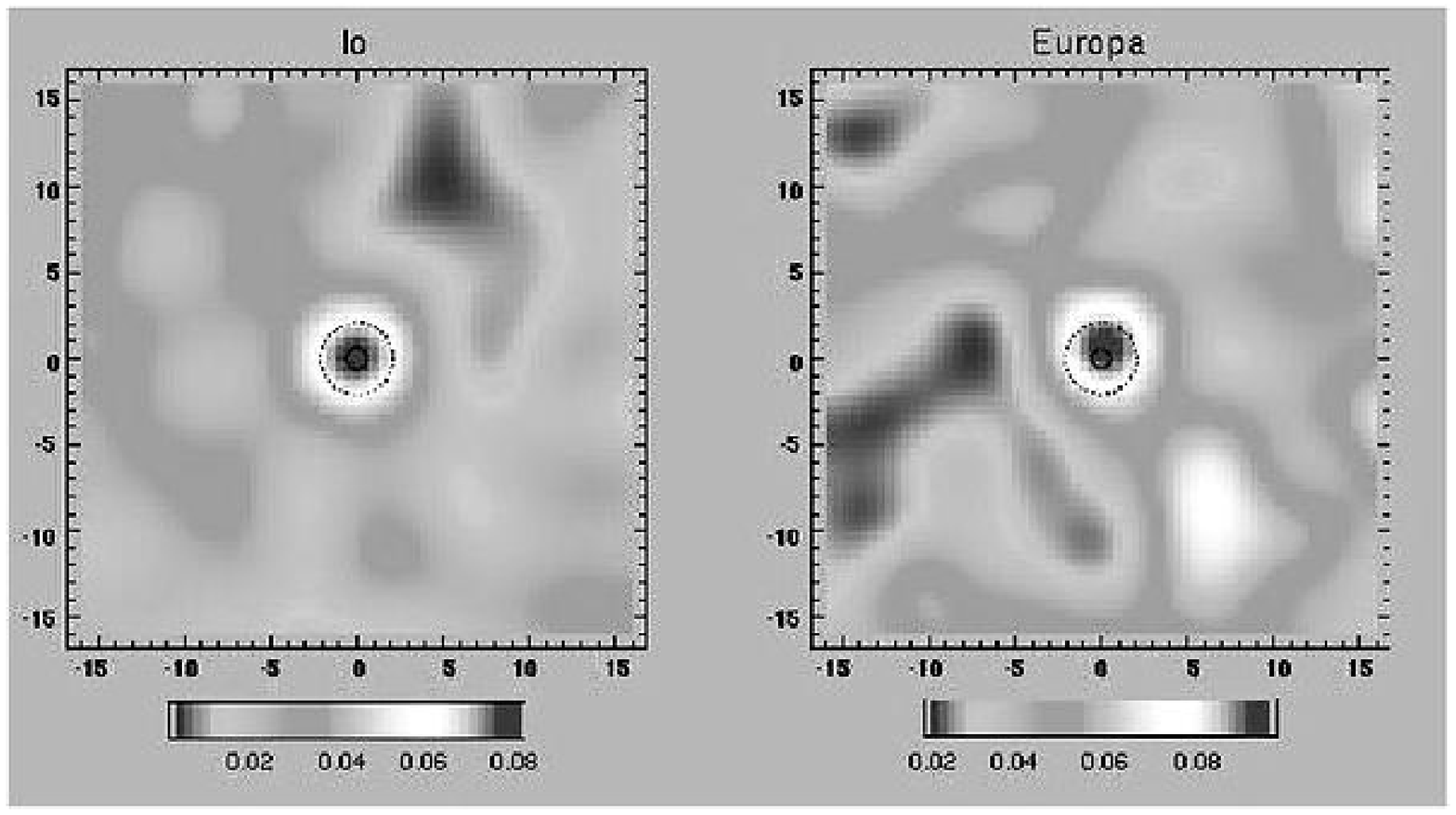} 
\caption{\chandra\ images of the Jovian moons Io and Europa. Courtesy R. Elsner.
\label{f:io_europa}}
\end{center}
\end{figure}

We begin with images of the X-ray emission from the planet Jupiter.
Figure~\ref{f:jupiter} shows hot spots at high (and unexpected) latitudes that
appear to pulsate at approximately a 45-minute period (Gladstone et al. 2002).
In this case the X-rays appear to be produced by particles bombarding the Jovian
atmosphere after precipitating along magnetic field lines. 
Figure~\ref{f:io_europa} continues the discoveries about the Jovian system and
shows the first detection of X-rays from two of the moons -- Io and
Europa (Elsner et al. 2002).

\begin{figure}
\begin{center} 
\epsfysize=8cm
\epsfbox{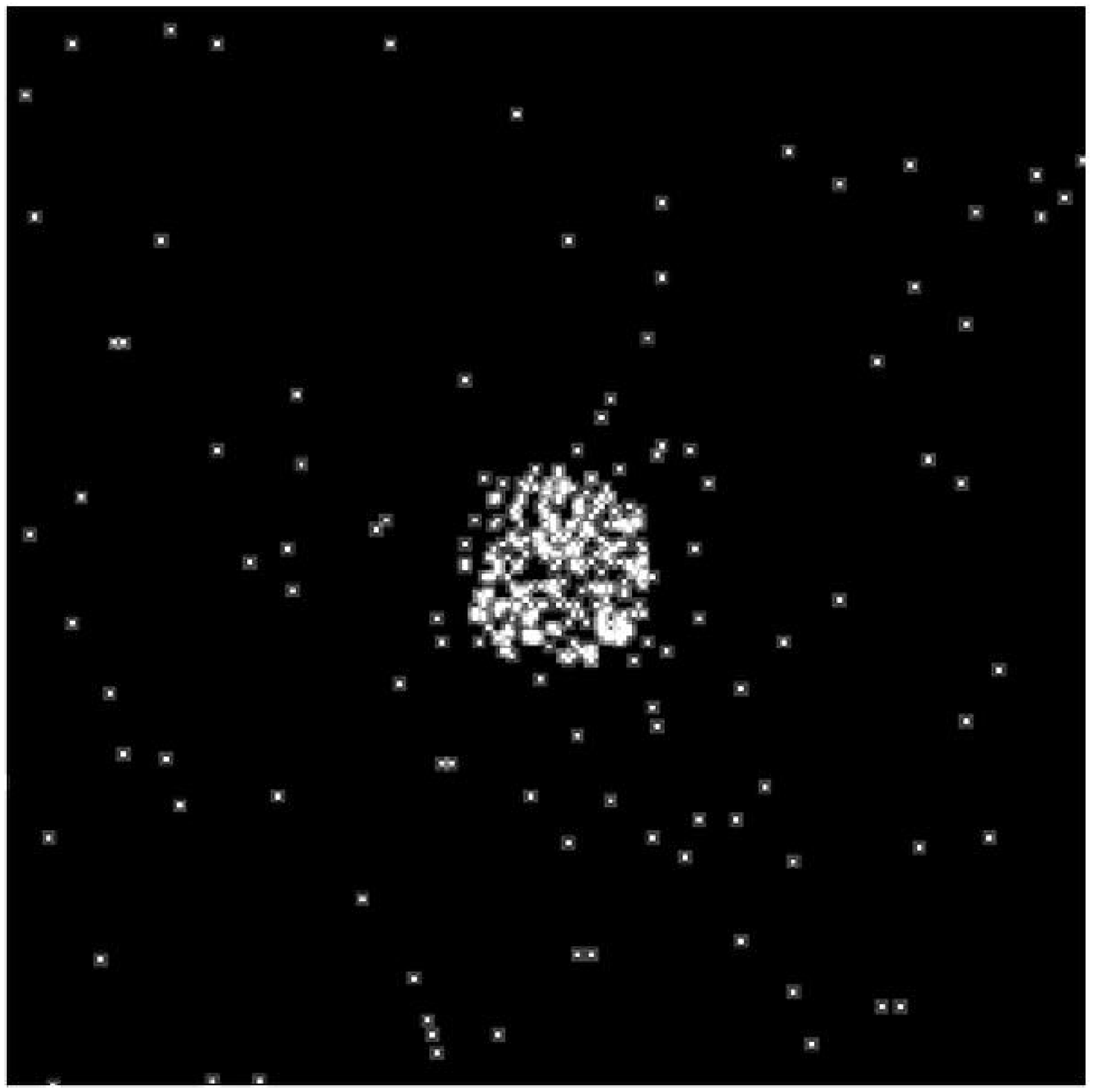} 
\caption{\chandra\ image of Mars. Image is 2 arcmin on a side and the Mars
disk is 20.3 arcsec in diameter. Courtesy NASA/CXC/MPE/K.Dennerl.
\label{f:mars}}
\end{center}
\end{figure}

In Figure~\ref{f:mars} we show a more recent detection (Dennerl, 2002) of
fluorescent scattering of solar X-rays in the upper atmosphere of Mars. 
The X-ray spectrum is dominated by a single narrow emission line, most likely
caused by oxygen K-shell fluorescence.
 
\begin{figure}
\begin{center} 
\epsfysize=8cm
\epsfbox{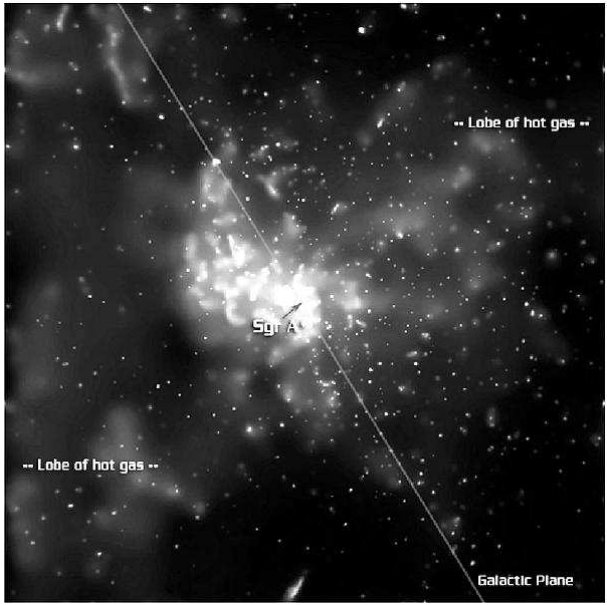} 
\caption{\chandra\ image of the Galactic Center. The image is 8.4 arcmin on a
side. Courtesy NASA/CXC/MIT/F.K.Baganoff et al.
\label{f:sgrA}}
\end{center}
\end{figure}

One of the most spectacular \chandra\ images is the one of the center of our own
galaxy (Baganoff et al. 2003) shown in Figure~\ref{f:sgrA}. 
Here we clearly see both point-like discrete sources (over 1000)
and diffuse extended emission. 
This large amount of hot X-ray emitting gas has been heated and chemically
enriched by numerous stellar explosions.

The final legacy of \chandra\ may ultimately be led by the spectroscopic data.
The energy resolution, enabled by the quality of the optics, is providing new
and extremely complex results. 
The broad bandpass of the grating spectrometers, combined with high resolution,
has proven equally important for astrophysical insights in situations where
spectral features are {\em not} present. 
For example, the remarkable line-free smoothness of the continuum of the
isolated neutron star RXJ~1856-3754 was a spectacular
surprise when revealed in detail using a 500~ks observation. 
This object is the nearest and brightest isolated neutron star candidate
(Walter, Wolk \& Neuh\"auser 1996), and it had been hoped that metal lines
formed in its outer atmosphere would provide a direct measurement of
gravitational redshift and insights into the equation of state of ultra-dense
matter.  
An early 55~ks exposure lacked the sensitivity to detect weak absorption
features (Burwitz et al.\ 2001), but the 500~ks LETG spectrum placed
stringent limits on the strengths of any absorption features (Drake et
al.\ 2002), prompting speculation that its outer layers might lack an
atmosphere and reside in a solid state (e.g.\ Turolla et al.\ 2004).

Observations with the gratings are not only providing new astrophysical results,
they also provide a challenge to atomic physicists. 
The heart of the LETG+HRC-S bandpass covers the historically relatively
uncharted part of the soft X-ray spectrum from 25-70~\AA .  
Prior to \chandra, only a small handful of astrophysical observations had
been made at anything approaching high spectral resolution in this range: these
were of the solar corona using photographic spectrometers (Widing \& Sandlin
1986; Freeman \& Jones 1970; Schweizer \& Schmidtke 1971; Behring, Cohen, \&
Feldman 1972; Manson 1972; Acton et al., 1985).  
In comparison, LETG spectra of similar X-ray sources -- the coronae
of the solar-like stars $\alpha$~Centuri A (G2~V) and B (K1~V) and of Procyon
(Raassen et al.\ 2002, 2003) -- in this range are at the same time both daunting
and revealing.  
The pseudo-continuum of solar coronal emission seen by the 1982 July
rocket-borne photographic spectrograph described by Acton et al.\ (1985) is
resolved into a dramatic forest of lines in the LETG spectra.  
This spectral range contains a superposition of ``L-shell'' emission of abundant
elements such as Mg, Si, S and Ar, providing a challenge to spectroscopists
hoping to understand this region in terms of individual atomic transitions.
Drake et al. (2004) have shown that current radiative loss models in common
usage by X-ray astronomers underestimate the line flux in the 25-70~\AA\ range
by factors of up to 5.  
Laboratory efforts prompted by \chandra\ spectra, and the need for a better
theoretical description of plasma radiative emission in this spectral region,
are just beginning to unravel the tangle of lines into their parent ions (e.g.\
Lepson et al.\ 2003 and references therein).

\begin{figure}
\begin{center} 
\epsfysize=8cm
\epsfbox{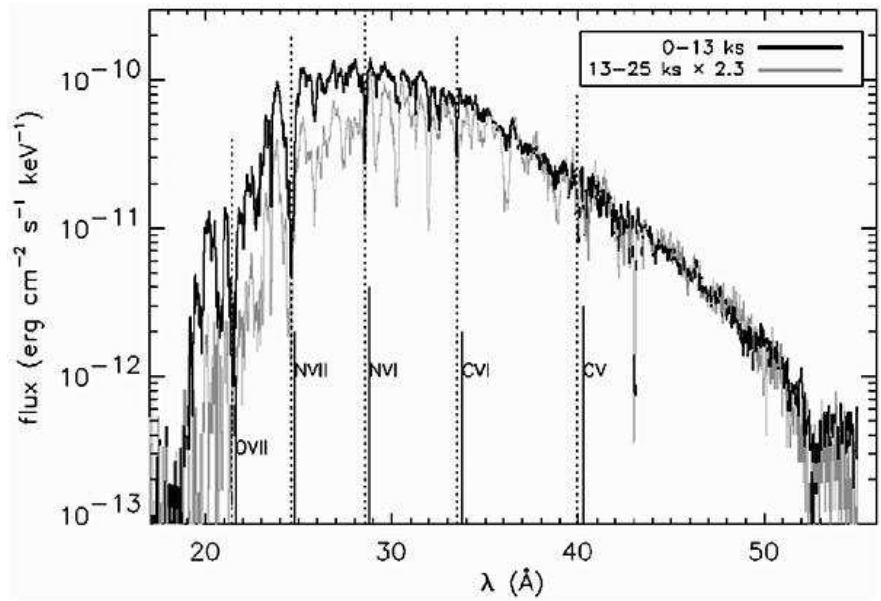} 
\caption{LETG+HRC-S spectra of the nova V4743~Sgr for two different segments of
an observation made on 2003, March 19.  
Identifications of resonance lines of C, N and O are indicated by vertical lines
(solid: rest wavelength; dotted: shifted by -2400 km s-1; from Ness et al.\
2003). 
\label{f:letg_v4743}}
\end{center}
\end{figure}

Absorption lines also challenge modelers of spectra.  
``Supersoft'' sources and X-ray novae with optically-thick atmospheres at
temperatures several $10^4$-$10^5$~K had been observed many times by ROSAT and
BeppoSAX (e.g.\ Krautter 2002, and references therein).  
The LETG has revealed that the soft X-ray continuum of these objects hosts a
rich array of metal absorption features.  
Perhaps the best example is the nova V4743~Sgr, which became the brightest X-ray
source in the sky at wavelengths above 25~\AA\ ($<0.5$~keV) in early 2003, and
was caught using the LETG+HRC-S as a target of opportunity.  
A preliminary analysis of its spectrum was presented by Ness et al.\ (2003); the
spectrum is shown in Figure~\ref{f:letg_v4743}.  
While prominent resonance lines of C, N and O can easily be identified, unlike
the case of low-density plasmas, the higher density environment of the nova
atmosphere in the gravity of its degenerate host can support substantial
absorption from excited states that are more difficult to identify. 
Determining the origin of the multitude of weaker lines ``gouging'' the
continuum will require more complex modeling using realistic model
atmospheres.  
It can be seen from this example how important the unique broad-band sensitivity
of the LETG is for disentangling the different parameters describing simple
models of these types of source.  
Coupled with the strong line absorption that modifies the apparent spectrum
continuum level is the absorption from intervening material in the
circumstellar environment and interstellar medium.  
The LETG clearly shows this latter attenuation down through two orders of
magnitude in flux, out to a wavelength of $\sim 60$~\AA\ (0.2~keV).

\begin{figure}
\begin{center} 
\epsfysize=8cm
\epsfbox{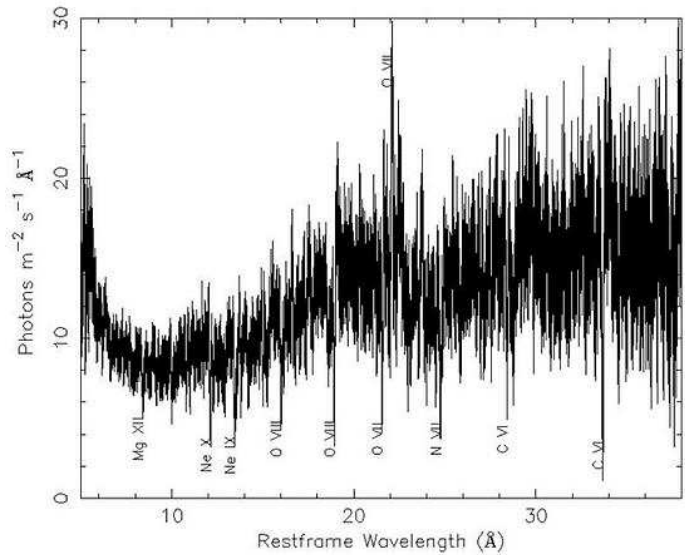} 
\caption{LETG/HRC spectrum of the Seyfert 1 galaxy NGC 5548 (Kaastra et al.
2000).
Several prominent absorption lines from H-like and He-like ions are marked, as
is the forbidden line of He-like oxygen.
\label{f:ngc5548}}
\end{center}
\end{figure}

\begin{figure}
\begin{center} 
\epsfysize=16cm
\epsfbox{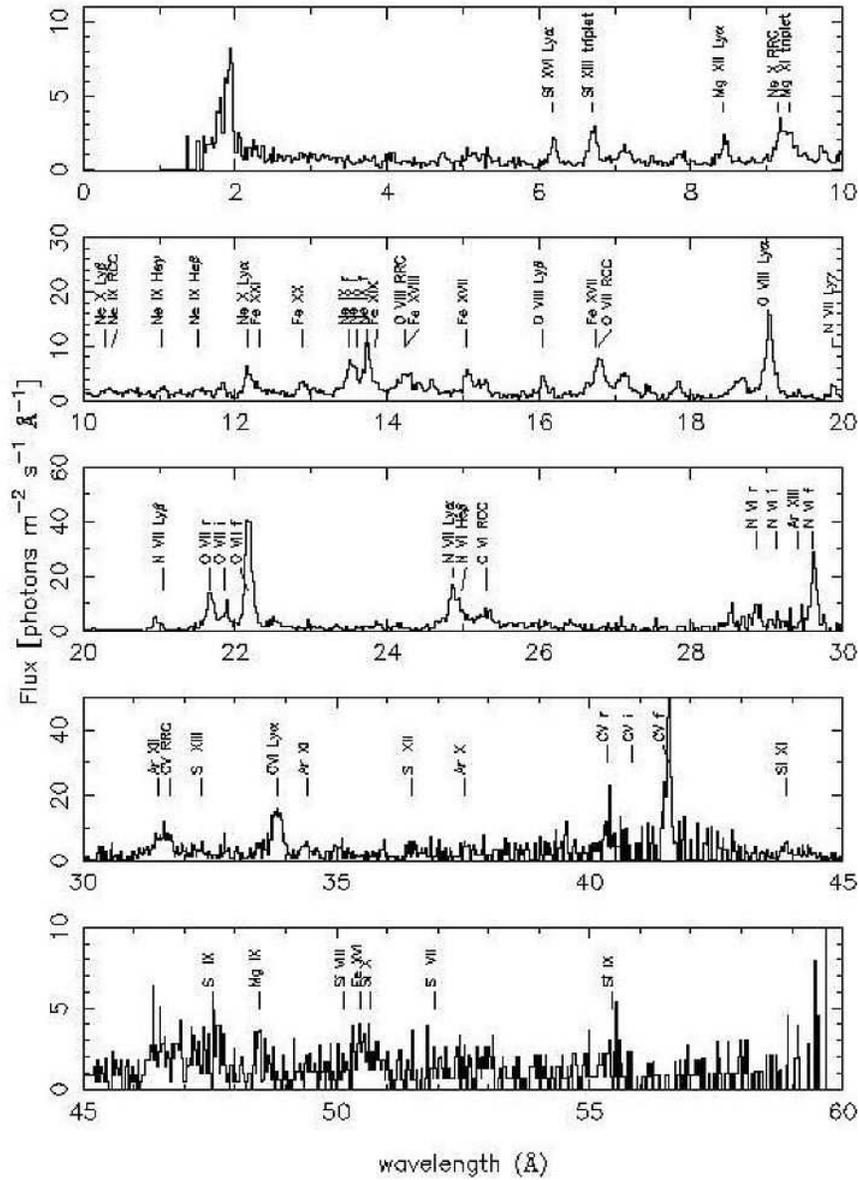} 
\caption{Emission-line spectrum from the Seyfert 2 galaxy NGC 1068. 
Kindly provided by A. Kinkhabwala.
\label{f:ngc1068}}
\end{center}
\end{figure}

High-resolution spectra of Seyfert galaxies are now providing new
details about the  physical and dynamical properties of material surrounding
the active nucleus.
For example, the Seyfert 1 active galaxy Mkn~478 was expected to exhibit
absorption lines at shorter wavelengths from a warm absorber that has
often been seen in the spectra of other Seyfert 1 galaxies, and emission lines
at wavelengths of $\sim 100$~\AA\ based on an analysis of EUVE spectra by Hwang
\& Bowyer (1997).  
Mkn~478 lies in a direction out of the galaxy that has a particularly low
neutral hydrogen column density, and so remains a strong source at these longer
wavelengths.  
Furthermore, for Seyfert-1s, whose signal is dominated by a bright X-ray
continuum  from the central engine, the partially ionized circum-source material
introduces prominent patterns of absorption lines and edges. 
Figure~\ref{f:ngc5548}, e.g. shows a LETG/HRC-S spectrum of NGC 5548. 
This spectrum has dozens of absorption lines (Kaastra et al. 2000).
For Seyfert 2's the strong continuum from the central engine is not seen
directly, so the surrounding regions are seen in emission.
Figure~\ref{f:ngc1068} provides an example of a LETG/HRC observation of the
Seyfert 2, NGC 1068 (Brinkman et al. 2002).

\begin{figure}
\begin{center} 
\epsfysize=6cm
\epsfbox{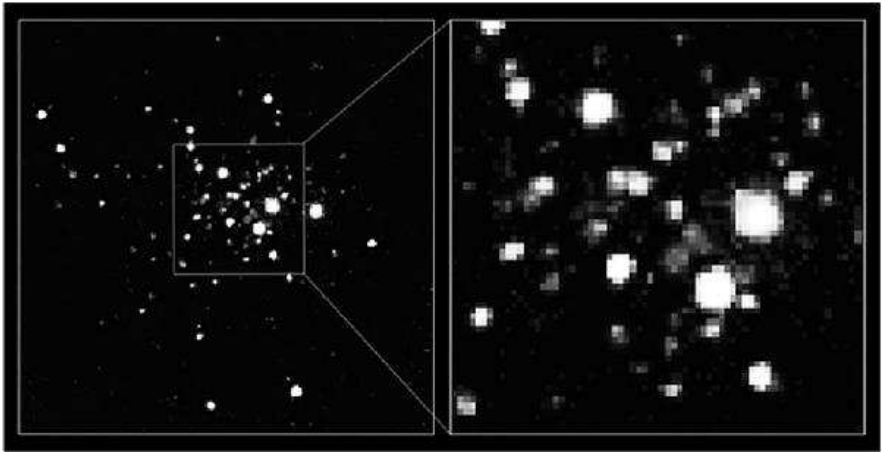} 
\caption{\chandra\ ACIS image of the globular cluster 47 Tucanae. The left panel
covers the central 2' x 2.5'. The central 35" x 35" are shown to the right.
Courtesy NASA/CfA/J.Grindlay et al.
\label{f:47tuc}}
\end{center}
\end{figure}

\begin{figure}
\begin{center} 
\epsfysize=6cm
\epsfbox{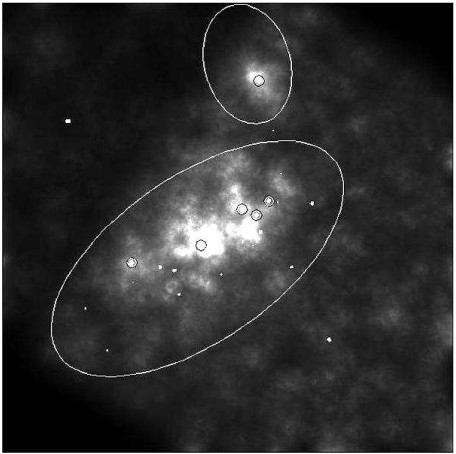} 
\caption{X-ray image of two interacting galaxies NGC 4490 and 4485. The image is
8-arcmin on a side. Large ovals represent the approximate boundaries of the
galaxies (NGC 4490 is the larger of the two). Small circles indicate the
brightest X-ray sources. Courtesy Doug Swartz. 
\label{f:ngc4490_4485}}
\end{center}
\end{figure}

One of the more important triumphs of the Observatory has been to use the
angular resolution and high sensitivity to perform detailed surveys of extended
objects such as globular clusters, galaxies, and clusters of galaxies. 
Figure~\ref{f:47tuc} shows one of the spectacular \chandra\ images of globular
clusters (Grindlay et al. 2001). 
A survey of two interacting galaxies is illustrated in
Figure~\ref{f:ngc4490_4485} where one sees emission from diffuse gas and bright
point sources.

\begin{figure}
\begin{center} 
\epsfysize=8cm
\epsfbox{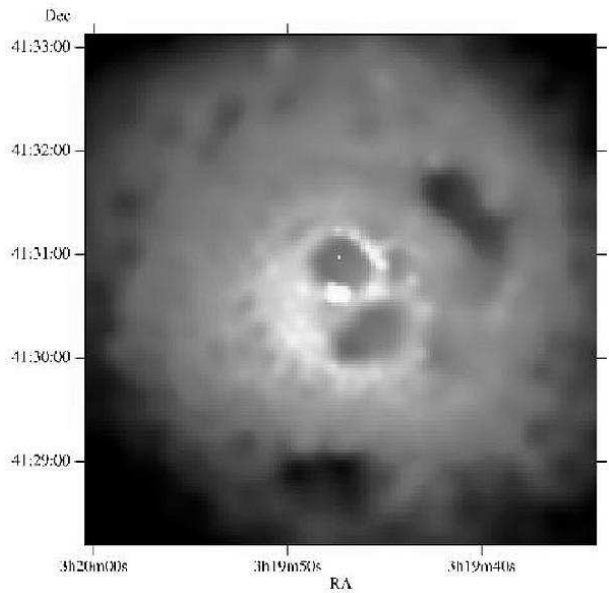} 
\caption{X-ray core of the Perseus  cluster. The image is about 3.5 arcmin on a
side. Courtesy NASA/IoA/A. Fabian et al.
\label{f:perseus}}
\end{center}
\end{figure}

\begin{figure}
\begin{center} 
\epsfysize=8cm
\epsfbox{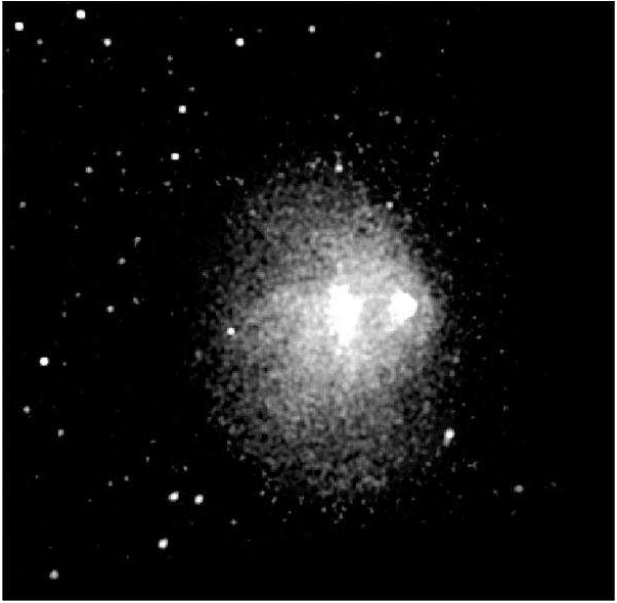} 
\caption{The \chandra\ image of the merging, hot galaxy cluster 1E 0657-56. The
image is about 9 arcmin on a side.
Courtesy NASA/SAO/CXC/M.Markevitch et al.
\label{f:1e0657}}
\end{center}
\end{figure}

\begin{figure}
\begin{center} 
\epsfysize=7cm
\epsfbox{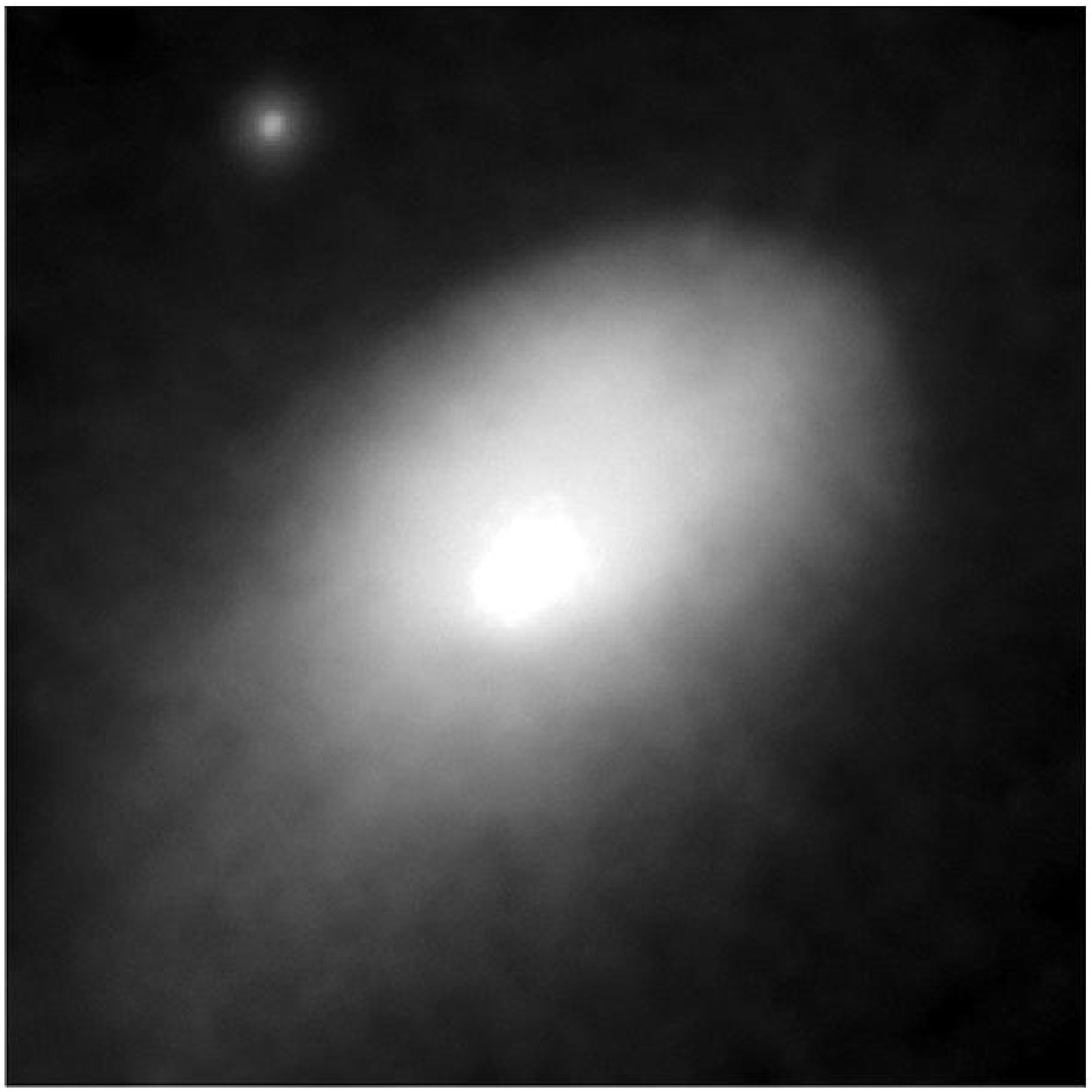} 
\caption{\chandra\ Image of the galaxy cluster Abell 2142. The image is 8.4
arcmin
on a side. The sharp border to the top right is an example of a cold front.
Courtesy NASA/CXC/SAO
\label{f:a2142}}
\end{center}
\end{figure}

\chandra\ observations of clusters of galaxies frequently exhibit previously
undetected structures with characteristic angular scales as small as a few arc
seconds. 
These include "bubbles" where there is strong radio emission, bow shocks, and
cold fronts. 
These phenomena are illustrated in the sequence of figures,
\ref{f:perseus}-\ref{f:a2142}. 
Figure~\ref{f:perseus} of the Perseus cluster (Fabian et al. 2000) is a
spectacular
example of bubbles produced in regions where there is strong radio emission.  
Figure~\ref{f:1e0657} shows a bow shock propagating in front of a bullet-like
gas cloud just exiting the disrupted cluster core. 
This observation provided the first clear example of such a shock
front (Markevitch et al. 2002).
In contrast, Figure~\ref{f:a2142} of Abell 2142 (Markevitch et al. 2000)
shows an example of a shockless cold front. 

A major triumph of \chandra\ (and XMM-Newton) high-resolution spectroscopic 
observations has been the discovery that that gas in the clusters is typically 
{\it not} cooling to below about 1-2 keV (see for example the discussion in
Fabian (2002) which indicates the presence of one (or more) heating
mechanisms).

\begin{figure}
\begin{center} 
\epsfysize=7cm
\epsfbox{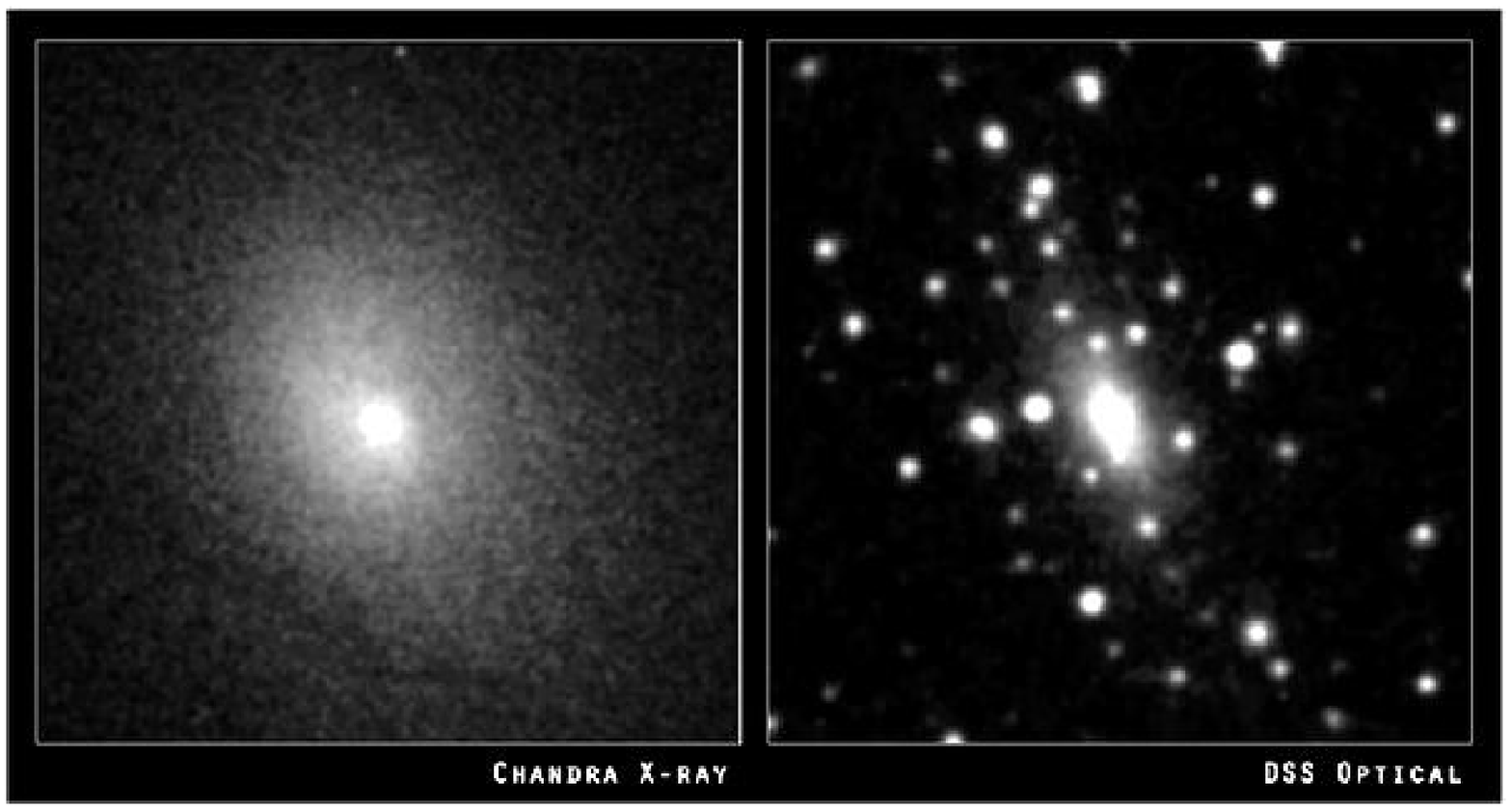} 
\caption{X-ray (left) and optical (right) images of the galaxy cluster Abell
2029. The images are 4-arcmin on a side. X-ray: NASA/CXC/UCI/A. Lewis et al.
Optical: Pal.Obs. DSS  
\label{f:abell2029}}
\end{center}
\end{figure}

Some clusters, such as Abell 2029 shown in Figure~\ref{f:abell2029}, do exhibit
a smoother relaxed structure. Here we see the thousands of galaxies inside the
cocoon of hot, X-ray-emitting gas. 
Measurement of the temperature and density profiles of the gas, inwards toward
the central, dominant galaxy, provides a map of the gravitational potential,
and hence the dark matter in the cluster.
The observers, Lewis, Buote, and Stocke (2003), showed that the dark
matter density increased toward the center in a manner consistent with cold
dark matter models. 

\begin{figure}
\begin{center} 
\epsfysize=7cm
\epsfbox{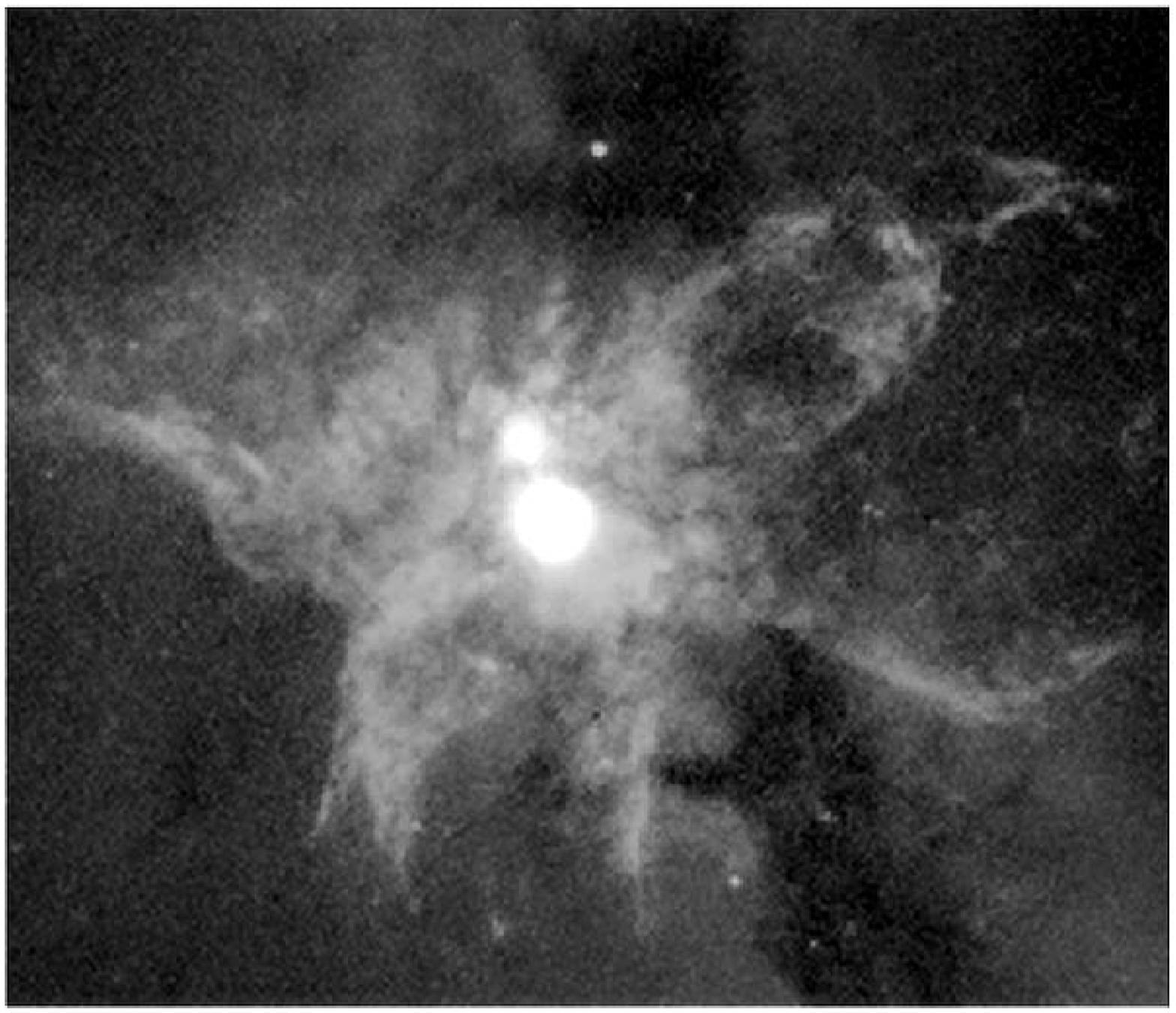} 
\caption{The \chandra\ image of NGC 6240, a butterfly-shaped galaxy that is the
product of the collision of two smaller galaxies, shows that the central region 
contains two active galactic nuclei. The image is 0.35 x 0.3 arcmin. Courtesy
NASA/CXC/MPE/S. Komossa et al.
\label{f:ngc6240}}
\end{center}
\end{figure}

The study of active galaxies is one of the centerpieces of studies with the
Observatory. 
Figure~\ref{f:ngc6240} illustrates a recent spectacular result --- the first
image
of a double quasar nucleus (Komossa et al. 2003).

\begin{figure}
\begin{center} 
\epsfysize=7cm
\epsfbox{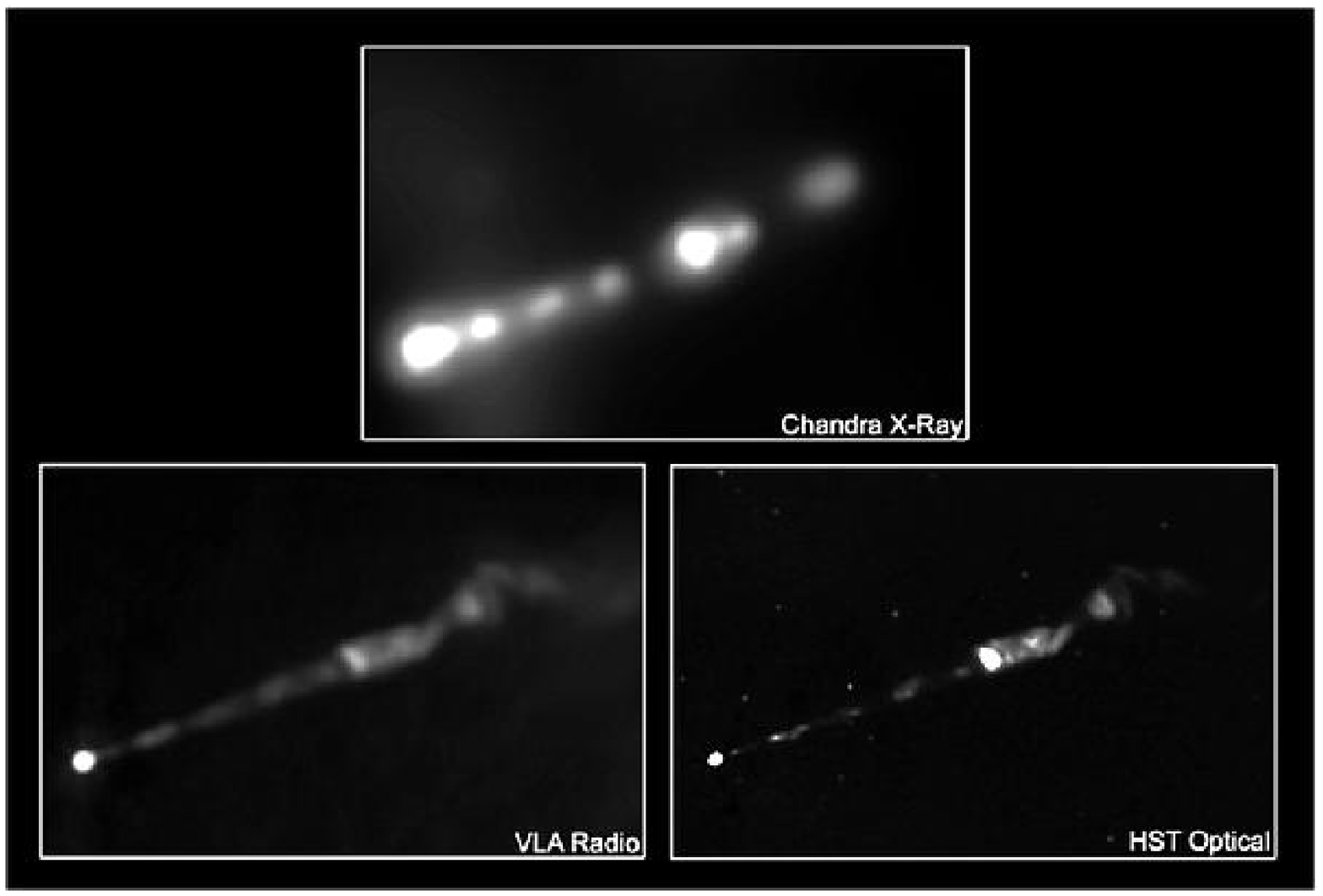} 
\caption{The X-ray jet emanating from the nucleus of the elliptical galaxy M87
as seen in three wavelength bands. Credits: X-ray: NASA/CXC/MIT/H. Marshall
et al. Radio: F. Zhou, F.Owen (NRAO), J.Biretta (STScI) Optical:
NASA/STScI/UMBC/E.Perlman et al. (2001).
\label{f:m87}}
\end{center}
\end{figure}

Figure~\ref{f:m87}\ illustrates multiwavelength observations of the jets from
active galaxies. 
The \chandra\ X-ray image (Marshall et al. 2002) shows an irregular, knotty
structure
similar to that seen at radio and optical (Perlman et al. 2001) wavelengths. 
However, the knots near the central core are much brighter in X-rays

\begin{figure}
\begin{center} 
\epsfysize=8cm
\epsfbox{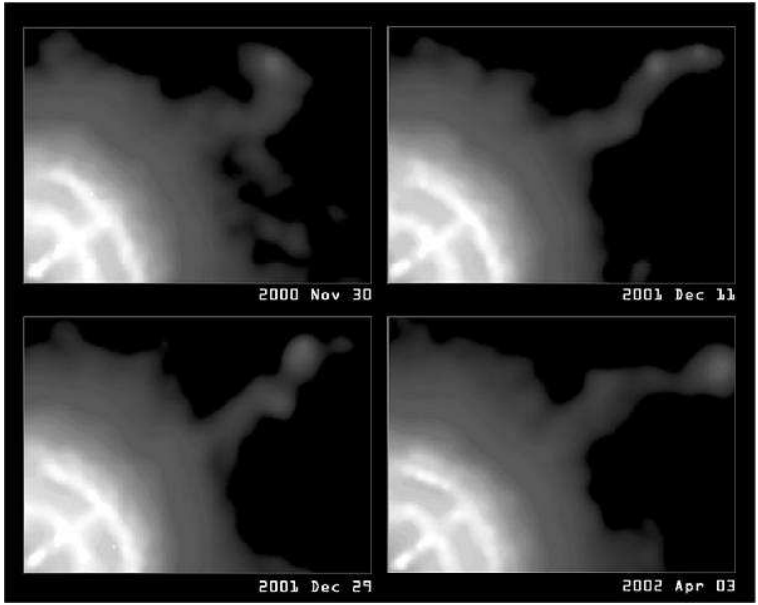} 
\caption{ Four observations of the Vela Pulsar and its outer jet. 
Each image is 1.6 x 1.2 arcmin  Courtesy NASA/CXC/PSU/G. Pavlov et al.
\label{f:vela}}
\end{center}
\end{figure}

The jet phenomenon now appears to be ubiquitous in astronomical settings,
especially with regards to X-ray emission. 
One of the most interesting recent \chandra\ discoveries has been the series of
observations of the outer jet of the Vela pulsar (Pavlov et al. 2003), a few of
which
are illustrated in Figure~\ref{f:vela} where we see the jet, always confined to
a narrow segment, but moving about at velocities of about 0.3-0.5 c.

\begin{figure}
\begin{center} 
\epsfysize=8cm
\epsfbox{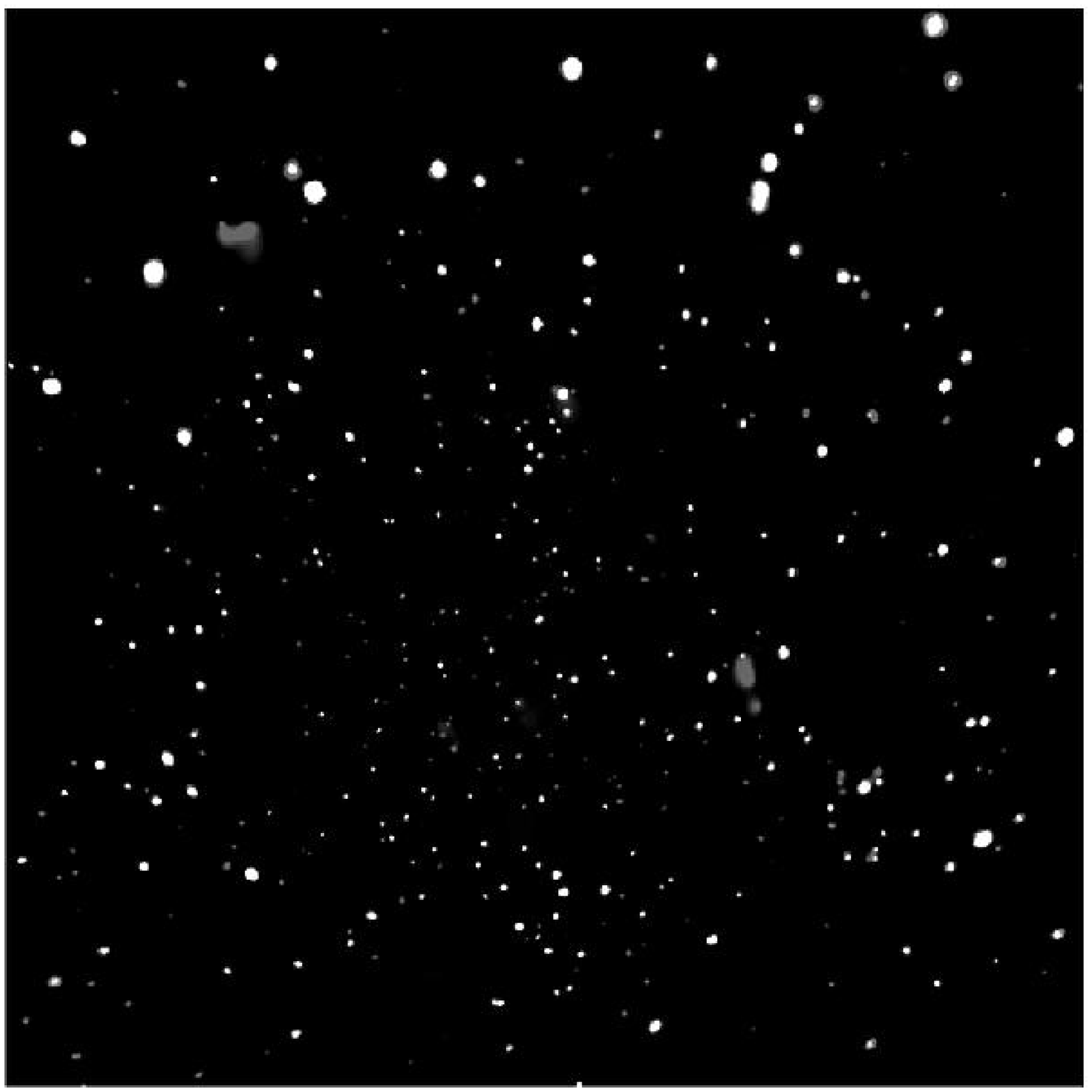} 
\caption{Two-million-second image of the \chandra\ Deep Field North. Courtesy
NASA/CXC/PSU/D.M. Alexander, F.E. Bauer, W.N. Brandt et al.
\label{f:cdfn}}
\end{center}
\end{figure}

No discussion of data taken with the Observatory is complete without a mention
of the deep surveys. 
These are deep exposures of particular regions of the sky to study the
populations of the objects detected, especially the faintest ones. 
This work is an outgrowth of the study the diffuse X-ray background, the nature
of which had been a puzzle for nearly 40 years, although the lack of distortion
of the spectrum of the Cosmic Microwave Background placed a strong upper limit
to the possibility of a truly diffuse component (Mather et al. 1990).
Observations with ROSAT at energies below 2 keV made a major step in resolving a
significant fraction (70-80\%) into discrete objects (Hasinger et al. 1998).
Currently two long exposures have been accomplished with the \chandra\ X-ray
Observatory - the \chandra\ Deep Field North (Alexander et al. 2003) depicted in
Figure~\ref{f:cdfn} with 2-Ms of exposure, and the \chandra\ deep field
south (Giacconi et al. 2001) with 1-Msec. 
These surveys have extended the study of the background to flux levels more than
an order of magnitude fainter than previously in the 0.5-2.0 keV band and have
resolved over 90\% of the background into a variety of discrete sources. 
The largest uncertainty in establishing the fraction is now in the knowledge of
the total level of the background itself.

\section{Conclusions and Acknowledgements}
Designed for a minium of three years of operation, the \chandra\ X-ray
Observatory has, in fact, been operating successfully for more than four and
one-half years.
The instrumentation is well understood and there have been few performance
surprises once the Observatory was comissioned for use.
The number of new discoveries has been outstanding and the Observatory has more
than lived up to its promise. 
The Observatory has had, and will continue to have, a profound influence on our
understanding of the universe, its constituents, and the physical processes
that take place within it.

We acknowledge the major contributions to the success of the Observatory and to
this paper by the scientists and engineers associated with the instrument
teams, the NASA Project at MSFC, the CXC and its Director H. Tananbaum,
and the various industrial contractor organizations.
TLA, RAC and JJD were supported by NASA contract NAS8-39073.
Finally, we acknowlege the tremendous contribution of Leon Van Speybroeck to the
\chandra\ Project and we mourn his passing. 

\section{ WORLD-WIDE WEB SITES}
The following lists several \chandra-related sites on the World-Wide Web:

http://chandra.harvard.edu/: \chandra\ X-ray Center (CXC), operated for NASA by
the Smithsonian Astrophysical Observatory.

http://wwwastro.msfc.nasa.gov/xray/axafps.html: \chandra\ Project Science, at
the
NASA Marshall Space Flight Center.

http://hea-www.harvard.edu/HRC/: \chandra\ High-Resolution Camera (HRC) team, at
the Smithsonian Astrophysical Observatory (SAO).

http://www.astro.psu.edu/xray/axaf/axaf.html: Advanced CCD Imaging Spectrometer
(ACIS) team at the Pennsylvania State University (PSU).

http://acis.mit.edu/: Advanced CCD Imaging Spectrometer (ACIS) team at the 
Massachusetts Institute of Technology.

http://www.sron.nl/missions/Chandra: \chandra\ Low-Energy Transmission Grating
(LETG) team at the Space Research Institute of the Netherlands.

http://wave.xray.mpe.mpg.de/axaf/: \chandra\ Low-Energy Transmission
Grating (LETG) team at the Max-Planck Instit\"ut f\"ur extraterrestrische Physik
(MPE).

http://space.mit.edu/HETG/: \chandra\ High-Energy Transmission Grating (HETG)
team, at the Massachusetts Institute of Technology

http://ifkki.kernphysik.uni-kiel.de/soho: EPHIN particle detector.

\end{document}